\pdfoutput=1
\documentclass[aps,prd,amsmath,floats,floatfix, onecolumn,
superscriptaddress,nofootinbib,showpacs,notitlepage]{revtex4-1}

\usepackage[T1]{fontenc}
\usepackage[utf8]{inputenc}
\usepackage{lmodern}

\usepackage{verbatim}

\usepackage[dvipsnames, usenames]{xcolor}
\definecolor{linkcolor}{rgb}{0.0,0.3,0.5}
\usepackage[hypertexnames=false, unicode, colorlinks=true, linkcolor=linkcolor,
citecolor=linkcolor, filecolor=linkcolor,urlcolor=linkcolor,
pdfusetitle]{hyperref}

\usepackage[all]{hypcap}
\usepackage{graphicx}
\usepackage{xspace}
\usepackage{amssymb}
\usepackage{mathrsfs}
\usepackage[normalem]{ulem} 
\usepackage{bm} 
\usepackage{enumitem,amssymb}
\usepackage[caption=false]{subfig}
\usepackage{orcidlink}

\usepackage{microtype}

\usepackage[english]{babel}
\usepackage{blindtext}

\graphicspath{%
  {figs/}%
}

\DeclareMathAlphabet{\mathpzc}{OT1}{pzc}{m}{it}

\newcommand{\wtR}{R_{\rm wt}}
\newcommand{\apjl}{Astrophys. J. Lett.}
\newcommand{\etal}{\textit{et al.\ }}

\newlist{todolist}{itemize}{2}
\setlist[todolist]{label=$\square$}

\begin{document}

\title{Einstein-Klein-Gordon system via Cauchy-characteristic evolution: Computation of memory and ringdown tail}

\newcommand{\Cornell}{\affiliation{Cornell Center for Astrophysics
    and Planetary Science, Cornell University, Ithaca, New York 14853, USA}}
\newcommand\CornellPhys{\affiliation{Department of Physics, Cornell
    University, Ithaca, New York 14853, USA}}
\newcommand\CornellParticlePhys{\affiliation{Laboratory for Elementary Particle Physics, Cornell University, Ithaca, New York 14853, USA}}
\newcommand\Caltech{\affiliation{TAPIR 350-17, California Institute of
    Technology, 1200 E California Boulevard, Pasadena, CA 91125, USA}}
\newcommand{\AEI}{\affiliation{Max Planck Institute for Gravitational Physics
    (Albert Einstein Institute), Am M\"uhlenberg 1, Potsdam 14476, Germany}} %
\newcommand{\UMassD}{\affiliation{Department of Mathematics,
    Center for Scientific Computing and Visualization Research,
    University of Massachusetts, Dartmouth, MA 02747, USA}}
\newcommand\Olemiss{\affiliation{Department of Physics and Astronomy,
    The University of Mississippi, University, MS 38677, USA}}
\newcommand{\Bham}{\affiliation{School of Physics and Astronomy and Institute
    for Gravitational Wave Astronomy, University of Birmingham, Birmingham, B15
    2TT, UK}}
\newcommand{\Perimeter}{\affiliation{Perimeter Institute for Theoretical Physics, Waterloo, ON N2L2Y5, Canada}}

\author{Sizheng Ma \orcidlink{0000-0002-4645-453X}}
\email{sma2@perimeterinstitute.ca }
\Perimeter

\author{Kyle C. Nelli \orcidlink{0000-0003-2426-8768}}
\Caltech

\author{Jordan Moxon \orcidlink{0000-0001-9891-8677}}
\Caltech

\author{Mark A. Scheel \orcidlink{0000-0001-6656-9134}}
\Caltech

\author{Nils Deppe \orcidlink{0000-0003-4557-4115}}
\CornellParticlePhys
\CornellPhys
\Cornell

\author{Lawrence E.~Kidder\orcidlink{0000-0001-5392-7342}}
\Cornell

\author{William Throwe \orcidlink{0000-0001-5059-4378}}
\Cornell

\author{Nils L.~Vu \orcidlink{0000-0002-5767-3949}}
\Caltech

\hypersetup{pdfauthor={Ma et al.}}

\date{\today}

\begin{abstract}
Cauchy-characteristic evolution (CCE) is a powerful method for accurately extracting gravitational waves at future null infinity. In this work, we extend the previously implemented CCE system within the numerical relativity code SpECTRE by incorporating a scalar field. This allows the system to capture features of beyond-general-relativity theories. We derive scalar contributions to the equations of motion, Weyl scalar computations, Bianchi identities, and balance laws at future null infinity. Our algorithm, tested across various scenarios, accurately reveals memory effects induced by both scalar and tensor fields and captures Price's power-law tail ($u^{-l-2}$) in scalar fields at future null infinity, in contrast to the $t^{-2l-3}$ tail at future timelike infinity.
\end{abstract}

\maketitle


\section{Introduction}
\label{sec:introduction}
The  LIGO, Virgo, and KAGRA collaboration \cite{TheLIGOScientific:2014jea,TheVirgo:2014hva,Somiya:2011np} has detected about 100  gravitational wave (GW) events \cite{TheLIGOScientific:2016pea, LIGOScientific:2018mvr, LIGOScientific:2020ibl, LIGOScientific:2021djp}, providing us with unprecedented opportunities to explore the strong gravitational fields surrounding compact binary systems. This marks a new era in testing general relativity (GR)~\cite{TheLIGOScientific:2016pea, LIGOScientific:2016lio, Will:2014kxa, Yunes:2013dva, Yunes:2016jcc, Berti:2015itd, LIGOScientific:2018dkp, Krolak:1995md, Yagi:2009zz, Ma:2019rei, Carson:2019fxr, Sampson:2014qqa, Scharre:2001hn, Will:2004xi, Berti:2005qd, Berti:2004bd, Yagi:2009zm, Arun:2012hf, Cardoso:2011xi, Yunes:2011aa, Berti:2012bp, Tuna:2022qqr}. 

To rigorously test GR, it is crucial to have precise GW predictions for both GR and alternative theories of gravity. Numerical relativity (NR) stands as the only {\it ab initio} method capable of producing complete inspiral-merger-ringdown waveforms to achieve this goal. Although two decades have passed since the first numerical simulation of a binary black hole (BBH) merger in GR \cite{Pretorius:2005gq}, evolving modified theories of gravity remains challenging, as reviewed in \cite{Ripley:2022cdh}. The primary difficulty lies in the potential modification of the principal part of the Einstein equations, \textcolor{black}{rendering the continuum equations unable to admit a well-posed initial value problem}. Several approaches have been proposed to address this issue. The first approach involves directly formulating well-posed schemes for evolving the full set of evolution equations in certain specific theories, such as weakly-coupled Horndeski gravity theories (e.g. Einstein-scalar-Gauss-Bonnet) \cite{East:2020hgw,East:2021bqk,East:2022rqi,Corman:2022xqg,Corman:2024vlk} based on the modified generalized harmonic formulation \cite{Kovacs:2020pns,Kovacs:2020ywu}, as well as Damour-Esposito-Farèse scalar-tensor theories \cite{Healy:2011ef,Barausse:2012da, Shibata:2013pra, Bezares:2021dma,Ma:2023sok}. The second approach is to expand the equations perturbatively in the coupling constant and solve them order by order. This order reduction scheme has been applied to the evolution of dynamical Chern-Simons \cite{Okounkova:2022grv,Okounkova:2017yby,Okounkova:2019zjf} and scalar Gauss-Bonnet gravity \cite{Witek:2018dmd,Okounkova:2020rqw}. The final treatment introduces auxiliary variables to fix high-energy (short length/timescale) degrees of freedom, while leaving low-energy parts unchanged, see \cite{Cayuso:2017iqc,Franchini:2022ukz,Lara:2024rwa}. Recently, Corman \etal \cite{Corman:2024cdr} compared the results of these three approaches under various physical setups. 

The progress mentioned above primarily centers on the Cauchy formalism, where the equations of motion are formulated as initial boundary value problems. Since it is not feasible to simulate an infinitely large space, the spatial Cauchy domain is typically truncated at a finite distance from the source, which prevents direct access to GWs at future null infinity. One approach to address this limitation is to measure waveform quantities at finite radii within the Cauchy domain and then extrapolate them to infinity \cite{Iozzo:2020jcu}. However, this extrapolation does not guarantee that the Einstein equations are satisfied, making it fail to capture memory effects \cite{Mitman:2024uss}. A more rigorous solution is to use Cauchy-characteristic evolution (CCE) and Cauchy-characteristic matching (CCM)\footnote{\textcolor{black}{Another method is hyperboloidal slicing, see~\cite{Peterson:2024bxk, PanossoMacedo:2024nkw} and references therein.} } \cite{Winicour:2008vpn}, which can faithfully extract GWs at future null infinity up to Bondi-Metzner-Sachs freedom. 
Recently, a CCE \cite{Moxon:2020gha,Moxon:2021gbv} and a CCM \cite{Ma:2023qjn} algorithm for GR have been built in a NR code \texttt{SpECTRE} \cite{deppe_2024_10967177}. \textcolor{black}{Well-posedness analysis
of CCE and CCM in Bondi gauges can be found in \cite{Frittelli:2004pk,Giannakopoulos:2020dih,Giannakopoulos:2023zzm,Gundlach:2024xmo}}.
However, a robust CCE/CCM code for beyond-GR theories is still missing.

In this paper, we take a step to let the SpECTRE CCE system capture beyond-GR features. Specifically, we choose to include a (massless) scalar field, which serves as an important ingredient in theories such as Horndeski \cite{Horndeski:1974wa} and dynamical Chern-Simons gravity \cite{Jackiw:2003pm, Alexander:2009tp}. We will be testing the accuracy of our code and demonstrating its ability to reveal ringdown tails \cite{PhysRevD.5.2419,PhysRevD.5.2439} in the scalar field, as well as the memory effects sourced by the scalar field.  

This paper is organized as follows. In Sec.~\ref{sec:KG_CCE}, we outline the details of CCE for an Einstein-Klein-Gordon system. In particular, we show how the equations of motion, Weyl scalar computations, and Bianchi identities are modified in the presence of a scalar field. Next in Sec.~\ref{sec:memory_effect}, we focus on the memory effects sourced by the scalar field and the corresponding balance laws. We then test our code by evolving scalar fields on two prescribed spacetime backgrounds (Sec.~\ref{sec:characteristic_tests}) and performing the full CCE procedure to compute tails and memory effects (Sec.~\ref{sec:cce_application}). Finally, we summarize the
results in Sec.~\ref{sec:conclusion}. Throughout this paper, complex conjugates are represented by overlines.
\section{Einstein-Klein-Gordon system via CCE}
\label{sec:KG_CCE}
The Einstein-Klein-Gordon action offers the most basic description for a system containing both scalar and tensor fields. For instance, after performing a conformal
transformation \cite{Damour:1992we}, Damour-Esposito-Farèse scalar-tensor theories \cite{bergmann1968comments,Wagoner:1970vr,Damour:1992we} in the \emph{Einstein frame} are governed by:
\begin{gather}
    S = \int \!  d^4x \sqrt{-g} \! \left [ \frac{R}{16\pi}
        -\! \frac{1}{2} \nabla_\mu \psi \, \nabla^\mu \psi + V(\psi)
        \right] \!  ,
\label{eq:einstein_action}
\end{gather}
where the real-valued scalar field $\psi$ is minimally coupled to the metric. The action leads to 
\begin{align}
    &R_{\mu\nu}=8\pi \nabla_\mu \psi \nabla_\nu \psi, && \Box \psi + \frac{\partial V}{\partial\psi} =0. \label{eq:box_psi}
\end{align}
Here, we see that the Einstein field equations obtain a source term, while the scalar field obeys the Klein-Gordon (KG) equation. In our following discussions, we assume $V(\psi)=0$ for simplicity (rendering $\psi$ massless), though these discussions can be easily extended to a complex-valued massive scalar field with an arbitrary potential. 

\subsection{Equations of motion}
\label{subsec:characteristic_equations_of_motion}
By adopting the Bondi-Sachs metric of an asymptotically flat spacetime \cite{1962RSPSA.269...21B,Barnich:2009se,Flanagan:2015pxa}
\begin{align}
    ds^2=&-e^{2\beta}(rW+1) du^2-2e^{2{\beta}}d{u}d{r}+r^2h_{AB}(dx^A-U^Adu)(dx^B-U^Bdu), \label{eq:BS_metric}
\end{align}
with $x^A=(\theta,\phi)$ standing for angular coordinates, the Einstein field equations can be converted to a set of hypersurface equations \cite{Barreto:2004fn,Tahura:2020vsa}
\begin{equation}
\begin{aligned}
    &\partial_r\beta=S_\beta(J)+2\pi r (\partial_r\psi)^2, 
    \qquad  \partial_r(r^2Q)=S_Q(J,\beta)+16\pi r^2 \partial_r\psi\eth \psi, \qquad \partial_rU=S_U(J,\beta,Q), \\
     &\partial_r(r^2W)=S_{W}(J,\beta,Q,U)+2\pi e^{2\beta}\left[J\left(\bar{\eth}\psi\right)^2+\bar{J}\left(\eth\psi\right)^2-2K\bar{\eth}\psi\eth\psi\right], \\
     &\partial_r(rH)+L_H(J,\beta,Q,U,W)H+L_{\bar{H}}(J,\beta,Q,U,W)\bar{H}=S_{H}(J,\beta,Q,U,W)+4\pi\frac{e^{2\beta}}{r} \left(\eth\psi\right)^2, \label{eq:metric_field_equation}
\end{aligned}
\end{equation}
with $U=U^Aq_A$, $Q=r^2 e^{-2\beta}q^Ah_{AB}\partial_rU^B$, $H=\partial_u J$, and $q^A=(-1,-i\csc\theta)$. The expressions of $S_\beta,\, S_Q,\, S_U,\, S_W,\, S_H,\,L_H$, and $L_{\bar{H}}$ as functions of $(J,\beta,Q,U,W)$ can be found in Sec.~IV of \cite{Moxon:2020gha}. The spin-weighted derivative operators $\eth$ and $\bar{\eth}$ are defined to be 
\begin{align}
    &\eth\psi=q^AD_A\psi, && \bar{\eth}\psi=\bar{q}^AD_A\psi \label{eq:def_eth}
\end{align}
where $D_A$ is the angular covariant derivative compatible with the unit sphere metric. On the other hand, the KG equation for the scalar field $\psi$ becomes \cite{Barreto:2004fn}
\begin{align}
    2r\partial_r\Pi+2\Pi =\frac{1}{r}\partial_r\left[r^2(rW+1)\partial_r\psi\right]+\frac{e^{2\beta}}{r}\left(N_{\psi 1}-N_{\psi 2} +N_{\psi 3}\right)-N_{\psi 5} -\frac{r}{2}N_{\psi 4} , \label{eq:evolution_psi_characteristic}
\end{align}
where we have defined an auxiliary variable $\Pi=\partial_u\psi$, and 
\begin{equation}
  \begin{aligned}
    & N_{\psi 1}= 2K\,{\rm Re}\left( \eth\beta\bar{\eth}\psi\right) +K\bar{\eth}\eth\psi, \qquad N_{\psi 2}={\rm Re} \left( \bar{J}\eth\eth\psi+2 \bar{\eth}\beta \bar{\eth}\psi J+ \bar{\eth}J\bar{\eth}\psi\right), \qquad N_{\psi 3}={\rm Re} \left( \eth K\bar{\eth}\psi \right), \\
    &N_{\psi 4}=2{\rm Re} \left(\bar{\eth}\psi  \partial_rU+ \eth \bar{U}\partial_r\psi+2 \bar{U}\eth\partial_r\psi\right), \qquad N_{\psi 5}=2{\rm Re} \left( U \Bar{\eth}\psi\right).
\end{aligned}  
\end{equation}

In Eq.~\eqref{eq:metric_field_equation}, the presence of $\psi$ does not alter the left-hand side of the evolution equations, making their hierarchical structure unchanged: the right-hand side of the $\beta$ equation solely comprises $J,\psi$ and their radial derivatives, the right-hand side of the $Q$ equation comprises solely $J,\psi,\beta$ and their derivatives, and likewise for the other evolved variables. To evolve the system, we provide initial data for $J$ and $\psi$ on the first $u={\rm const}$ null slice, then radially integrate Eqs.~\eqref{eq:metric_field_equation} and \eqref{eq:evolution_psi_characteristic} by order to determine $\beta,Q,U,W,H$, and finally $\Pi$. Following this, we advance $J$ and $\psi$ to the subsequent null slice based on the values of $H=\partial_uJ$ and $\Pi=\partial_u\psi$.

\subsection{SpECTRE CCE system}
\label{subsec:spectre_cce}
A spectral CCE system has been implemented for GR in \texttt{SpECTRE} \cite{Moxon:2020gha,Moxon:2021gbv}. To facilitate spectral implementations, the authors introduced numerically adapted coordinates $(\Breve{u},\Breve{y},\Breve{x}^{\Breve{A}})$:
\begin{align}
    \Breve{u}=u, \quad \Breve{y}=1-\frac{2\wtR}{r}, \quad \Breve{\theta}=\theta, \quad \Breve{\phi}=\phi, \label{eq:numerically_adapted_coordinates}
\end{align}
where $\wtR$ is the Bondi radius of a worldtube for CCE, and it is not to be confused with the Ricci scalar $R$. On a $\Breve{u}={\rm const}$ null slice, the simulation domain spans radially from the worldtube at $\Breve{y}=-1\,(r=\wtR)$ to future null infinity at $\Breve{y}=1\,(r=\infty)$. With this new coordinate system, the time $(\Breve{u})$ derivative of $J$, $\Breve{H}=\partial_{\Breve{u}}J$, is related to the original one $H$, via \cite{Moxon:2020gha}
\begin{align}
    &\Breve{H}=H+(1-\Breve{y})\frac{\partial_{\Breve{u}}\wtR}{\wtR}\partial_{\Breve{y}}J.
\end{align}
The second term arises as the Jacobian of the coordinate transformation. Similarly, the time derivative of the scalar field, $\Breve{\Pi}=\partial_{\Breve{u}}\psi$, transforms in the same way: 
\begin{align}
    &\Breve{\Pi}=\Pi+(1-\Breve{y})\frac{\partial_{\Breve{u}}\wtR}{\wtR}\partial_{\Breve{y}}\psi. \label{eq:breve_pi_pi}
\end{align}
The values of other variables are not affected by the coordinate transformation, namely $\Breve{F}=F$, for all $F\in \{\psi,J,\beta,Q,U,W\}$.

Inserting Eq.~\eqref{eq:breve_pi_pi} into Eq.~\eqref{eq:evolution_psi_characteristic} yields the hypersurface equation for $\Breve{\Pi}$:
\begin{align}
    &(1-\Breve{y})\partial_{\Breve{y}}\Breve{\Pi}+\Breve{\Pi}=-{\rm Re}\,\left( \Breve{U} \Bar{\eth}\Breve{\psi}\right)+(1-\Breve{y})\left[\Lambda_{\Breve{\Pi}}+\frac{e^{2\Breve{\beta}}}{4\wtR}\left(\Breve{N}_{\psi 1}-\Breve{N}_{\psi 2} +\Breve{N}_{\psi 3}\right)\right], \label{eq:hypersurface_eq_for_pi}
\end{align}
with
\begin{align}
    &\Lambda_{\Breve{\Pi}}=\frac{1}{2}(1-\Breve{y})\partial_{\Breve{y}} \Breve{W}\partial_{\Breve{y}}\Breve{\psi}+\frac{(1-\Breve{y})^2}{4\wtR}\partial_{\Breve{y}}^2\Breve{\psi}+\frac{\partial_{\Breve{u}}\wtR}{\wtR}(1-\Breve{y})\partial_{\Breve{y}}^2\Breve{\psi}+\frac{1}{2}(1-\Breve{y})\Breve{W}\partial_{\Breve{y}}^2\Breve{\psi}+\frac{1}{2}\Breve{W}\partial_{\Breve{y}}\Breve{\psi} \notag \\
    &-\frac{1}{2}{\rm Re}\,\left(\bar{\eth}\Breve{\psi}\partial_{\Breve{y}} \Breve{U}+\eth \bar{U}\partial_{\Breve{y}}\Breve{\psi}+2\Breve{U}\bar{\eth}\partial_{\Breve{y}}\Breve{\psi} +2\frac{\eth \wtR}{\wtR}\bar{\Breve{U}}\partial_{\Breve{y}}\Breve{\psi}\right).\label{eq:rhs_lambda_for_pi}
\end{align}
Here we have used two identities:
\begin{align}
    &\eth\partial_r=\frac{(1-\Breve{y})^2}{2\wtR}\left(\eth \partial_{\Breve{y}} +\frac{\Breve{\eth}\wtR}{\wtR}\partial_{\Breve{y}}\right), & \partial_r=\frac{(1-\Breve{y})^2}{2\wtR} \partial_{\Breve{y}}. \label{eq:two_identities_coordinate_transform}
\end{align}
Notice that in Eqs.~\eqref{eq:hypersurface_eq_for_pi}, \eqref{eq:rhs_lambda_for_pi}, and \eqref{eq:two_identities_coordinate_transform} the spin-weighted derivative $\eth$ is still evaluated with respect to the original angular system $x^A$. It is related to the numerically adapted version $\Breve{\eth}$ by the following relation \cite{Moxon:2020gha}:
\begin{align}
    \eth\Breve{\psi}=\Breve{\eth}\Breve{\psi}-(1-\Breve{y})\frac{\Breve{\eth}\wtR}{\wtR}\partial_{\Breve{y}}\Breve{\psi}.
\end{align}
To compute $\Breve{\eth}\Breve{\psi}$, we follow \cite{Moxon:2021gbv} and use {\sc libsharp} routines \cite{Reinecke:2013, Libsharp} to obtain a modal decomposition of $\Breve{\psi}$ in terms of spin-weighted spherical harmonics; subsequently, we apply the differential matrix.

The transformation of the metric sector in Eq.~\eqref{eq:metric_field_equation} has been extensively discussed in \cite{Moxon:2021gbv}. Here we simply present additional terms contributed by $\Breve{\psi}$ 
\begin{equation}
\label{eq:eom_metric}
\begin{aligned}
    &\partial_{\Breve{y}}\Breve{\beta}=S_{\Breve{\beta}}(\Breve{J})+2\pi (1-\Breve{y})(\partial_{\Breve{y}}\Breve{\psi})^2, \\
    &(1-\Breve{y})\partial_{\Breve{y}}\Breve{Q}+2\Breve{Q}=S_{\Breve{Q}}(\Breve{J},\Breve{\beta}) +16\pi (1-\Breve{y}) \eth\Breve{\psi}\partial_{\Breve{y}}\Breve{\psi},\\
    &(1-\Breve{y})\partial_{\Breve{y}}\Breve{W}+2\Breve{W}=S_{\Breve{W}}(\Breve{J},\Breve{\beta},\Breve{Q},\Breve{U})+\pi e^{2\Breve{\beta}} \frac{(1-\Breve{y})}{\wtR}\left(\Breve{J}(\bar{\eth}\Breve{\psi})^2+\bar{\Breve{J}}({\eth}\Breve{\psi})^2-2\Breve{K}\eth\Breve{\psi}\bar{\eth}\Breve{\psi}\right),\\
    &(1-\Breve{y})\partial_{\Breve{y}}\Breve{H}+\Breve{H}+L_{\Breve{H}}(\Breve{J},\Breve{\beta},\Breve{Q},\Breve{U},\Breve{W}) \Breve{H} +L_{\bar{\Breve{H}}}(\Breve{J},\Breve{\beta},\Breve{Q},\Breve{U},\Breve{W})\bar{\Breve{H}}=S_{\Breve{H}}(\Breve{J},\Breve{\beta},\Breve{Q},\Breve{U},\Breve{W})+\frac{2\pi (1-\Breve{y}) e^{2\Breve{\beta}}}{\wtR}\left(\eth\Breve{\psi}\right)^2.
\end{aligned}
\end{equation}

\subsection{Worldtube transformation}
\label{subsec:summary_wt_transformation}
In order to integrate the hypersurface equation for $\Breve{\Pi}$ in Eq.~\eqref{eq:hypersurface_eq_for_pi}, we need to supply a boundary condition at a worldtube $\wtR$. Normally it comes from a separate Cauchy evolution of the KG equation, e.g., see \cite{Scheel:2003vs}. Since the two simulation methods adopt different gauges, a worldtube transformation is needed to construct the value of $\Breve{\Pi}$ at the worldtube out of Cauchy variables.

It turns out that the transformation has a concise expression: 
\begin{align}
    &\Breve{\Pi}|_{\rm wt}=\partial_{t^\prime}\psi+{\rm Re}\left(\mathcal{U}^{(0)}\bar{\Breve{\eth}}\psi\right). \label{eq:final_wt_transformation}
\end{align}
Derivation details can be found in Appendix \ref{app:worldtube_transformation}. In Eq.~\eqref{eq:final_wt_transformation}, $\partial_{t^\prime}\psi$ is the Cauchy time derivative of $\psi$; $\bar{\Breve{\eth}}\psi$ is the angular derivative of $\psi$ [see Eq.~\eqref{eq:def_eth}] with respect to the numerically adapted coordinates; and $\mathcal{U}^{(0)}$ is defined in Eq.~(34b) of \cite{Moxon:2020gha}, with a spin weight of 1. Physically speaking, the numerically adapted angular  coordinates $\Breve{x}^{\Breve{A}}$ on the worldtube evolve over time with respect to the Cauchy coordinates, via [Eq.~(28) of \cite{Moxon:2020gha}] 
\begin{align}
    \partial_{t^\prime}\Breve{x}^{\Breve{A}}=-\frac{1}{2}\left(\mathcal{U}^{(0)}\bar{\Breve{q}}^{\Breve{A}}+\bar{\mathcal{U}}^{(0)}\Breve{q}^{\Breve{A}}\right), \label{eq:dtprime_xA_num}
\end{align}
where the quantity $\mathcal{U}^{(0)}$ captures the rate of the coordinate motion. Therefore, $\mathcal{U}^{(0)}$ corresponds to the shift vector pulled back\footnote{Under the embedding map.} to the 3D timelike worldtube. On the other hand, the numerically adapted time $\Breve{u}$ is chosen to flow at the same rate as Cauchy's time $t^\prime$ --- the lapse is manually set to 1 \cite{Moxon:2020gha}. \textcolor{black}{As a result, the worldtube transformation in Eq.~\eqref{eq:final_wt_transformation} represents the Lie derivative of $\psi$ along $\partial_{\Breve{u}}$, expressed in terms of Cauchy coordinates.}

Since $\psi$ is an evolved variable of the characteristic system, we can directly compute the value of $\bar{\Breve{\eth}}\psi$  on the characteristic grid using the {\sc libsharp} routines, without taking information from the Cauchy side. Meanwhile, the spin-weighted variable $\mathcal{U}^{(0)}$ is already available while evolving the GR system. Therefore, in order to perform the worldtube transformation in Eq.~\eqref{eq:final_wt_transformation}, we simply need to communicate $\partial_{t^\prime}\psi$ from the Cauchy to the characteristic system.

After setting the boundary condition $\Breve{\Pi}|_{\rm wt}$, we integrate the hypersurface equation in Eq.~\eqref{eq:hypersurface_eq_for_pi} and evolve the scalar field forward in time. On the worldtube, the scalar field computed by the Cauchy system, $\psi_{\rm Cauchy}$, should agree with the one obtained from the characteristic system $\psi_{\rm Char}$. This yields a worltube constraint
\begin{align}
    C_{\rm wt}\equiv\psi_{\rm Char}-\psi_{\rm Cauchy}. \label{eq:wt_constraint}
\end{align}
In actual numerical simulations, the constraint is normally nonvanishing since the two fields are evolved independently; and it should converge to 0 in the continuum limit. Therefore, tracking its
evolution provides a diagnosis of our simulations. In fact, $C_{\rm wt}$ offers a specific examination for the worldtube transformation in Eq.~\eqref{eq:final_wt_transformation}, since the characteristic evolution of  $\psi_{\rm Char}$ is fully controlled by the boundary condition $\Breve{\Pi}|_{\rm wt}$  --- a wrong transformation would lead to a wrong characteristic evolution of $\psi_{\rm Char}$, which drives it away from $\psi_{\rm Cauchy}$.

\subsection{Computation of Weyl scalars at future null infinity}
\label{subsec:weyl_scalars_scri}
The \texttt{SpECTRE} CCE system \cite{Moxon:2020gha,Moxon:2021gbv} uses the Newman-Penrose (NP) formalism to compute  Weyl scalars. The relevant equations are\footnote{Chandrasekhar's \cite{Chandrasekhar:BHs} metric signature $(+{}-{}-{}-)$ differs from ours $(-{}+{}+{}+)$. However, the NP equations are not affected by the convention.}\cite{Chandrasekhar:BHs} 
\begin{subequations}
\label{eq:NR_Weyl_Scalars}
   \begin{align}
    \Psi_0=& D\sigma-\delta\kappa-\sigma(\rho+\bar{\rho}+3\epsilon-\bar{\epsilon})+\kappa(\tau-\bar{\pi}_{\rm NP}+\bar{\alpha}+3\beta_{\rm NP}), \label{eq:NR_Weyl_Scalars_Psi0}\\
    \Psi_1=&D\tau -\Delta \kappa-(\tau+\bar{\pi}_{\rm NP})\rho-(\bar{\tau}+\pi_{\rm NP})\sigma-(\epsilon-\bar{\epsilon})\tau+(3\gamma+\bar{\gamma})\kappa-\Phi_{01}, \\
    \Psi_2=&D\mu -\delta \pi_{\rm NP} -(\bar{\rho}\mu+\sigma\lambda)-\pi_{\rm NP}(\bar{\pi}_{\rm NP}-\bar{\alpha}+\beta_{\rm NP}) +\mu (\epsilon+\bar{\epsilon})+\nu \kappa -2\Lambda, \\
    \Psi_3=&\bar{\delta}\mu-\delta\lambda+\nu(\rho-\bar{\rho})+\pi_{\rm NP}(\mu-\bar{\mu})+\mu(\alpha+\bar{\beta}_{\rm NP})+\lambda (\bar{\alpha}-3\beta_{\rm NP})+\Phi_{21}.
\end{align} 
\end{subequations}
where $D=l^a\nabla_a,\Delta=n^a\nabla_a,\delta=m^a\nabla_a$ are the derivative operators used in the NP formalism. The tetrad $(l^a,n^a,m^a)$ is chosen to be \cite{Moxon:2020gha}
\begin{equation}
\label{eq:CCE_tetrad}
   \begin{aligned}
    l^a=\frac{1}{\sqrt{2}}\partial_r^a,
    \quad n^a=\sqrt{2}e^{-2\beta}\left(\partial_u^a-\frac{V}{2r}\partial_r^a+\frac{1}{2}\bar{U}q^a+\frac{1}{2}U\bar{q}^a\right),
    \quad m^a=-\frac{1}{\sqrt{2}r}\left(\sqrt{\frac{K+1}{2}}q^a-\frac{J}{\sqrt{2(1+K)}}\bar{q}^a\right).
\end{aligned} 
\end{equation}
Compared with their vacuum GR counterparts, the equations gain additional source terms $\Phi_{01},\Lambda$, and $\Phi_{21}$. They are related to the scalar field $\psi$ through \cite{Chandrasekhar:BHs} 
\begin{equation}
    \begin{aligned}
    &\Phi_{01}=\frac{1}{2}R_{ml}=4\pi \delta\psi D\psi,
    \quad \Phi_{21}=\frac{1}{2}R_{n\bar{m}}=4\pi \bar{\delta}\psi \Delta \psi, 
    \quad \Lambda=\frac{1}{24}R=-\frac{1}{12}(R_{ln}-R_{m\bar{m}})=-\frac{2\pi}{3}\left[D\psi\Delta\psi-\delta\psi \bar{\delta}\psi\right].
\end{aligned}
\end{equation}

The asymptotic expansion of these quantities at future null infinity $\mathcal{I}^+$ is of particular interest. Hereafter, we denote the terms in the expansion with the following notation:
\begin{align}
    f=\frac{f^{(n)}}{r^n}+\frac{f^{(n+1)}}{r^{n+1}}+\mathcal{O}\left(\frac{1}{r^{n+2}}\right), \label{eq:asymp_expaonsion_order}
\end{align}
where $f$ represents any variable under investigation, and $n$ is an integer. As for the massless scalar field $\psi$, its value at $\mathcal{I}^+$, namely $\psi^{(0)}$, is a constant and can be set to 0 without changing physical aspects of the system. Therefore, its radial falloff obeys\footnote{\textcolor{black}{Here we assume that the scalar field does not develop logarithmic divergences, as studied in \cite{Gasperin:2024bfc}.}} $\psi\sim r^{-1}$. Substituting the asymptotic behavior of $\psi$ into Eq.~\eqref{eq:CCE_tetrad} leads to
\begin{align}
    D\psi = -\frac{1}{\sqrt{2}}\frac{\psi^{(1)}}{r^2}+\mathcal{O}\left(\frac{1}{r^{3}}\right), \quad \Delta\psi=\sqrt{2}e^{-2\beta^{(0)}}\frac{\Pi^{(1)} }{r}+\mathcal{O}\left(\frac{1}{r^{2}}\right), \quad \delta\psi=-\frac{1}{\sqrt{2}}\frac{\eth\psi^{(1)}}{r^2}+\mathcal{O}\left(\frac{1}{r^{3}}\right),
\end{align}
where we have used the fact that $U^{(0)}=J^{(0)}=0$, $K^{(0)}=1$, and $V\sim r$. The radial falloff rate for $\Phi_{01},\Lambda$, and $\Phi_{21}$ can be obtained accordingly:
\begin{align}
    \Phi_{01}=2\pi\frac{\psi^{(1)}\eth\psi^{(1)}}{r^4}+\mathcal{O}\left(\frac{1}{r^5}\right), \quad \Lambda=\frac{2\pi}{3} e^{-2\beta^{(0)}}\frac{\Pi^{(1)}\psi^{(1)}}{r^3} +\mathcal{O}\left(\frac{1}{r^4}\right), \quad \Phi_{21}=-4\pi e^{-2\beta^{(0)}}\frac{\Pi^{(1)}\bar{\eth}\psi^{(1)}}{r^3}+\mathcal{O}\left(\frac{1}{r^4}\right).
\end{align}
In the code, we compute the value of $\Pi^{(1)},\psi^{(1)}$, and $\eth \psi^{(1)}$ via
\begin{align}
    &\Pi^{(1)}= -2\wtR \left.\left(\partial_{\Breve{y}}\Breve{\Pi}+\frac{\partial_{\Breve{u}}\wtR}{\wtR}\partial_{\Breve{y}}\Breve{\psi}\right)\right|_{\Breve{y}=1}, \quad \psi^{(1)}= -2\wtR \partial_{\Breve{y}}\Breve{\psi}, \quad \eth \psi^{(1)}=-2\left.\left(\frac{\eth\wtR}{\wtR} \partial_{\Breve{y}}\Breve{\psi}+\eth\partial_{\Breve{y}}\Breve{\psi}\right)\right|_{\Breve{y}=1}.
\end{align}
The asymptotic expansions of Wely scalars in GR have been shown in Eq.~(93) of \cite{Moxon:2020gha}. With the presence of $\psi$, their expressions are modified to
\begin{align}
    &\Psi_1^{(4)}=\Psi_1^{(4) \rm GR} -\Phi_{01}^{(4)}, \quad \Psi_2^{(3)}=\Psi_2^{(3) \rm GR} -2\Lambda^{(3)}, \quad \Psi_3^{(2)}=\Psi_3^{(2) \rm GR}. \label{eq:weyl_scalars_asympt}
\end{align}
Notice that the expression for $\Psi_3^{(2)}$ is the same as its GR counterpart, since the additional source term $\Phi_{21}\sim \mathcal{O}(r^{-3})$ decays faster than that of $\Psi_3$. The computation of $\Psi_0$ requires more attention. Although the formula in Eq.~\eqref{eq:NR_Weyl_Scalars_Psi0} is valid geometrically and does not rely on the detail of a gravity theory, the asymptotic expansion in \cite{Moxon:2020gha} [see their Eq.~(93a)] implicitly assumes the vanishing of the stress-energy tensor. Instead, here we directly expand Eq.~\eqref{eq:NR_Weyl_Scalars_Psi0} and obtain:
\begin{align}
    \Psi_0^{(5)}=\beta^{(2)}J^{(1)}+\frac{\bar{J}^{(1)} J^{(1) 2}}{4}-\frac{3}{2}J^{(3)}.
\end{align}
The value of $\beta^{(2)}$ can be obtained by expanding the $rr$ component of Einstein's equations [see Eq.~(2.9c) in \cite{Flanagan:2015pxa} or Eq.~(2.9a) in \cite{Tahura:2020vsa}]
\begin{align}
    &\beta^{(2)}=-\frac{J^{(1)}\bar{J}^{(1)}}{16}-\pi \psi ^{(1)2}. \label{eq:beta_scalars_asympt}
\end{align}
Finally, the Weyl scalar $\Psi_4^{(1)}$ and the Bondi-Sachs News take the same forms as in GR, as shown in Eq.~(93e) and (42) of \cite{Moxon:2020gha}, respectively.

\subsection{Bianchi identities at future null infinity}
\label{subsec:bianchi_identities}
One way to validate our CCE code is to examine Bianchi identities at future null infinity, which establish connections between Weyl scalars $\Psi_{0\ldots4}^{(5)\ldots(1)}$ and the strain $h^{(1)}$. Their expressions in GR can be found in, e.g. \cite{Iozzo:2020jcu}. To extend the identities to our case, we consider the following NP equations \cite{stephani2009exact}:
\begin{subequations}
    \label{eq:NP_equations_Bianchi}
    \begin{align}
    &-\Delta\lambda+\bar{\delta}\nu -\lambda(\mu+\bar{\mu}) - (3\gamma-\bar{\gamma})\lambda+(3\alpha+\bar{\beta}_{\rm NP}+\pi_{\rm NP}-\bar{\tau})\nu -\Psi_4=0, \\
    &-\Delta\Psi_3+\delta\Psi_4+(4\beta_{\rm NP}-\tau)\Psi_4-2(2\mu+\gamma)\Psi_3+3\nu\Psi_2-\bar{\delta}\Phi_{22}+\Delta\Phi_{21}+(\bar{\tau}-2\bar{\beta}_{\rm NP}-2\alpha)\Phi_{22}+2(\gamma+\bar{\mu})\Phi_{21} \notag \\
    &+2\lambda\Phi_{12}-2\nu\Phi_{11}-\bar{\nu}\Phi_{20}=0, \\
    &-\Delta\Psi_2+\delta\Psi_3+2\nu\Psi_1-3\mu\Psi_2+2(\beta_{\rm NP}-\tau)\Psi_3+\sigma\Psi_4 -D\Phi_{22}+\delta\Phi_{21}+2(\bar{\pi}_{\rm NP}+\beta_{\rm NP})\Phi_{21}-2\mu\Phi_{11}-\bar{\lambda}\Phi_{20} \notag  \\
    &+2\pi_{\rm NP}\Phi_{12}+(\bar{\rho}-2\epsilon-2\bar{\epsilon})\Phi_{22}-2\Delta\Lambda =0, \\
    &-\Delta\Psi_1+\delta\Psi_2+\nu\Psi_0+2(\gamma-\mu)\Psi_1-3\tau\Psi_2+2\sigma\Psi_3 +\Delta\Phi_{01}-\bar{\delta}\Phi_{02}+2(\bar{\mu}-\gamma)\Phi_{01}-2\rho\Phi_{12}-\bar{\nu}\Phi_{00}+2\tau \Phi_{11} \notag\\
    &+(\bar{\tau}-2\bar{\beta}_{\rm NP}+2\alpha)\Phi_{02}+2\delta\Lambda =0, \\
    &-\Delta \Psi_0 +\delta \Psi_1 + (4\gamma-\mu)\Psi_0 -2(2\tau+\beta_{\rm NP})\Psi_1+3\sigma\Psi_2 -D\Phi_{02}+\delta\Phi_{01}+2(\bar{\pi}_{\rm NP}-\beta_{\rm NP})\Phi_{01}-2\kappa\Phi_{12}-\bar{\lambda}\Phi_{00}+2\sigma\Phi_{11} \notag \\
    &+(\bar{\rho}+2\epsilon-2\bar{\epsilon})\Phi_{02}=0.
\end{align}
\end{subequations}
where the source terms $\Phi_{22},\Phi_{00},\Phi_{11},\Phi_{02}$, and $\Phi_{20}$ are given by
\begin{equation}
    \begin{aligned}
    &\Phi_{22}=\frac{1}{2}R_{nn}=4\pi (\Delta\psi)^2=8\pi\frac{\dot{\psi}^{(1)\,2}}{r^2}+\mathcal{O}\left(\frac{1}{r^3}\right),     \quad \Phi_{00}=\frac{1}{2}R_{ll}=4\pi (D\psi)^2=2\pi \frac{\psi^{(1)\,2}}{r^4} +\mathcal{O}\left(\frac{1}{r^5}\right), \\
     &\Phi_{11}=\frac{1}{4}(R_{ln}+R_{m\bar{m}})=2\pi(D\psi\Delta\psi+\delta\psi\bar{\delta}\psi)=-2\pi\frac{\psi^{(1)}\dot{\psi}^{(1)}}{r^3} +\mathcal{O}\left(\frac{1}{r^4}\right), \\
    &\bar{\Phi}_{20}=\Phi_{02}=\frac{1}{2}R_{mm}=4\pi (\delta\psi)^2=2\pi \frac{ (\eth\psi^{(1)})^2}{r^4} +\mathcal{O}\left(\frac{1}{r^5}\right). \label{eq:bianchi_source_terms}
\end{aligned}
\end{equation}
In Eq.~\eqref{eq:bianchi_source_terms} we have used 
\begin{align}
    \delta= -\frac{1}{\sqrt{2}r}\eth +\mathcal{O}\left(\frac{1}{r^2}\right), \qquad \Delta = \sqrt{2} \partial_u+\mathcal{O}\left(\frac{1}{r}\right),  \label{eq:delta_eth_expansion}
\end{align}
to obtain the asymptotic expansions. The overhead dots denote the time derivative $\partial_u$. Meanwhile, the Bianchi identities involve NP spin coefficients, whose expressions in terms of Bondi-Sachs variables can be found in Eq.~(86) of \cite{Moxon:2020gha}. Near future null infinity, their asymptotic behaviors read\footnote{Our results are consistent with those in \cite{Adamo:2009vu,Scholtz:2013toa} after switching the convention and performing a tetrad transformation.}
\begin{equation}
    \label{eq:NP_expansion_scri}
    \begin{aligned}
    &J=\frac{\bar{h}^{(1)}}{r}+\mathcal{O}\left(\frac{1}{r^2}\right), \quad \sigma=\frac{\bar{h}^{(1)}}{2\sqrt{2}r^2}+\mathcal{O}\left(\frac{1}{r^3}\right), \quad \mu=-\frac{1}{\sqrt{2}r}+\mathcal{O}\left(\frac{1}{r^2}\right), \quad \beta_{\rm NP}=\frac{\cot\theta}{2\sqrt{2}r} +\mathcal{O}\left(\frac{1}{r^2}\right),\\
    &\lambda^{(1)}=\frac{1}{\sqrt{2}}\dot{h}^{(1)}+\mathcal{O}\left(\frac{1}{r^2}\right), \quad \rho = -\frac{1}{\sqrt{2}r}, \quad \kappa=0, \quad \tau \sim \mathcal{O}\left(\frac{1}{r^2}\right), \quad \epsilon \sim \mathcal{O}\left(\frac{1}{r^2}\right), \quad \pi_{\rm NP} \sim \mathcal{O}\left(\frac{1}{r^2}\right),\\
    & \nu \sim \mathcal{O}\left(\frac{1}{r^2}\right), \quad \gamma \sim \mathcal{O}\left(\frac{1}{r^2}\right).
\end{aligned}
\end{equation}
Plugging Eqs.~\eqref{eq:bianchi_source_terms}, \eqref{eq:delta_eth_expansion}, and \eqref{eq:NP_expansion_scri} into Eq.~\eqref{eq:NP_equations_Bianchi}, we obtain the following Bianchi identities at future null infinity
\begin{subequations}
\label{eq:Bianchi_identities_scri_spectre}
\begin{align}
    &\Psi_4^{(1)}=-\ddot{h}^{(1)}, \\
    &\dot{\Psi}_3^{(2)}=-\frac{1}{2}\eth\Psi_4^{(1)}, \\
    &\dot{\Psi}_2^{(3)}=-\frac{1}{2}\eth\Psi_3^{(2)}+\frac{1}{4}\bar{h}^{(1)}\Psi_4^{(1)} +\frac{8\pi}{3}\dot{\psi}^{(1)\, 2}-\frac{4\pi}{3}\psi^{(1)}\ddot{\psi}^{(1)} ,  \\
    &\dot{\Psi}_1^{(4)}=-\frac{1}{2}\eth\Psi_2^{(3)}+\frac{1}{2}\bar{h}^{(1)}\Psi_3^{(2)} -\frac{8\pi}{3}\dot{\psi}^{(1)}\eth\psi^{(1)}+\frac{4\pi}{3}\psi^{(1)}\eth\dot{\psi}^{(1)} , \label{eq:Bianchi_identities_scri_spectre_psi1} \\
    &\dot{\Psi}_0^{(5)}= -\frac{1}{2}\eth \Psi_1^{(4)}  +\frac{3}{4}\bar{h}^{(1)}\Psi_2^{(3)} +2\pi(\eth\psi^{(1)})^2-\pi\psi^{(1)}\eth\eth\psi^{(1)}-\pi\dot{\bar{h}}^{(1)} \psi^{(1)\,2}-\sqrt{2}\pi\bar{h}^{(1)}\psi^{(1)}\dot{\psi}^{(1)},
\end{align}
\end{subequations}

We will use the $\texttt{PYTHON}$ package $\texttt{scri}$ \cite{scri,mike_boyle_2020_4041972,Boyle:2013nka,Boyle:2014ioa,Boyle:2015nqa} to verify the identities in our following calculations. The package adopts a different convention (hereafter MB) \cite{Boyle:2015nqa,Lehner:2007ip} compared to SpECTRE CCE. The conversion factors read \cite{Iozzo:2020jcu}
\begin{align}
    \Psi_n^{\rm [MB]}=(-\sqrt{2})^{2-n}\Psi_n, \qquad \eth^{\rm [MB]}=\frac{1}{\sqrt{2}}\eth, \qquad \sigma^{(2)\rm MB}=\sqrt{2}\sigma^{(2)}. \label{eq:MB_convention}
\end{align}
Then Eqs.~\eqref{eq:Bianchi_identities_scri_spectre} become
\begin{subequations}
    \begin{align}
    & {\Psi}_4^{(1) \rm[MB]}=-\ddot{\bar{\sigma}}^{(2) \rm[MB]}, \label{eq:Bianchi_identities_scri_scri_psi4}\\
    &\dot{\Psi}_3^{(2) \rm[MB]}=\eth^{\rm [MB]}\Psi_4^{(1)\rm[MB]},\label{eq:Bianchi_identities_scri_scri_psi3} \\
    &\dot{\Psi}_2^{(3)\rm [MB]}=\eth^{\rm [MB]}\Psi_3^{(2)\rm [MB]}+\sigma^{(2)\rm [MB]}\Psi_4^{(1)\rm [MB]} +\frac{8\pi}{3}\dot{\psi}^{(1)\, 2}-\frac{4\pi}{3}\psi^{(1)}\ddot{\psi}^{(1)} , \\
    &\dot{\Psi}_1^{(4)\rm [MB]}=\eth^{\rm [MB]}\Psi_2^{(3)\rm [MB]}+2\sigma^{(2)\rm [MB]}\Psi_3^{(2)\rm [MB]} +\frac{16\pi}{3}\dot{\psi}^{(1)}\eth^{\rm [MB]}\psi^{(1)}-\frac{8\pi}{3}\psi^{(1)}\eth^{\rm [MB]}\dot{\psi}^{(1)} ,  \\
    &\dot{\Psi}_0^{(5)\rm [MB]}= \eth^{\rm [MB]} \Psi_1^{(4)\rm [MB]}  +3\sigma^{(2)\rm [MB]}\Psi_2^{(3)\rm [MB]} +8\pi(\eth^{\rm [MB]}\psi^{(1)})^2-4\pi\psi^{(1)}\eth^{\rm [MB]}\eth^{\rm [MB]}\psi^{(1)}-4\pi\dot{\sigma}^{(2)\rm [MB]}\psi^{(1)\,2} \notag \\
    &-4\pi\sigma^{(2)\rm [MB]}\psi^{(1)}\dot{\psi}^{(1)}.
\end{align}
\label{eq:Bianchi_identities_scri_scri}
\end{subequations}

\section{Scalar-induced memory effects}
\label{sec:memory_effect}
A massless scalar field can propagate to future null infinity \cite{helfer1993null} and leave nontrivial imprints on memory effects. Here we extend the discussions in \cite{Mitman:2020pbt, Mitman:2024uss} and derive the memory contribution from the scalar field. Since most of our discussions in this section adopt only the leading-order asymptotic behavior near future null infinity, we suppress the superscript defined in Eq.~\eqref{eq:asymp_expaonsion_order} for conciseness, unless stated otherwise. To be consistent with previous discussions, we still use the convention of SpECTRE CCE, as opposed to MB [Eq.~\eqref{eq:MB_convention}].

The computation of memory effects can be achieved via the balance laws \cite{Flanagan:2015pxa} [see their Eq.~(2.11a) and (C6)]
\begin{subequations}
    \begin{align}
    &\dot{m}=-4\pi \dot{\psi}^2-\frac{1}{8}N_{AB}N^{AB}+\frac{1}{4}D_AD_B N^{AB}, \label{eq:balance_law_electric}\\
    &\dot{\hat{N}}_A=-6\pi\dot{\psi}D_A\psi+2\pi\psi\partial_A\dot{\psi}+4\pi u D_A \dot{\psi}^2+\frac{u}{8}D_A(N_{BC}N^{BC})-\frac{3}{8}N_{AB}D_CC^{BC}+\frac{3}{8}C_{AB}D_CN^{BC}+\frac{1}{8}N^{BC}D_BC_{AC}\notag \\
    &-\frac{1}{8}C^{BC}D_BN_{AC}+\frac{1}{4}D_BD_AD_CC^{BC}-\frac{1}{4}D^2D^CC_{AC}-\frac{1}{4}uD_AD_BD_CN^{BC}, \label{eq:balance_law_magnetic}
\end{align}
\end{subequations}
where $m=-\frac{1}{2} W^{(2)}$ is the Bondi mass aspect; $\hat{N}_A$ is related to the angular momentum aspect $N_A$ via
\begin{align}
    \hat{N}_A=N_A-uD_Am-\frac{1}{16}D_A(C_{BC}C^{BC})-\frac{1}{4}C_{AB}D_CC^{BC}. \label{eq:Nhat_N_def}
\end{align}
Finally, $C_{AB}$ is the $\mathcal{O}(r^{-1})$ order of the angular metric $h_{AB}$ defined in Eq.~\eqref{eq:BS_metric}, and $N_{AB}=\partial_uC_{AB}$ is the Bondi News tensor.

Following \cite{Flanagan:2015pxa,Mitman:2020pbt}, we decompose $C_{AB}$ into an electric $(\Phi)$ and a magnetic $(\Psi)$ potential 
\begin{align}
    C_{AB}=\left(D_AD_B-\frac{1}{2}q_{AB}D^2\right)\Phi+\epsilon_{C(A}D_{B)}D^C\Psi, \label{eq:decomposition_C_AB_ele_mag}
\end{align}
where $\epsilon_{CA}=\frac{i}{2} q_C\wedge \bar{q}_A$ is the volume form compatible with the unit sphere metric. 

\subsection{Electric memory}
We insert Eq.~\eqref{eq:decomposition_C_AB_ele_mag} into Eq.~\eqref{eq:balance_law_electric} and obtain \cite{Mitman:2020pbt}
\begin{align}
    \Phi=\mathfrak{D}^{-1}\left[m+\int \left(4\pi \dot{\psi}^2+\frac{1}{4}\dot{h}\dot{\bar{h}}\right)du+\alpha(\theta,\phi)\right].
\end{align}
Here $\alpha(\theta,\phi)$ is an arbitrary function on a two-sphere and $\mathfrak{D}\equiv\frac{1}{8}D^2(D^2+2)$ is an angular differential operator. We see that the null (nonlinear) memory obtains an additional contribution from the scalar energy flux $\sim \int \dot{\psi}^2 du$, compared to its GR counterpart \cite{Mitman:2020pbt}. On the other hand, the Bondi mass aspect contributes to the ordinary (linear) memory, and is given by \cite{Scholtz:2013toa}
\begin{align}
    m=-{\rm Re}\left(\Psi_2+\frac{1}{4}\dot{h}\bar{h}\right)-\frac{4\pi}{3}\psi\dot{\psi}.
\end{align}    
Notice that the scalar ordinary memory vanishes in a nonradiative regime $(\dot{\psi}\sim0)$. 

The potential $\Phi$ is related to the electric memory $J^E$ via \cite{Mitman:2020pbt}
\begin{align}
    J^E=\frac{1}{2}\bar{\eth}^2\Phi=J^{ET}_{o}+J^{ET}_{\rm null}+J^{ES}_o+J^{ES}_{\rm null}, \label{eq:JE_decomposition}
\end{align}
where the scalar pieces read
\begin{align}
    J^{ES}_o=\frac{1}{2}\bar{\eth}^2\mathfrak{D}^{-1}\left(-\frac{4\pi}{3}\psi\dot{\psi}\right), \quad 
    J^{ES}_{\rm null}=\frac{1}{2}\bar{\eth}^2\mathfrak{D}^{-1}\left(\int 4\pi \dot{\psi}^2du\right). \label{eq:JES_decomposition_LNL}
\end{align}
The tensor contributions $J^{ET}_o$ and $J^{ET}_{\rm null}$ are the same as GR, and we provide their expressions below for completeness
\begin{align}
    J^{ET}_o=\frac{1}{2}\bar{\eth}^2\mathfrak{D}^{-1}\left[-{\rm Re}\left(\Psi_2+\frac{1}{4}\dot{h}\bar{h}\right)\right], \quad 
    J^{ET}_{\rm null}=\frac{1}{2}\bar{\eth}^2\mathfrak{D}^{-1}\left[\int \frac{1}{4}\dot{h}\dot{\bar{h}} du+\alpha(\theta,\phi)\right]. \label{eq:JET_decomposition_LNL}
\end{align}

\subsection{Magnetic memory}
The magnetic memory can be obtained by projecting both sides of Eq.~\eqref{eq:balance_law_magnetic} with $\epsilon^{AE}D_E$, and the corresponding expression for $\Psi$ in GR is provided in Eq.~(45) of \cite{Mitman:2020pbt}. With the presence of the scalar field, a convenient way to compute $\Psi$ is by replacing $\dot{\hat{N}}=q^A\dot{\hat{N}}_A$ in the GR version with the following combination:
\begin{align}
    \dot{\hat{N}}\to \dot{\hat{N}}+6\pi\dot{\psi}\eth\psi-2\pi\psi\eth\dot{\psi},
\end{align}
which yields
\begin{align}
    D^2\mathfrak{D}\Psi= -{\rm Im}\left[\bar{\eth}\dot{\hat{N}}+8\pi\bar{\eth}\dot{\psi}\eth\psi+\frac{1}{8}\eth\left(3h\bar{\eth}\dot{\bar{h}}-3\dot{h}\bar{\eth}\bar{h}+\dot{\bar{h}}\bar{\eth}h-\bar{h}\bar{\eth}\dot{h}\right)\right]. \label{eq:magnetic_memory_psi}
\end{align}
To obtain $\hat{N}$, we first contract Eq.~\eqref{eq:Nhat_N_def} with $q^A$ and get
\begin{align}
     \hat{N}=N-u\eth m-\frac{1}{8}\eth(h\bar{h})-\frac{1}{4}\bar{h}\eth h.  \label{eq:Nhat_N_projected_def}
\end{align}
In GR, the angular momentum aspect $N$ is solely determined by $\Psi_1$, e.g., see Appendix B of \cite{Mitman:2020pbt}. As for the Einstein-Klein-Gordon system, an extra term shows up. To see this, we use Eqs.~(B2), (B5) and (B6) in \cite{Mitman:2020pbt}, namely
\begin{align}
U^{(3)}=-\frac{1}{3}(Q^{(2)}-\bar{h}\bar{Q}^{(1)}), \qquad U^{(3)}=-\frac{2}{3}N+\frac{1}{8}\eth (h\bar{h})+\frac{1}{2}\bar{h}\eth {h}, \qquad \bar{Q}^{(1)}=\eth h,
\end{align}
together with the asymptotic expansion for $\Psi_1$ in Eq.~\eqref{eq:weyl_scalars_asympt} and the one for $\beta$ in Eq.~\eqref{eq:beta_scalars_asympt}, then we arrive at
\begin{align}
    N=2\Psi_1-2\pi\psi \eth\psi. \label{eq:N_Nhat_Psi1}
\end{align}
Plugging Eqs.~\eqref{eq:Nhat_N_projected_def} and \eqref{eq:N_Nhat_Psi1} into Eq.~\eqref{eq:magnetic_memory_psi}, and by virtue of the Bianchi identity in Eq.~\eqref{eq:Bianchi_identities_scri_spectre_psi1}, we find the scalar contribution to $\Psi$ vanishes identically. The corresponding magnetic memory $J^B=-\frac{i}{2}\bar{\eth}^2\Psi$ reads \cite{Mitman:2020pbt}
\begin{align}
    J^{B}= J^{B}_o+J^{B}_{\rm null}, \label{eq:JM_decomposition}
\end{align}
with
\begin{align}
    J^{B}_o= \frac{i}{2} \bar{\eth}^2\mathfrak{D}^{-1}D^{-2}{\rm Im}\left\{\bar{\eth}\left[-\eth\Psi_2 + \frac{1}{4}(\bar{h}\eth \dot{h}-\dot{\bar{h}}\eth h) \right] \right\}, \quad
    J^{B}_{\rm null}= \frac{i}{2} \bar{\eth}^2\mathfrak{D}^{-1}D^{-2}{\rm Im}\left\{\frac{1}{8}\eth\left(3h\bar{\eth}\dot{\bar{h}}-3\dot{h}\bar{\eth}\bar{h}+\dot{\bar{h}}\bar{\eth}h-\bar{h}\bar{\eth}\dot{h}\right)\right\}. \label{eq:JM_decomposition_LNL}
\end{align}
The expression can be further simplified by replacing the term $\bar{\eth}(\bar{h}\eth \dot{h}-\dot{\bar{h}}\eth h)$ in $J^{B}_o$ with its negative complex conjugate\footnote{Notice that only the imaginary part contributes to $J^B$.}. After combing with $J^B_{\rm null}$, we obtain
\begin{align}
    J^{B}= -\frac{i}{2} \bar{\eth}^2\mathfrak{D}^{-1}D^{-2}{\rm Im}\,\bar{\eth}\eth\left[\Psi_2 +\frac{1}{4} \dot{h}\bar{h}\right]= \frac{i}{8} \bar{\eth}^2\mathfrak{D}^{-1}{\rm Im}\,  \eth^2 h, \label{eq:magnetic_memory_simplification}
\end{align}
where we have used the fact that the Bondi mass aspect is real-valued, namely \cite{Adamo:2009vu}
\begin{align}
    {\rm Im}\, \left(\Psi_{2}+\frac{1}{4}\eth^2 h+\frac{1}{4}\dot{h}\bar{h}\right)=0. \label{eq:imaginary_of_psi2_vanishes}
\end{align}
Notice that the right-hand side of Eq.~\eqref{eq:magnetic_memory_simplification} is linear in $h$, as opposed to the electric memory in Eq.~\eqref{eq:JE_decomposition} that depends on energy fluxes, the balance law for the angular momentum aspect in Eq.~\eqref{eq:balance_law_magnetic} in fact defines a projection operator for the magnetic component of $h$. In particular, Eq.~\eqref{eq:magnetic_memory_simplification} implies that
\begin{align}
    J^B_{lm}\sim h_{lm} - (-1)^m \bar{h}_{l,-m}.
\end{align}
It corresponds to the radiative current moment, see e.g., Eq.~(30) in \cite{Schnittman:2007ij}, which plays an important role in building up gravitational recoils \cite{Schnittman:2007ij,Boyle:2014ioa,Ma:2021znq} and the excitation of quadratic quasinormal modes \cite{Ma:2024qcv,Bourg:2024jme}. Under parity conjugation, $J^B_{lm}$ transforms as follows 
\begin{align}
    J^B_{lm} \to (-1)^{l+1} J^B_{lm},
\end{align}
where we have used Eq.~(C10d) in \cite{Boyle:2014ioa}. Therefore, $J^B_{lm}$ carries odd parity. In the context of Schwarzschild perturbation theory, the dynamics of $J_B$ is described by the Regge-Wheeler equation \cite{Baumgarte_Shaprio:NumRel}. 

\section{Testing the characteristic evolution}
\label{sec:characteristic_tests}
Having outlined CCE for the Einstein-Klein-Gordon system, we now proceed to test our code built in SpECTRE. Given that the metric evolution in GR has been thoroughly examined in \cite{Moxon:2021gbv}, our tests primarily focus on the new features introduced by the scalar field. This involves two main parts: solving the KG equation in Eq.~\eqref{eq:hypersurface_eq_for_pi} and conducting the full CCE procedure for the coupled (scalar+tensor) system. We address these two parts separately in this and the subsequent sections.

Below in this section, we test the implementation of the KG equation by evolving scalar fields on two prescribed spacetime backgrounds (namely without scalar-induced backreaction into the metric sector). We pay particular attention to two aspects: (a) the correct volume integration of the KG equation, and (b) the accurate implementation of the worldtube transformation in  Eq.~\eqref{eq:final_wt_transformation}. For (a), we compare the computed scalar modes at future null infinity to expected (analytical) behaviors. For (b), we monitor the worldtube constraint defined in Eq.~\eqref{eq:wt_constraint}, as discussed in Sec.~\ref{subsec:summary_wt_transformation}.





\subsection{Bouncing Schwarzchild BH} 
Following \cite{Barkett:2019uae,Moxon:2021gbv}, we consider a Schwarzchild BH in an oscillating coordinate system --- a time-dependent transformation is applied to Kerr-Schild coordinates  $\{t_{\rm KS}, x_{\rm KS}, y_{\rm KS}, z_{\rm KS}\}$ \cite{Barkett:2019uae,Moxon:2021gbv}:
\begin{subequations}
\label{eq:bouncing_bh_coordinates}
  \begin{align}
   & t^\prime = t_{\rm KS}, \qquad x^\prime = x_{\rm KS} - a\sin^4\left(\frac{2\pi t_{\rm KS}}{b}\right), \qquad  y^\prime = y_{\rm KS},\qquad  z^\prime = z_{\rm KS},
\end{align}  
\end{subequations}
where $a$ and $b$ are two constants. Since the transformation is a pure gauge effect, one expects no gravitational waves at future null infinity, even though each metric component evolves with time $t^\prime$ nontrivially. The whole characteristic system must be properly established to cancel out the gauge effect at future null infinity. This makes the bouncing BH system the most demanding test \cite{Moxon:2021gbv}.

For our testing purposes, it is preferable to construct a scalar profile that can be controlled analytically, allowing us to compare numerical results with the expected analytical expression. To do this, we first work in the outgoing Eddington–Finkelstein coordinate system $\{u,r,\theta,\phi\}$. The retarded time $u$ is related to the Kerr-Schild time $t_{\rm KS}$ via
\begin{align}
    u=t_{\rm KS}-r-4M\ln\left(\frac{r}{2M}-1\right), \label{eq:bouncing_bh_u_tks}
\end{align}
with $M$ being the mass of the BH. The KG equation $\Box\psi=0$ in the Eddington–Finkelstein frame reads
\begin{align}
    2\partial_{r}\partial_{u}\left(r \psi\right)=\partial_{r}\left[\left(1-\frac{2M}{r}\right)\partial_{r}\left(r \psi\right)\right]-\frac{2M}{r^3}\left(r \psi\right)+\frac{\eth\bar{\eth}\left(r \psi\right)}{r^2}, \label{eq:KG_Schwarschild_BH}
\end{align}
where $\eth$ and $\bar{\eth}$ are derivatives associated with the outgoing Eddington–Finkelstein angular coordinates $\{\theta,\phi\}$. Considering a spherical profile, namely $\eth\bar{\eth}\left(r \psi\right)=0$, the solution to the KG equation has the following form 
\begin{align}
    &r\psi=\sum_{n=0}^{\infty}\frac{\psi_n(u)}{r^n}. \label{eq:KG_sol_pert}
\end{align}
The coefficients $\psi_n$'s are functions of the retarded time $u$. Plugging Eq.~\eqref{eq:KG_sol_pert} into \eqref{eq:KG_Schwarschild_BH} yields a three-term recurrence relation
\begin{align}
    & \frac{d\psi_1}{du}=0, &&2\frac{d \psi_{n+1}}{du}=\frac{2M}{n+1}n^2\psi_{n-1}-n\psi_n,\quad  n>0.
\end{align}
It has a solution of (assuming $M=1$)
\begin{align}
    &\psi_0=\sin u,\quad \psi_1=0,\quad \psi_2=-\frac{1}{2}\cos u,\quad \psi_3=\frac{1}{2}\sin u, \quad\psi_4=\frac{3}{4}\cos u-\frac{9}{8}\sin u, \quad \psi_5=-\frac{77}{20}\cos u-\frac{3}{2}\sin u,\notag \\
    &\psi_6=\frac{15}{16}\cos u+\frac{51}{4}\sin u, \quad \psi_7=\frac{1287}{28}\cos u-\frac{1809}{80}\sin u,\quad \psi_8=-\frac{12579}{80}\cos u-\frac{19857}{128}\sin u, \notag \\
    &\psi_9=-\frac{73557}{160}\cos u +\frac{133813}{140}\sin u,
    \quad \psi_{10}=\frac{49797063}{8960} \cos u + \frac{1272267}{1600} \sin u, \notag \\
    &\psi_{11}=-\frac{116136241}{24640}\cos u-\frac{57286503}{1792}\sin u, \quad \psi_{12}=-\frac{16472195091}{89600}\cos u +\frac{419653067}{5120}\sin u,\notag \\
    &\psi_{13}=\frac{197057851611}{232960} \cos u + \frac{517853793843}{492800} \sin u, \quad \psi_{14}=\frac{23027722022071}{3942400} \cos u - \frac{2420206444191}{313600} \sin u, \notag \\
    &\psi_{15}=-\frac{166944468961581}{2464000} \cos u - \frac{218435295225339}{7321600} \sin u.
\end{align}
Here we truncate the solution at $n=15$, which is sufficient for our following tests. At future null infinity, the solution is given by
\begin{align}
    r\psi|_{\mathcal{I}^+}=\psi_0(u)=\sin u.
\end{align}

As mentioned in Sec.~\ref{subsec:summary_wt_transformation}, we need to provide a boundary condition for $\partial_{t^\prime}\psi$ at a worldtube to perform CCE. This can be obtained by differentiating Eq.~\eqref{eq:KG_sol_pert}:
\begin{align}
    \partial_{t^\prime}\psi=\sum_{n=0}^{\infty}\frac{\dot{\psi}_n(u)}{r^{n+1}} \frac{du}{dt^\prime}-\sum_{n=0}^{\infty}(n+1)\frac{\psi_n(u)}{r^{n+2}}\frac{dr}{dt^\prime}, \label{eq:bouncing_bh_dtprime_psi}
\end{align}
with
\begin{align}
    &\frac{du}{dt^\prime}=1-\frac{r+2M}{r-2M}\frac{dr}{dt^\prime},
    &\frac{dr}{dt^\prime}=\frac{8\pi a}{br}\left[x^\prime+a\sin^4\left(\frac{2\pi t^\prime}{b}\right)\right]\sin^3\left(\frac{2\pi t^\prime}{b}\right)\cos\left(\frac{2\pi t^\prime}{b}\right),
\end{align}
where we have used Eq.~\eqref{eq:bouncing_bh_u_tks} and 
\begin{align}
    &r^2=\left[x^\prime+a\sin^4\left(\frac{2\pi t^\prime}{b}\right)\right]^2+y^{\prime\,2}+z^{\prime\,2}.
\end{align}

\begin{figure}[tb]
        \subfloat[$\psi$ at $\mathcal{I}^+$\label{fig:bouncing_bh_inf}]{\includegraphics[width=0.49\columnwidth,clip=true]{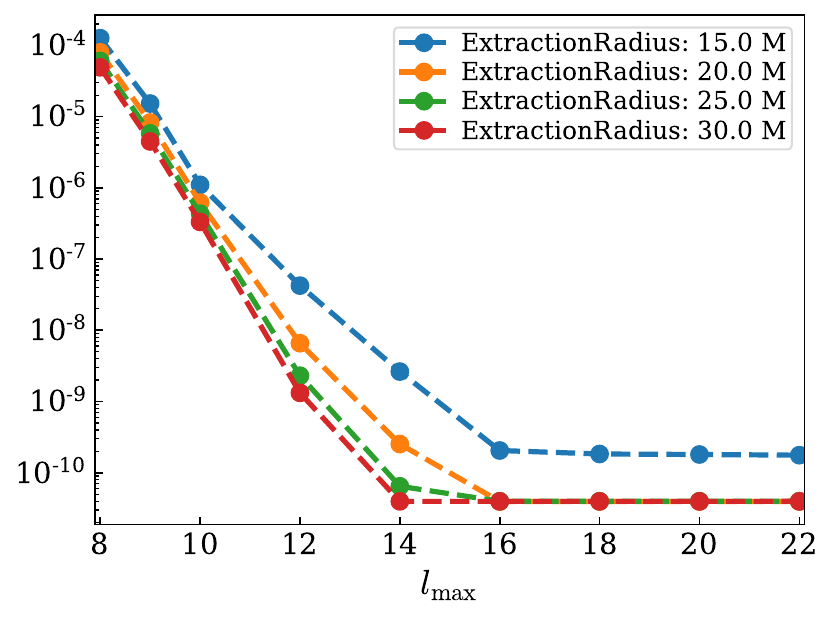}}
        \subfloat[Worldtube constraint\label{fig:bouncing_bh_wt_contraint}]{\includegraphics[width=0.49\columnwidth,clip=true]{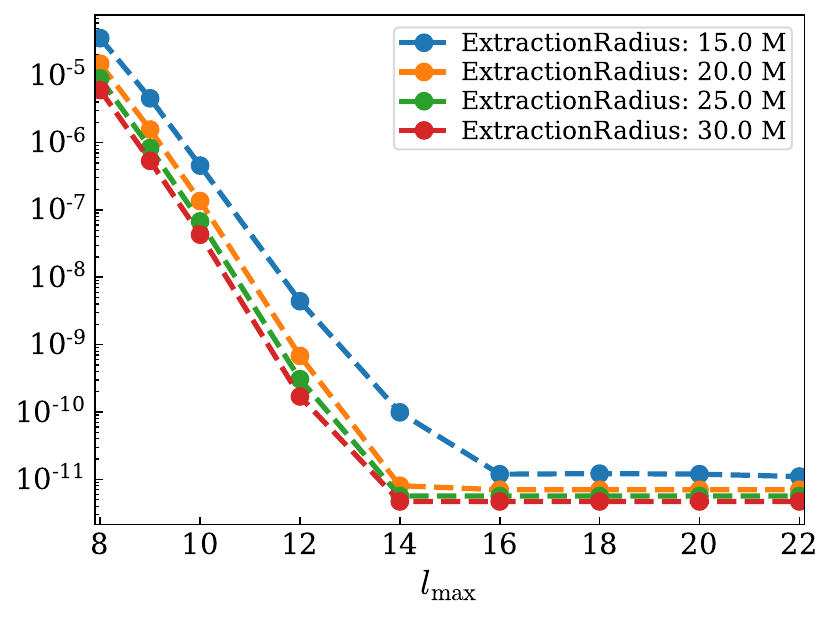}}
\caption{The convergence of numerical error with angular resolution $(l_{\max})$ in the bouncing Schwarzschild BH test. The left panel shows the difference between the computed $\psi$ at future null infinity and the expected value $\psi_0(u)=\sin u$, while the right panel displays the worldtube constraint as defined in Eq.~\eqref{eq:wt_constraint}.}
 \label{fig:bouncing_bh_wt_constraint}
\end{figure}

In our tests, we follow \cite{Moxon:2021gbv} and set the oscillation period $b$ to $40M$ and the amplitude $a$ to $2M$. Four worldtube radii, $15M,\, 20M,\, 25M,\, 30M$, are used for CCE. We choose the absolute tolerance of \texttt{SpECTRE}'s time stepper residual to be $10^{-12}$ to ensure that differences between our results and the expected one $\psi_0(u)=\sin u$ are dominated by spatial resolution. Figure \ref{fig:bouncing_bh_inf} provides the differences with angular resolution $l_{\rm max}$ ranging from 8 to 22. We can see that they converge exponentially to the level of $10^{-11}-10^{-10}$. Similarly, as shown in Fig.~\ref{fig:bouncing_bh_wt_contraint}, the worldtube constraint defined in Eq.~\eqref{eq:wt_constraint} follows a similar convergence behavior and finally levels off at $\sim 10^{-12}$.

\subsection{Linearized Bondi-Sachs metric}
In the second test, we choose the linearized Bondi-Sachs metric \cite{Bishop:2004ug} as our spacetime background. The corresponding worldtube data read \cite{Barkett:2019uae}
\begin{subequations}
  \begin{align}
    &J_{{\rm lin}\, lm}=\sqrt{\frac{(l+2)!}{(l-2)!}} \ _{2}Z_{lm} {\rm Re}\, \left[J_l(r) e^{i\nu u}\right], &
    U_{{\rm lin}\, lm}=\sqrt{l(l+1)} \ _{1}Z_{lm} {\rm Re}\, \left[U_l(r) e^{i\nu u}\right],\\
    &\beta_{{\rm lin}\, lm}= \ _{0}Z_{lm} {\rm Re}\, \left[\beta_l(r) e^{i\nu u}\right],
    & W_{{\rm lin}\, lm}= \ _{0}Z_{lm} {\rm Re}\, \left[W_l(r) e^{i\nu u}\right].
\end{align}  
\end{subequations}
Following \cite{Moxon:2021gbv}, we consider a $l=m=2$ angular profile, with $_s Z_{lm}$ given by\footnote{Eq.~\eqref{eq:linearized_bs_angular} only holds for $m>0$ modes, see Eq.~(137) in \cite{Barkett:2019uae} for other scenarios.}  
\begin{align}
    _s Z_{lm}(\theta,\phi)=\frac{1}{\sqrt{2}}\left[_s Y_{lm}(\theta,\phi) + (-1)^m _s Y_{l-m}(\theta,\phi)\right] . \label{eq:linearized_bs_angular}
\end{align}
The expression of $J_l(r),U_l(r), \beta_l(r)$ and $W_l(r)$ can be found in Eqs.~(139) and (140) of \cite{Barkett:2019uae}. In particular, we have 
\begin{align}
    J_{l=2}(r)=\frac{24B_2+3i\nu C_{2a}-i\nu^3 C_{2b}}{36}+\frac{C_{2a}}{4r}-\frac{C_{2b}}{12r^3}.
\end{align}
Here $B_{2}, C_{2a}$ and $C_{2b}$ are three free complex constants. Below we set $B_2$ to 0 for simplicity, which subsequently yields $3C_{2a}=\nu^2 C_{2b}$ due to the asymptotic flatness condition $J_{l=2}\sim \mathcal{O}(r^{-1})$. Therefore, the size of the linearized Bondi-Sachs wave is controlled by a single parameter $C_{2a}$. As for the scalar sector, we choose the worldtube data to be 
\begin{align}
    \psi(u)|_{\rm worldtube}=\frac{\sin u}{r},
    \qquad \partial_{t^\prime}\psi|_{\rm worldtube}=\frac{\cos u}{r}.
\end{align}

When the linearized  Bondi-Sachs wave vanishes $(C_{2a}=0)$, the scalar field propagates through flat spacetime. The consequent wavefunction at future null infinity is simply given by $r\psi(u)|_{\mathcal{I}^+}=\sin u$. However, for a finite $C_{2a}$, an exact analytical solution is not available. Instead, we can consider a perturbative expansion:
\begin{align}
    r\psi(u,\theta,\phi)|_{\mathcal{I}^+}=\sin u + C_{2a} \psi^{(1)}(u,\theta,\phi) + C^2_{2a} \psi^{(2)}(u,\theta,\phi) + \mathcal{O}(C_{2a}^3). \label{eq:Linearized_BS_scri_perturbative_expansion}
\end{align}
To the first order in $C_{2a}$, $\psi^{(1)}(u,\theta,\phi)$ is generated by the coupling between metric components (characterized by the angular profile $_{s} Z_{l=m=2}$) and the leading-order evolution of $\psi$, namely $\sin u$ (characterized by the angular profile $_{0} Z_{l=m=0}$). The angular selection rule immediately leads to 
\begin{align}
    \psi^{(1)}(u,\theta,\phi)  \sim \ _0Z_{22}(\theta,\phi).
\end{align}
In other words, the amplitude of the $l=m=2$ harmonic of the scalar field is expected to scale linearly with $C_{2a}$. By contrast, other harmonics of the scalar field are generated by nonlinear couplings, making their amplitudes depend at least quadratically on $C_{2a}$.

\begin{figure}[tb]
        \subfloat[$\psi$ at $\mathcal{I}^+$\label{fig:linearized_bs_alpha}]{\includegraphics[width=0.49\columnwidth,clip=true]{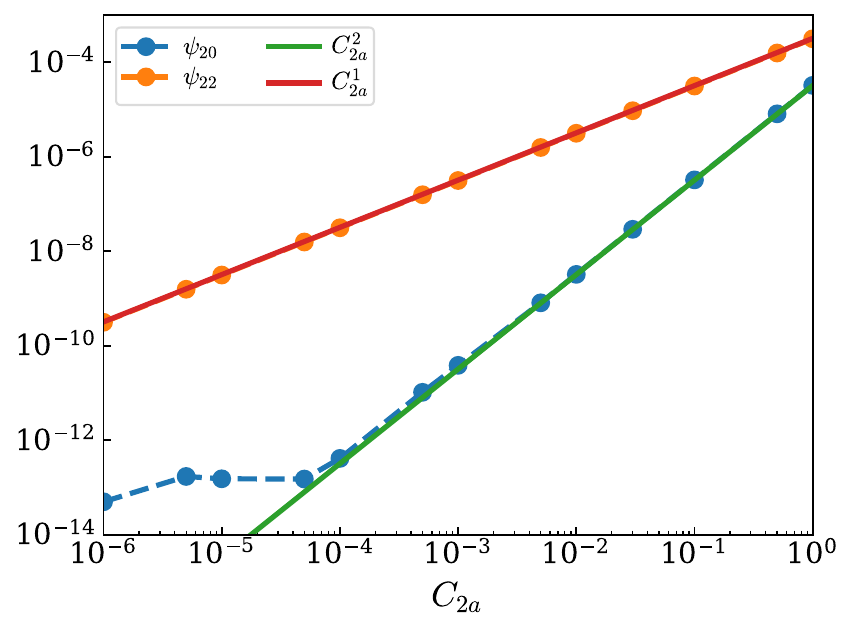}}
        \subfloat[Worldtube constraint\label{fig:linearized_bs_alpha_wt_con}]{\includegraphics[width=0.49\columnwidth,clip=true]{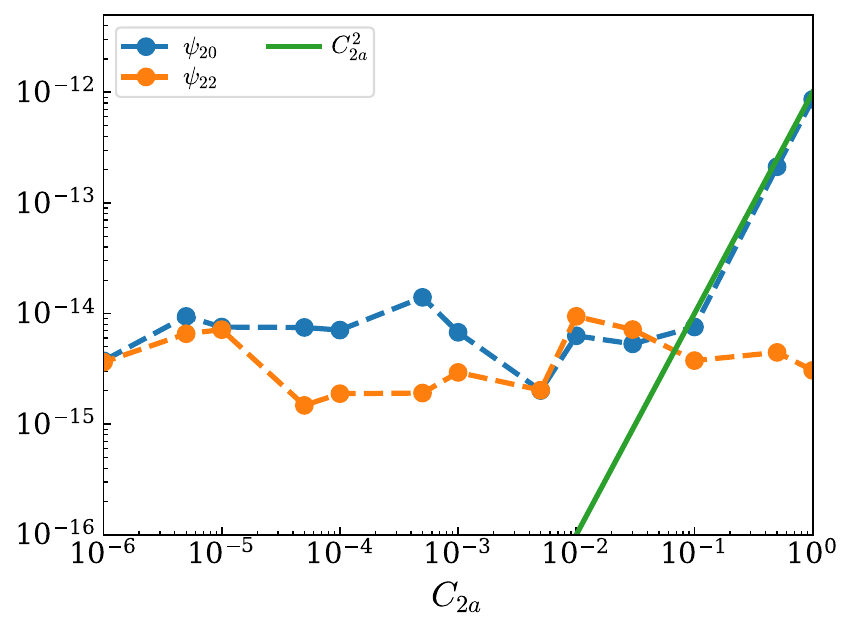}}
\caption{The linearized Bondi-Sachs test. The left panel shows the amplitude of $\psi_{l=2,m=0}$ (blue dots) and $\psi_{l=m=2}$ (orange dots) as a function of $C_{2a}$. They depend quadratically and linearly on $C_{2a}$, as expected. The right panel provides the worldtube constraint as defined in Eq.~\eqref{eq:wt_constraint}. }
\end{figure}

To perform our tests, we fix the frequency of the linearized Bondi-Sachs wave at $\nu=0.2$, while varying the wave amplitude $C_{2a}$ from $10^{-6}$ to $1$. To ensure the accuracy of our analysis, we still set the absolute tolerance of the time stepper to $10^{-12}$. Meanwhile, we find different choices of angular resolution $l_{\max}$ lead to only a $\sim10^{-12}$ difference in final products. Figure \ref{fig:linearized_bs_alpha} shows the amplitude of $\psi_{l=2,m=0}$ (blue dots) and $\psi_{l=m=2}$ (orange dots) as a function of $C_{2a}$, measured at future null infinity. As expected, they depend quadratically and linearly on $C_{2a}$, respectively. In Fig.~\ref{fig:linearized_bs_alpha_wt_con} we provide the worldtube constraint [Eq.~\eqref{eq:wt_constraint}] associated with $(l=2,m=0)$ and $(l=m=2)$. We see that the constraint violation for $\psi_{l=2,m=2}$ is always on the order of $10^{-15}$, whereas for $\psi_{l=2,m=2}$ it scales quadratically with $C_{2a}$, consistent with the behavior at future null infinity.

\section{Cauchy characteristic evolution}
\label{sec:cce_application}
After examining the characteristic evolution of the KG equation in Eq.~\eqref{eq:hypersurface_eq_for_pi}, we now proceed to test the full CCE procedure for the Einstein-Klein-Gordon system, which involves both Cauchy and characteristic evolutions. Here we consider a simple setup where a scalar pulse strikes a BH. To perform the simulation, we first construct initial data by solving the extended conformal thin-sandwich equations with the spectral elliptic solver in SpEC \cite{Pfeiffer:2002wt, SpECwebsite}. Next, we evolve the system nonlinearly using a first-order generalized harmonic formulation \cite{Lindblom:2005qh}. Our Cauchy evolution for the scalar field follows the method in \cite{Scheel:2003vs}. Finally, we use the dumped worldtube data for CCE, as outlined in \cite{Moxon:2020gha,Moxon:2021gbv} and earlier in this paper. For conciseness, we always set the initial BH mass to unity and use it as our code unit.

Below we will consider two distinct features for tests. (a) It is expected that the scalar emission from a perturbed BH could undergo a tail phase at late times, where the scalar field decays as a power law \cite{PhysRevD.5.2419,PhysRevD.5.2439}. In particular, the power law has different exponents at timelike infinity and null infinity \cite{PhysRevD.5.2419,PhysRevD.5.2439,PhysRevD.34.384,PhysRevD.49.883,PhysRevD.49.890}. The late-time tail at timelike infinity was investigated by Scheel \etal \cite{Scheel:2003vs}. Below in Sec.~\ref{subsec:cce_tail}, we focus on the behavior at null infinity to demonstrate the accuracy of our CCE system. (b) As discussed in Sec.~\ref{sec:memory_effect}, the GW (tensor emission) of a BH stirred by a scalar field consists of memory effects. They are governed by the balance laws in Eqs.~\eqref{eq:JE_decomposition} and \eqref{eq:JM_decomposition}. In the second part of this section (Sec.~\ref{subsec:cce_memory}), we will examine the deviations from the balance laws, as well as the Bianchi identities derived in Sec.~\ref{subsec:bianchi_identities}.

\textcolor{black}{Throughout this section, we initialize the scalar field with a linear ansatz:
\begin{align}
    \psi=\frac{A}{r},
\end{align}
where the coefficient $A$ is determined from the worldtube data obtained via the Cauchy evolution. We note that this initial data is constructed fully empirically. Developing an \textit{ab initio} initial-data solver is beyond the scope
of this paper and is left for future work.}

\subsection{Late-time tail}
\label{subsec:cce_tail}
In the first test, we place a Schwarzschild BH at the coordinate center. Initially, the scalar field profile is constructed as follows:
\begin{align}
    \psi=A\frac{e^{-(r^\prime-r^\prime_0)^2/w^2}}{r^\prime}Y_{l=m=1}(\theta^\prime,\phi^\prime), \label{eq:cauchy_psi_id}
\end{align}
with $A=0.1, r_0^\prime=50$ and $w=5$. The angular dependence is set to be the $l=m=1$ spherical harmonic. Meanwhile, we choose $\Pi^\prime = \partial_{r^\prime}\psi$ at $t^\prime=0$, where\footnote{The Cauchy variable $\Pi^\prime$ is not to be confused with the characteristic variable $\Pi$ in Eq.~\eqref{eq:evolution_psi_characteristic}. } 
\begin{align}
    \Pi^\prime \equiv - \frac{1}{\alpha^\prime}(\partial_{t^\prime}\psi-\beta^{i^\prime}\partial_{i^\prime}\psi), \label{eq:Cauchy_Pi_prime}
\end{align}
is an auxiliary variable introduced by Scheel \etal \cite{Scheel:2003vs}. The choice of $\Pi^\prime$ represents an infalling scalar pulse into the BH.  

\begin{figure}[tb]
        \subfloat[Null infinity versus timelike infinity\label{fig:tail_time_vs_null_comparison}]{\includegraphics[width=0.49\columnwidth,clip=true]{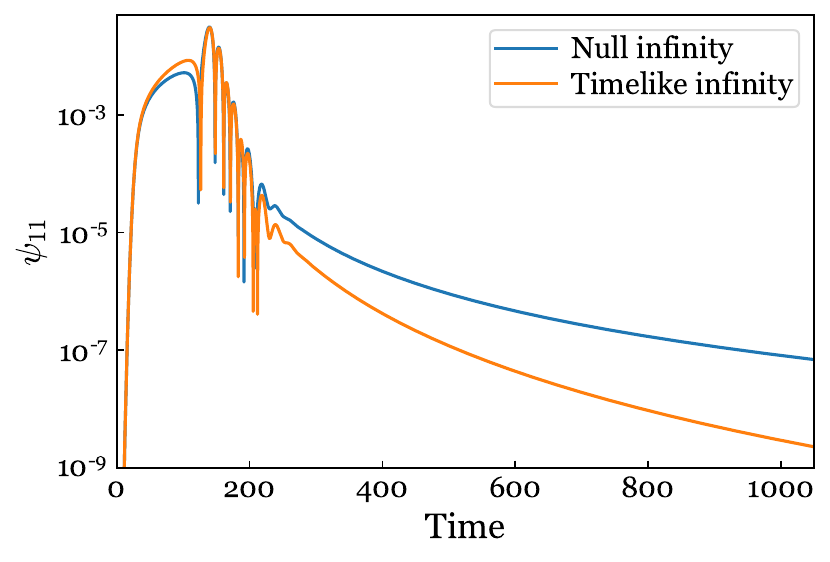}} \\
        \subfloat[Null infinity\label{fig:tail_nulllike}]{\includegraphics[width=0.49\columnwidth,clip=true]{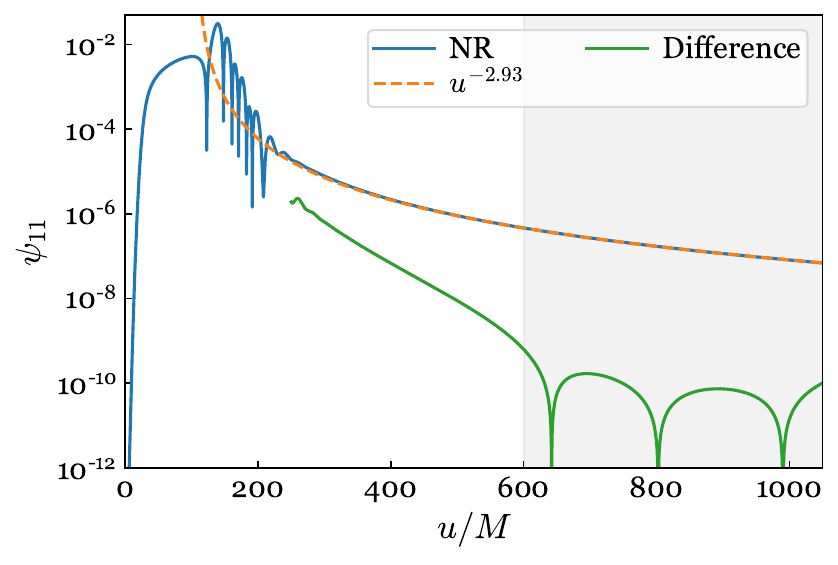}}
        \subfloat[Timelike infinity\label{fig:tail_timelike}]{\includegraphics[width=0.49\columnwidth,clip=true]{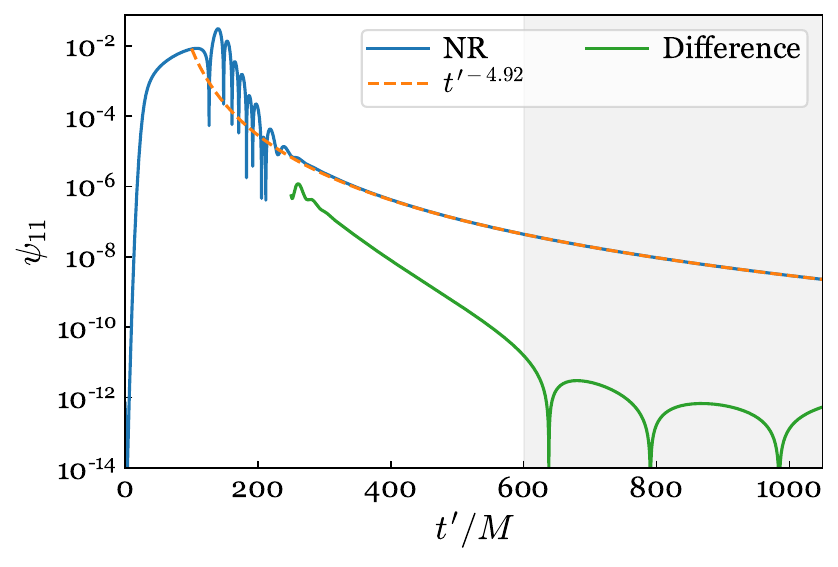}}
\caption{Resolving ringdown tail in the scalar emission from a Schwarzschild BH. The BH is struck by a scalar field composed of a $l=m=1$ spherical harmonic [Eq.~\eqref{eq:cauchy_psi_id}]. Figure \ref{fig:tail_time_vs_null_comparison} shows the extracted $\psi_{11}$ at null infinity (in blue, computed with CCE) and at an approximate timelike infinity (in orange, measured at a fixed Cauchy radius). In Figs.~\ref{fig:tail_nulllike} and \ref{fig:tail_timelike}, the late-time portion (in gray) of $\psi_{11}$ is fitted to a power law. The best fits are plotted as two orange dashed curves. The green curves represent the difference between $\psi_{11}$ and the best fits. }
\end{figure}

As shown in Fig.~\ref{fig:tail_time_vs_null_comparison}, the blue curve corresponds to the $l=m=1$ spherical harmonic of the extracted scalar field as a function of the retarded time $u$, with the CCE worltube placed at a radius of $75$. The evolution consists of three stages: the excitation phase $({\rm time}\lesssim 125)$, the quasinormal ringing phase $(125 \lesssim {\rm time}\lesssim 250)$, and the tail phase $({\rm time}\gtrsim 250)$. We note that the scalar field is extracted at future null infinity, where the Bondi radius $r$ approaches $\infty$ while the retarded time $u$ remains fixed. By comparison, we also obtain the scalar radiation at an approximate timelike infinity by measuring late-Cauchy time $\psi$ at a fixed Cauchy radius, as done by Scheel \etal \cite{Scheel:2003vs}. The orange curve in Fig.~\ref{fig:tail_time_vs_null_comparison} shows the corresponding $l=m=1$ scalar harmonic. To make a fair comparison with CCE, our extraction radius is still at $75$. 

We can see that $\psi_{11}$ at both timelike and null infinity evolves similarly during the excitation and ringdown phases, whereas their late-time tails exhibit distinct decay rates. Specifically, in Figs.~\ref{fig:tail_nulllike} and \ref{fig:tail_timelike} we fit the late-time portion of $\psi_{11}$ ($600<{\rm time}< 1050$, i.e., the gray-shaded regions) to a power law $A(t+t_0)^{-\mu}$, with $A,t_0$, and $\mu$ being three constants. We consider the following cost function for the fit:
\begin{align}
    \sum_{t_i} \left[\log |\psi_{11}(t_i)| - \log A +\mu \log(t_i+t_0) \right]^2,
\end{align}
where $t_i$'s are time samples. To improve the fit performance, we notice that for a given nonlinear parameter $t_0$, two linear parameters $\log A$ and $\mu$ can be determined uniquely using ordinary least squares linear regression. This reduces the fit to a one-dimensional minimization problem, and we deal with it using the Nelder–Mead method built in \texttt{SciPy} \cite{2020SciPy-NMeth}. We have checked that the results are insensitive to the initial guess. The best fits are plotted as orange dashed lines in Figs.~\ref{fig:tail_nulllike} and \ref{fig:tail_timelike}. We find the tail exponents at null infinity $(\mu_{\rm n})$ and timelike infinity $(\mu_{\rm t})$ to be 2.93 and 4.92, respectively. They are consistent with the predictions of BH perturbation theory \cite{PhysRevD.5.2419,PhysRevD.5.2439,PhysRevD.34.384,PhysRevD.49.883,PhysRevD.49.890}: $\mu_{\rm n}=l+2$ and $\mu_{\rm t}=2l+3$.



\subsection{Memory effects}
\label{subsec:cce_memory}
In the second test, we consider a Kerr BH with a dimensionless spin $\chi=0.7$, initially surrounded by a scalar field \textcolor{black}{with $\Pi^\prime=0$} [defined in Eq.~\eqref{eq:Cauchy_Pi_prime}]. The initial profile of $\psi$ is chosen to be the same as Eq.~\eqref{eq:cauchy_psi_id}, with $A=1, r_0^\prime=20$ and $w=2$. The CCE worldtube is placed at $r_{\rm wt}^\prime=100M$.

\begin{figure}[!htb]
        \includegraphics[width=0.49\columnwidth,clip=true]{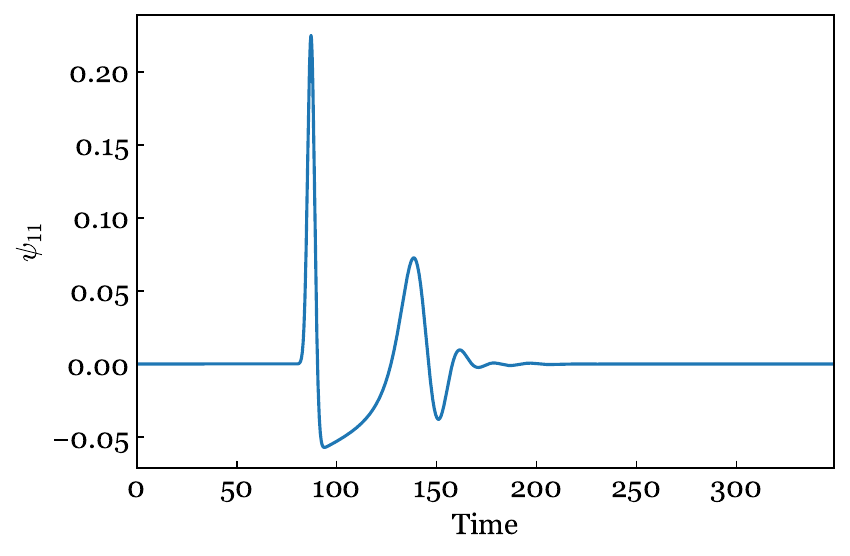}
\caption{The $l=m=1$ spherical harmonic of the scalar emission from a Kerr BH, extracted at future null infinity.}
 \label{fig:memory_scalar_11}
\end{figure}

Figure \ref{fig:memory_scalar_11} shows the $l=m=1$ harmonic of the scalar field measured at future null infinity. \textcolor{black}{The scalar field initially contains an ingoing and an outgoing component in the radial direction.} The outgoing piece reaches the CCE worldtube and appears at future null infinity at a time of $r_{\rm wt}^\prime-r_{0}^\prime=80M$; whereas the ingoing wave first strikes the Kerr BH at $r_{0}^\prime=20M$, the excited quasinormal modes are then transmitted to future null infinity by CCE at $r_{\rm wt}^\prime+r_{0}^\prime=120M$. Due to the no-hair theorem \cite{1971ApJ...166L..35T, Sotiriou:2011dz,
Hawking:1972qk, Healy:2011ef, Yunes:2011aa}, a Kerr BH does not carry a scalar charge. Consequently, the scalar field vanishes when the BH settles to stationary. 

\begin{figure}[tb]
        \subfloat[$h_{20}$\label{fig:scalar_memory_20}]{\includegraphics[width=0.49\columnwidth,clip=true]{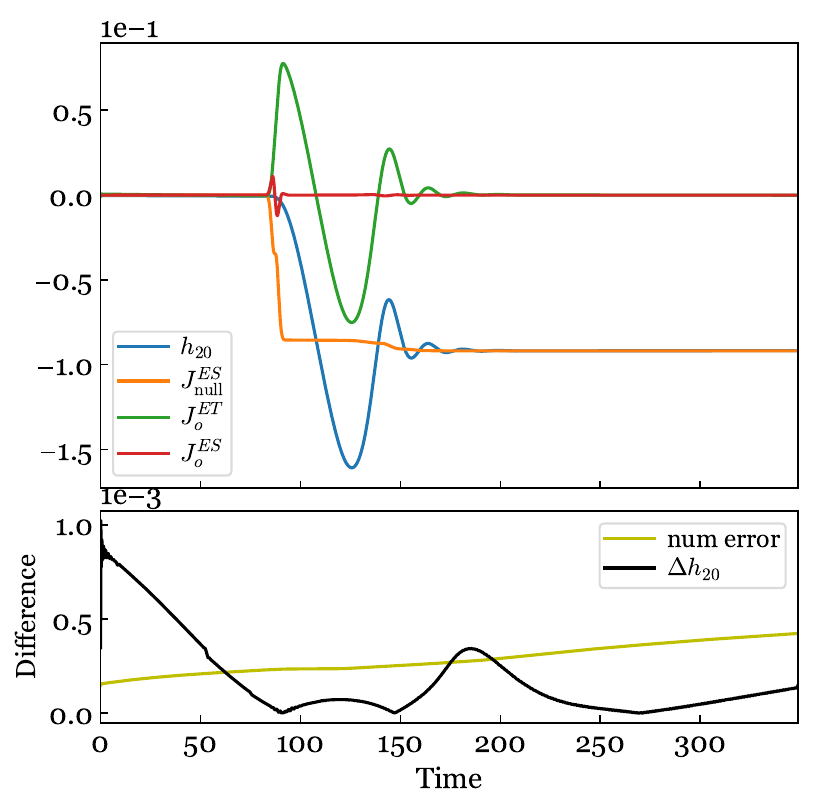}}
        \subfloat[$h_{22}$\label{fig:scalar_memory_22}]{\includegraphics[width=0.49\columnwidth,clip=true]{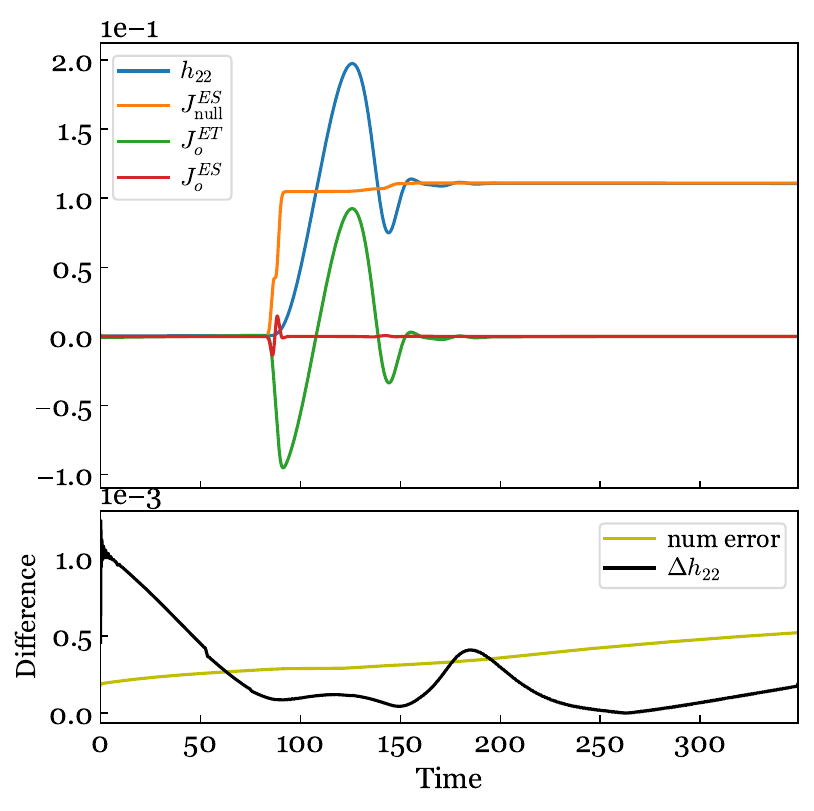}} \\
        \subfloat[$h_{32}$\label{fig:scalar_memory_32}]{\includegraphics[width=0.49\columnwidth,clip=true]{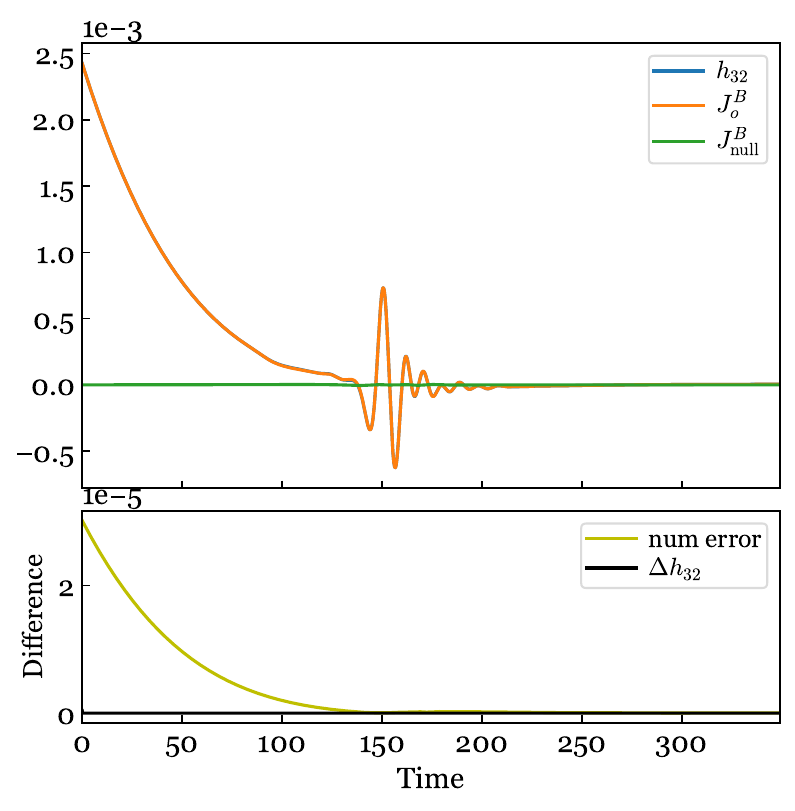}}
        \subfloat[$\int h_{32}du$\label{fig:scalar_memory_int_32}]{\includegraphics[width=0.49\columnwidth,clip=true]{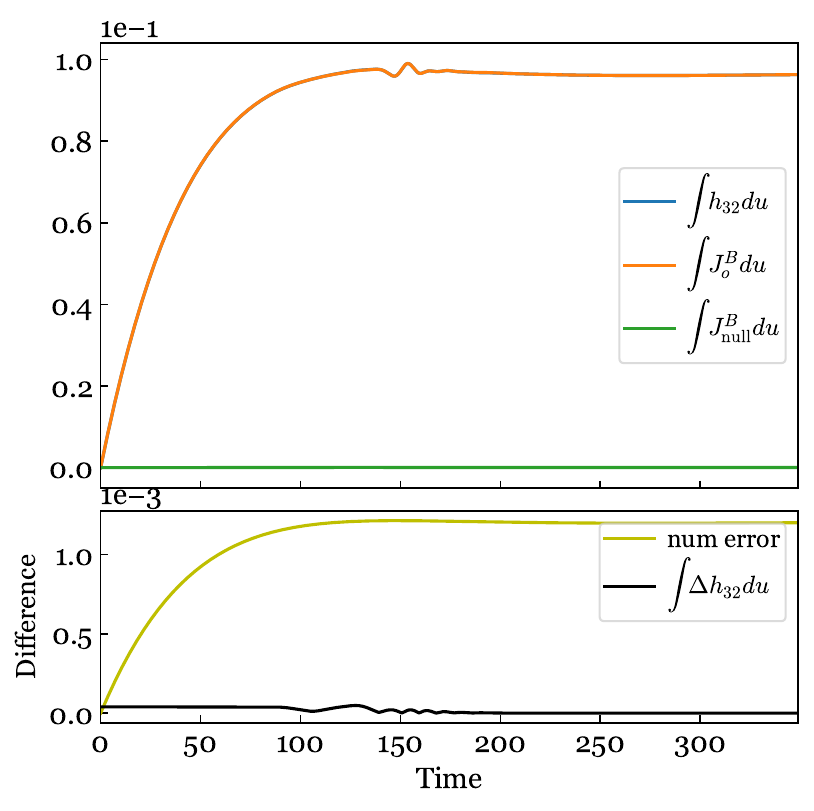}}
\caption{Memory effects in the GWs emitted by a Kerr BH, perturbed by a scalar field. Figures \ref{fig:scalar_memory_20} and \ref{fig:scalar_memory_22}: the blue curves show the $(l=2,m=0)$ and $(l=m=2)$ harmonics of the full strain. They are compared to the three dominant electric memory components (in orange, green, and red), as evaluated based on Eqs.~\eqref{eq:JES_decomposition_LNL} and \eqref{eq:JET_decomposition_LNL}. Two lower panels display the difference between the total strain and the sum of all electric memory contributions (in black), which is further compared to numerical error (in yellow). Figures \ref{fig:scalar_memory_32} and \ref{fig:scalar_memory_int_32}: the blue curves (overlapped with the orange curves) illustrate the $(l=3,m=2)$ harmonic of the full strain and its time integration (spin memory). They are solely contributed by the magnetic ordinary memory $J_o^B$ (in green), as evaluated based on Eq.~\eqref{eq:JM_decomposition_LNL}.  } 
\end{figure}

During the dynamical process, the scalar field stirs spacetime nonlinearly [Eq.~\eqref{eq:metric_field_equation}] and excites GWs. The blue curves in Figs.~\ref{fig:scalar_memory_20} and \ref{fig:scalar_memory_22} represent the corresponding $(l=2,m=0)$ and $(l=m=2)$ harmonics, respectively. Both curves exhibit a permanent jump after the passage of the scalar wave, with a quasinormal-mode ringing superimposed. As discussed in Sec.~\ref{sec:memory_effect}, a strain can be decomposed into an electric and a magnetic memory. In this case, we find that the magnetic sector vanishes. Three dominant electric components [see Eq.~\eqref{eq:JE_decomposition}] are plotted in orange, green, and red in  Figs.~\ref{fig:scalar_memory_20} and \ref{fig:scalar_memory_22}. The scalar-driven null memory $J^{ES}_{\rm null}$ contributes to the overall jump, while the tensor-driven ordinary piece $J^{ET}_{o}$ contributes to the quasinormal-mode ringing. The scalar-driven ordinary memory $J^{ES}_{o}$ only results in a small pulse when the initial outgoing scalar wave reaches future null infinity. Unlike in BBH systems, where the tensor-driven null component $J^{ET}_{\rm null}$ induces a strong memory effect \cite{Mitman:2020pbt}, its contribution is negligible in this process. Two lower panels of Figs.~\ref{fig:scalar_memory_20} and \ref{fig:scalar_memory_22} present the difference between the total strain and the sum of all the individual memory contributions. We see that the difference is overall smaller than the numerical error (in yellow) except for the initial junk-radiation regime, thereby justifying our CCE code. To examine the magnetic memory, we consider the $(l=3,m=2)$ harmonic in Fig.~\ref{fig:scalar_memory_32}. For the current system, it is solely contributed by the ordinary memory associated with the angular momentum aspect. The difference between $J^B_o$ and the total strain is again smaller than the numerical error. Finally, Figure \ref{fig:scalar_memory_int_32} shows the $(l=3,m=2)$ harmonic  of the spin memory \cite{Pasterski:2015tva}, as a time integration of $h_{32}$.

\begin{figure}[tb]
        \includegraphics[width=0.49\columnwidth,clip=true]{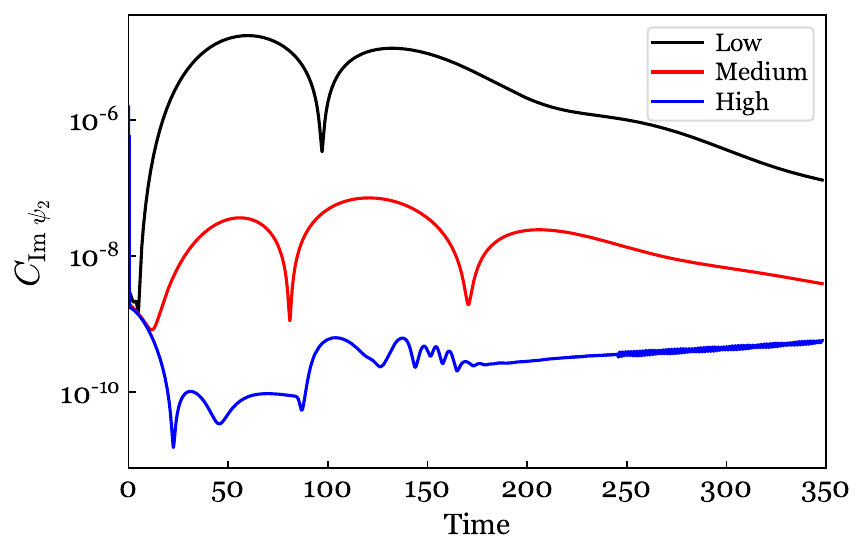}
        \includegraphics[width=0.49\columnwidth,clip=true]{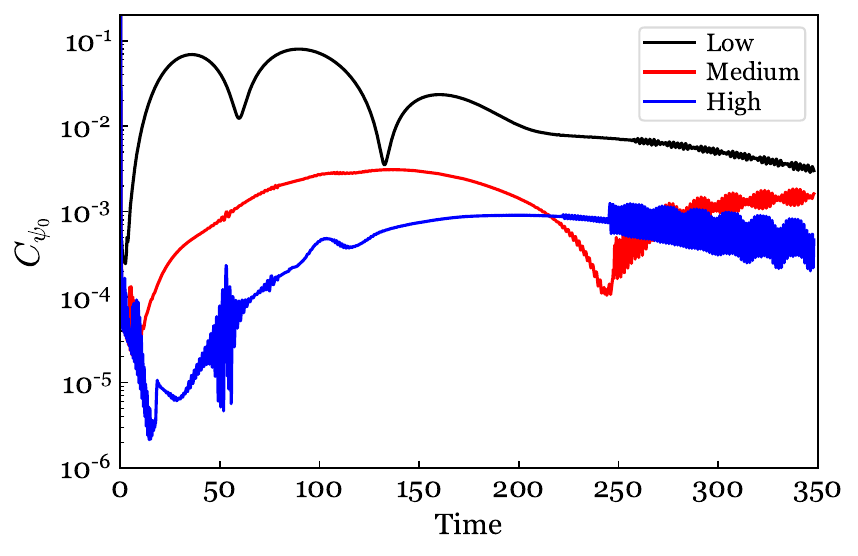} \\
        \includegraphics[width=0.49\columnwidth,clip=true]{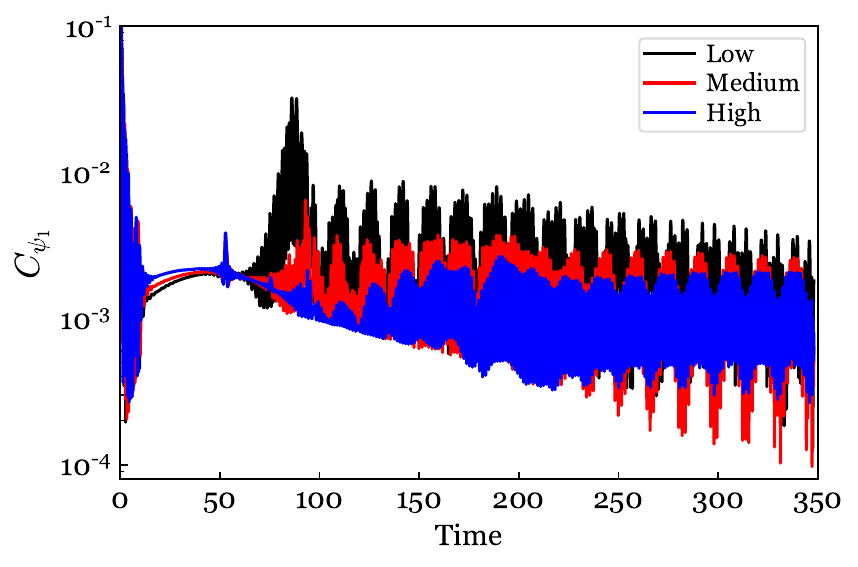}
        \includegraphics[width=0.49\columnwidth,clip=true]{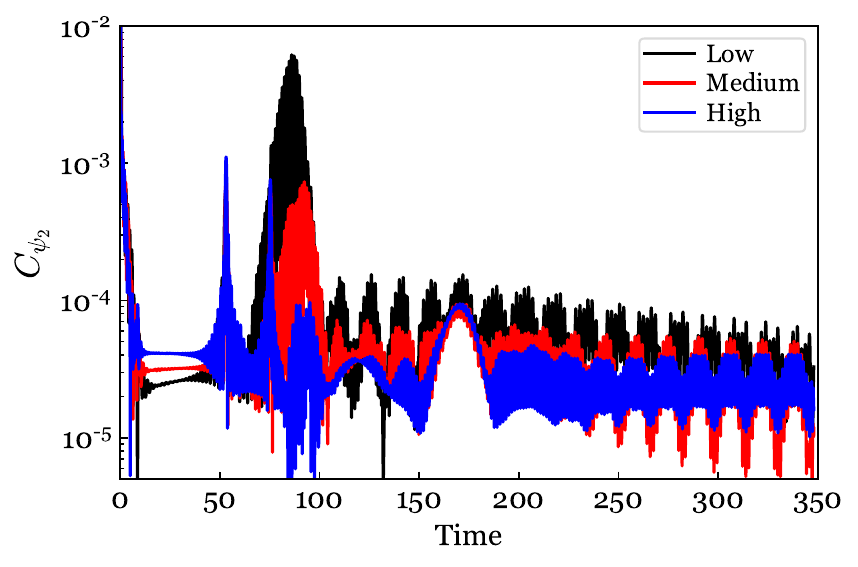} \\
        \includegraphics[width=0.49\columnwidth,clip=true]{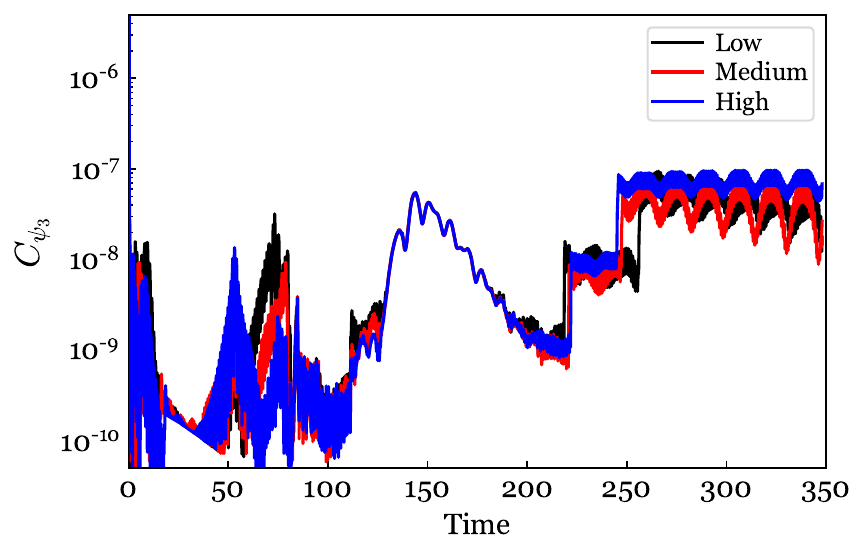}
        \includegraphics[width=0.49\columnwidth,clip=true]{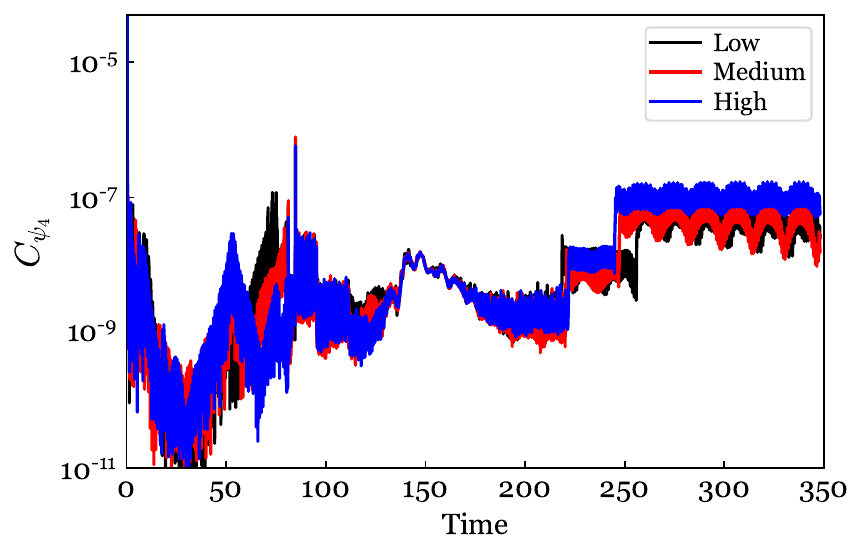}
\caption{The $L^2-$norms of the constraint violations given in Eq.~\eqref{eq:imaginary_of_psi2_vanishes} ($C_{{\rm Im} \psi_2}$, top left) and Eq.~\eqref{eq:Bianchi_identities_scri_scri}, computed at three numerical resolutions.}
 \label{fig:constraints_scri}
\end{figure}

As discussed in Sec.~\ref{subsec:bianchi_identities}, the Bianchi identities in Eq.~\eqref{eq:Bianchi_identities_scri_scri} yield relations between the Weyl scalars, the strain, and the scalar field. Meanwhile, Eq.~\eqref{eq:imaginary_of_psi2_vanishes} imposes a further constraint between $\Psi_2$ and $h$. Figure \ref{fig:constraints_scri} shows the $L^2-$norms of deviations from these constraints, computed at three numerical resolutions. In the first two rows, the convergence of the deviations with increasing resolution validates our CCE procedure. In the bottom row, the constraint violations for $\Psi_4$ [\eqref{eq:Bianchi_identities_scri_scri_psi4}] and $\Psi_3$ [\eqref{eq:Bianchi_identities_scri_scri_psi3}] do not converge, presumably because they are already small $(\sim 10^{-7})$.

\section{Conclusion}
\label{sec:conclusion}
In this paper, we have implemented a CCE algorithm for the Einstein-Klein-Gordon system. Compared to GR, we presented the additional terms contributed by the scalar field in the equations of motion (Sec.~\ref{subsec:characteristic_equations_of_motion}), the computation of Weyl scalars (Sec.~\ref{subsec:weyl_scalars_scri}), the Bianchi identities (Sec.~\ref{subsec:bianchi_identities}), as well as the memory effects (Sec.~\ref{sec:memory_effect}). In addition, we reformulated the characteristic form of the KG equation using the numerically adapted coordinates [Eq.~\eqref{eq:numerically_adapted_coordinates}], facilitating its implementation in our numerical relativity code \texttt{SpECTRE}. We also derived a concise worldtube transformation to construct the boundary condition for the characteristic KG equation from a generic Cauchy evolution (Sec.~\ref{subsec:summary_wt_transformation}). 

To evaluate the accuracy of our code, we designed various test systems. We first focused on the implementation of the KG equation and the worldtube transformation by evolving scalar fields on two prescribed spacetime backgrounds. As expected, we found that the numerical error decreases exponentially with increasing simulation resolution. We then examined the full CCE procedure by striking a BH with a scalar pulse. During the late-time evolution, we observed that the scalar field decays as a power law $u^{-l-2}$, consistent with Price's law at future null infinity. Furthermore, we verified that the tensor emissions extracted with CCE are consistent with the balance laws and the Bianchi identities.

The tests demonstrate that our CCE code can effectively reveal scalar-induced memory effects. In the future, a direct application of this work is to extract GWs emitted by binary mergers in alternative theories of gravity. For example, a recent study simulated a black hole - neutron star merger in scalar-tensor theory \cite{Ma:2023sok}, where the neutron star underwent spontaneous scalarization. Working within the Einstein frame, the current CCE algorithm can faithfully compute memory effects in both tensor and breathing modes \cite{Du:2016hww,Koyama:2020vfc,Heisenberg:2023prj}, allowing us to make comparisons with the post-Newtonian approximation \cite{Lang:2013fna,Lang:2014osa,Tahura:2021hbk}. These numerical studies will improve our understanding of asymptotic symmetries \cite{Hou:2020tnd,Hou:2020wbo,Hou:2020xme,Tahura:2020vsa}. In addition, our implementation lays a foundation for performing CCE simulations in other modified theories of gravity, e.g., \cite{Corman:2024vlk,Lara:2024rwa}, paving the way for studying the associated memory effects \cite{Heisenberg:2023prj}.

Our simulations show tails in scalar fields. A possible avenue for future work is to look for tails in breathing modes (e.g. \cite{Ma:2023sok}) and tensor modes \cite{DeAmicis:2024not,Islam:2024vro,Carullo:2023tff} emitted by binary systems. The accurate simulation of these tails is sensitive to the choice of boundary conditions \cite{Allen:2004js}. Incorrect boundary conditions can introduce nonoscillatory numerical artifacts during the ringdown phase, which might be mistaken for physical tails. This issue can be avoided by either positioning the outer boundaries far enough away to remain causally disconnected from the system or by adopting CCM \cite{Ma:2023qjn}.

Finally, it would also be interesting to apply our CCE algorithm to extract scalar emissions from intermediate-mass-ratio inspirals. Recently, a worldtube excision method \cite{Wittek:2023nyi,Wittek:2024gxn} was developed to simulate a small scalar charge orbiting around a BH. By matching a perturbative description around the scalar charge to a Cauchy simulation, the method enables the evolution of a binary system in the intermediate-mass-ratio regime. The accuracy of this method was evaluated by comparing scalar modes extracted at a finite radius to perturbative calculations. Our CCE algorithm could further improve this comparison. Additionally, the algorithm allows for verifying balance laws at future null infinity.
\begin{acknowledgments}
S.M. would like to thank Vijay Varma and Leo C. Stein for useful discussion.
We thank the anonymous referees for their insightful and heuristic comments and suggestions.
Research at Perimeter Institute is supported in part by the Government of Canada through the Department of Innovation, Science and Economic Development and by the Province of Ontario through the Ministry of Colleges and Universities. 
This material is based upon work supported by the National Science Foundation under Grants No. PHY-2407742, No. PHY- 2207342, and No. OAC-2209655 at Cornell. Any opinions, findings, and conclusions or recommendations expressed in this material are those of the author(s) and do not necessarily reflect the views of the National Science Foundation. This work was supported by the Sherman Fairchild Foundation at Cornell.
\end{acknowledgments}  

\appendix

\section{Worldtube transformation}
\label{app:worldtube_transformation}
As discussed in \cite{Moxon:2020gha}, in SpECTRE, five coordinate systems are involved in worldutbe transformations, including:
\begin{itemize}
    \item Cauchy coordinates $r^\prime, x^{\prime A^\prime}, t^\prime$.
    \item Null-radius coordinates $\underline{\lambda}, \underline{x}^{\underline{A}},\underline{u}$
    \item Bondi-like coordinates $r,x^A,u$.
    \item Partially flat Bondi-like coordinates $\hat{r},\hat{x}^{\hat{A}},\hat{u}$.
    \item Numerically adapted coordinates $\Breve{y},\Breve{x}^{\Breve{A}},\Breve{u}$
\end{itemize}
The numerically adapted coordinates are used for characteristic evolution, namely our target system. The Jacobians between these coordinate systems are summarized in Ref.~\cite{Ma:2023qjn}. Here we shall not go into full details. Instead, we list only necessary components in the following discussions. 

\subsection{Obtaining $\Breve{\Pi}|_{\rm wt}$: from numerically adapted to partially flat Bondi-like coordinates}
\label{subsec:wt_transformation_numerically_adapted}
To obtain the boundary condition $\Breve{\Pi}|_{\rm wt}$, we need the following  Jacobian [see Eq.~(46b) of \cite{Moxon:2020gha}]
\begin{align}
    \partial_{\hat{u}} = \partial_{\Breve{u}}-(1-\Breve{y})\frac{\partial_{\Breve{u}}\hat{R}_{\rm wt}}{\hat{R}_{\rm wt}}\partial_{\Breve{y}},
\end{align}
where $\hat{R}_{\rm wt}$ is the radius of the worldtube in the partially flat Bondi-like coordinate system. Applying the Jacobian to $\psi$ on the worldtube $\Breve{y}=-1$, we obtain
\begin{align}
\Breve{\Pi}|_{\rm wt}=\left(\partial_{\Breve{u}}\psi\right)_{\Breve{y}}=\left(\partial_{\hat{u}}\psi\right)_{\hat{r}}+\frac{\partial_{\Breve{u}}\hat{R}_{\rm wt}}{\hat{R}_{\rm wt}}\partial_{\Breve{y}}\psi.\label{eq:numtheta_wt}
\end{align}
The second term $\frac{\partial_{\Breve{u}}\hat{R}_{\rm wt}}{\hat{R}_{\rm wt}}\partial_{\Breve{y}}\psi$ is already available in the current characteristic system. Therefore, we are left to obtain the value of $\left(\partial_{\hat{u}}\psi\right)_{\hat{r}}$ from the Cauchy system. Below we will transform $\left(\partial_{\hat{u}}\psi\right)_{\hat{r}}$ to the rest of coordinates.

\subsection{Obtaining $\left(\partial_{\hat{u}}\psi\right)_{\hat{r}}$: from partially flat Bondi-like to Bondi-like coordinates}
\label{subsec:wt_transformation_partiall_flat}
To obtain the value of $\left(\partial_{\hat{u}}\psi\right)_{\hat{r}}$ from the Bondi-like coordinates, the necessary Jacobian is
\begin{align}
    \left(\partial_u\right)_r&=\left(\partial_{\hat{u}}\right)_{\hat{r}}+\left(\partial_u\hat{x}^{\hat{A}}\right)_r\partial_{\hat{A}}+\hat{r}\frac{\partial_u\hat{\omega}}{\hat{\omega}}\partial_{\hat{r}}=\left(\partial_{\hat{u}}\right)_{\hat{r}}-{\rm Re}\,\mathcal{U}^{(0)} \bar{\hat{\eth}}+(1-\Breve{y})\frac{\partial_u\hat{\omega}}{\hat{\omega}}\partial_{\Breve{y}}, \label{eq:jacobian_partially_flat_bondi_like}
\end{align}
where the first line comes from Eq.~(4.20) of \cite{Ma:2023qjn}. The second line uses the evolution equation of $\hat{x}^{\hat{A}}$:
\begin{align}
    \left(\partial_u\hat{x}^{\hat{A}}\right)_r=-\frac{1}{2}\left(\mathcal{U}^{(0)}\bar{\hat{q}}^{\hat{A}}+\bar{\mathcal{U}}^{(0)}\hat{q}^{\hat{A}}\right),
\end{align}
see Eq.~\eqref{eq:dtprime_xA_num} with $t^\prime$ replaced by $u$ and $\Breve{x}^{\Breve{A}}$ by $\hat{x}^{\hat{A}}$; as well as an identity [Eq.~(46a) of \cite{Moxon:2020gha}]
\begin{align}
    \hat{r}\partial_{\hat{r}}=(1-\Breve{y})\partial_{\Breve{y}}.
\end{align}
Finally, $\hat{\omega}$ in Eq.~\eqref{eq:jacobian_partially_flat_bondi_like} is the conformal factor of the angular transformation $\hat{x}^{\hat{A}}(x^A)$:
\begin{align}
    \hat{q}_{\hat{A}\hat{B}}\partial_A\hat{x}^{\hat{A}}\partial_B\hat{x}^{\hat{B}}=\frac{1}{\hat{\omega}^2}q_{AB},
\end{align}
where $q_{AB}\,(\hat{q}_{\hat{A}\hat{B}})$ is the unit sphere metric of the (partially flat) Bondi-like coordinates.

Applying the Jacobian in Eq.~\eqref{eq:jacobian_partially_flat_bondi_like} to $\psi$ and restricting it to the worldtube, we obtain
\begin{align}
    \left(\partial_u\psi\right)_r=\left(\partial_{\hat{u}}\psi\right)_{\hat{r}}-{\rm Re}\,\mathcal{U}^{(0)} \bar{\hat{\eth}}\psi+2\frac{\partial_u\hat{\omega}}{\hat{\omega}}\partial_{\Breve{y}}\psi. \label{eq:du_psi_1}
\end{align}
In practice, our CCE system does not directly provide the value of $\partial_u\hat{\omega}$. Instead, it gives us $\partial_{\hat{u}}\hat{\omega}$. The two are related by applying the Jacobian in Eq.~\eqref{eq:jacobian_partially_flat_bondi_like} to $\hat{\omega}$, which yields
\begin{align}
    \left(\partial_u\hat{\omega}\right)_r=\left(\partial_{\hat{u}}\hat{\omega}\right)_{\hat{r}}-{\rm Re}\,\mathcal{U}^{(0)} \bar{\hat{\eth}}\hat{\omega}. \label{eq:du_omega_hat}
\end{align}
Note that we have used the fact that the conformal factor $\hat{\omega}$ does not depend on $\Breve{y}$, i.e., $\partial_{\Breve{y}}\hat{\omega}=0$. Plugging Eq.~\eqref{eq:du_omega_hat} into Eq.~\eqref{eq:du_psi_1}, we get
\begin{align}
    \left(\partial_{\hat{u}}\psi\right)_{\hat{r}}=\left(\partial_u\psi\right)_r+{\rm Re}\,\mathcal{U}^{(0)} \bar{\hat{\eth}}\psi-2\frac{\left(\partial_{\hat{u}}\hat{\omega}\right)_{\hat{r}}-{\rm Re}\,\mathcal{U}^{(0)} \bar{\hat{\eth}}\hat{\omega}}{\hat{\omega}}\partial_{\Breve{y}}\psi. \label{eq:duhat-du}
\end{align}
The only remaining unknown variable on the right-hand side is $\left(\partial_u\psi\right)_r$. Therefore, our next task is to compute its value from Cauchy variables.

\subsection{Obtaining $\left(\partial_u\psi\right)_r$: from Bondi-like to Cauchy coordinates}
\label{subsec:wt_transformation_bondi_like}
To compute $\left(\partial_u\psi\right)_r$, we need the following Jacobians
\begin{align}
    &(\partial_{\underline{u}})_{\underline{\lambda}}=(\partial_u)_{r}+(\partial_{\underline{u}}r)\partial_{r} \label{eq:jac_underline_bondi_like}, 
    &&(\partial_{t^\prime})_{r^\prime}=(\partial_{\underline{u}})_{\underline{\lambda}},
\end{align}
see Eqs.~(4.6) and (4.3) of \cite{Ma:2023qjn}. Restricting them to the worldtube, we obtain
\begin{align}
    (\partial_u\psi)_r&=(\partial_{t^\prime}\psi)_{r^\prime}-2\frac{\partial_{u}\wtR}{\wtR}\partial_{\Breve{y}}\psi, \label{eq:du-dtprime}
\end{align}
where $R_{\rm wt}$ is the radius of the worldtube in the Bondi-like coordinate system. It is related to $\hat{R}_{\rm wt}$ via 
\begin{align}
    R_{\rm wt}=\frac{\hat{R}_{\rm wt}}{\hat{\omega}}.
\end{align}
The first term $(\partial_{t^\prime}\psi)_{r^\prime}$ in Eq.~\eqref{eq:du-dtprime} is a dynamical variable of the Cauchy system, and the second term $\frac{\partial_{u}\wtR}{\wtR}\partial_{\Breve{y}}\psi$ is already available in the CCE system. Therefore, once we have collected the value of $(\partial_{t^\prime}\psi)_{r^\prime}$ from a Cauchy evolution, we can retrace our derivations presented in Secs.~\ref{subsec:wt_transformation_numerically_adapted}, \ref{subsec:wt_transformation_partiall_flat} and \ref{subsec:wt_transformation_bondi_like} to construct the boundary condition $\Breve{\Pi}|_{\rm wt}$.

\subsection{Simplication}
Combining Eqs.~\eqref{eq:numtheta_wt}, \eqref{eq:duhat-du} and \eqref{eq:du-dtprime}, we obtain a lengthy expression for $\Breve{\Pi}|_{\rm wt}$:
\begin{align}
\Breve{\Pi}|_{\rm wt}&=(\partial_{t^\prime}\psi)_{r^\prime}-2\frac{\partial_{u}\wtR}{\wtR}\partial_{\Breve{y}}\psi+{\rm Re}\,\mathcal{U}^{(0)} \bar{\hat{\eth}}\psi-2\frac{\left(\partial_{\hat{u}}\hat{\omega}\right)_{\hat{r}}-{\rm Re}\,\mathcal{U}^{(0)} \bar{\hat{\eth}}\hat{\omega}}{\hat{\omega}}\partial_{\Breve{y}}\psi+\frac{\partial_{\Breve{u}}\hat{R}_{\rm wt}}{\hat{R}_{\rm wt}}\partial_{\Breve{y}}\psi.
\end{align}
A pivotal step in simplifying it is to leverage an identity, see Eq.~(15b) of \cite{Moxon:2021gbv}\footnote{Note that there is a typo.}:
\begin{align}
    \frac{\partial_{\hat{u}}\hat{R}_{\rm wt}}{\hat{R}_{\rm wt}}=\frac{\partial_u\wtR}{\wtR}+\frac{\partial_{\hat{u}}\hat{\omega}}{\hat{\omega}}+{\rm Re}\,\mathcal{U}^{(0)}\frac{\hat{\eth}\wtR}{\wtR}. 
\end{align}
which leads to
\begin{align}
    &\Breve{\Pi}|_{\rm wt}=(\partial_{t^\prime}\psi)_{r^\prime}+ {\rm Re}\left[\mathcal{U}^{(0)}\left(\bar{\hat{\eth}}\psi+2\partial_{\Breve{y}}\psi\frac{\bar{\hat{\eth}}\hat{R}_{\rm wt}}{\hat{R}_{\rm wt}}\right)\right]. \label{eq:Pi_wt_before_simplification}
\end{align}
The expression becomes even more simplified when we transform the covariant angular derivative $\hat{\eth}$ from the partially flat Bondi-like coordinates to the numerically adapted ones $\Breve{\eth}$, via [Eq.~(46c) of \cite{Moxon:2020gha}]:
\begin{align}
\bar{\Breve{\eth}}\psi=\bar{\hat{\eth}}\psi+2\partial_{\Breve{y}}\psi\left(\frac{\bar{\hat{\eth}} \hat{R}_{\rm wt}}{\hat{R}_{\rm wt}}\right), \label{eq:eth_psi_eth_hat_psi}
\end{align}
where we have used the fact $\Breve{y}=-1$ and $\bar{\hat{\eth}} \hat{R}_{\rm wt}=\bar{\Breve{\eth}} \hat{R}_{\rm wt}$. Plugging Eq.~\eqref{eq:eth_psi_eth_hat_psi} into \eqref{eq:Pi_wt_before_simplification} leads to our final result given in Eq.~\eqref{eq:final_wt_transformation}.

\def\bibsection{\section*{References}}
\bibliography{References}

\begin{thebibliography}{126}%
\makeatletter
\providecommand \@ifxundefined [1]{%
 \@ifx{#1\undefined}
}%
\providecommand \@ifnum [1]{%
 \ifnum #1\expandafter \@firstoftwo
 \else \expandafter \@secondoftwo
 \fi
}%
\providecommand \@ifx [1]{%
 \ifx #1\expandafter \@firstoftwo
 \else \expandafter \@secondoftwo
 \fi
}%
\providecommand \natexlab [1]{#1}%
\providecommand \enquote  [1]{``#1''}%
\providecommand \bibnamefont  [1]{#1}%
\providecommand \bibfnamefont [1]{#1}%
\providecommand \citenamefont [1]{#1}%
\providecommand \href@noop [0]{\@secondoftwo}%
\providecommand \href [0]{\begingroup \@sanitize@url \@href}%
\providecommand \@href[1]{\@@startlink{#1}\@@href}%
\providecommand \@@href[1]{\endgroup#1\@@endlink}%
\providecommand \@sanitize@url [0]{\catcode `\\12\catcode `\$12\catcode `\&12\catcode `\#12\catcode `\^12\catcode `\_12\catcode `\%12\relax}%
\providecommand \@@startlink[1]{}%
\providecommand \@@endlink[0]{}%
\providecommand \url  [0]{\begingroup\@sanitize@url \@url }%
\providecommand \@url [1]{\endgroup\@href {#1}{\urlprefix }}%
\providecommand \urlprefix  [0]{URL }%
\providecommand \Eprint [0]{\href }%
\providecommand \doibase [0]{http://dx.doi.org/}%
\providecommand \selectlanguage [0]{\@gobble}%
\providecommand \bibinfo  [0]{\@secondoftwo}%
\providecommand \bibfield  [0]{\@secondoftwo}%
\providecommand \translation [1]{[#1]}%
\providecommand \BibitemOpen [0]{}%
\providecommand \bibitemStop [0]{}%
\providecommand \bibitemNoStop [0]{.\EOS\space}%
\providecommand \EOS [0]{\spacefactor3000\relax}%
\providecommand \BibitemShut  [1]{\csname bibitem#1\endcsname}%
\let\auto@bib@innerbib\@empty
\bibitem [{\citenamefont {Aasi}\ \emph {et~al.}(2015)\citenamefont {Aasi} \emph {et~al.}}]{TheLIGOScientific:2014jea}%
  \BibitemOpen
  \bibfield  {author} {\bibinfo {author} {\bibfnamefont {J.}~\bibnamefont {Aasi}} \emph {et~al.} (\bibinfo {collaboration} {LIGO Scientific}),\ }\href {\doibase 10.1088/0264-9381/32/7/074001} {\bibfield  {journal} {\bibinfo  {journal} {Class. Quant. Grav.}\ }\textbf {\bibinfo {volume} {32}},\ \bibinfo {pages} {074001} (\bibinfo {year} {2015})},\ \Eprint {http://arxiv.org/abs/1411.4547} {arXiv:1411.4547 [gr-qc]} \BibitemShut {NoStop}%
\bibitem [{\citenamefont {Acernese}\ \emph {et~al.}(2015)\citenamefont {Acernese} \emph {et~al.}}]{TheVirgo:2014hva}%
  \BibitemOpen
  \bibfield  {author} {\bibinfo {author} {\bibfnamefont {F.}~\bibnamefont {Acernese}} \emph {et~al.} (\bibinfo {collaboration} {Virgo}),\ }\href {\doibase 10.1088/0264-9381/32/2/024001} {\bibfield  {journal} {\bibinfo  {journal} {Class. Quant. Grav.}\ }\textbf {\bibinfo {volume} {32}},\ \bibinfo {pages} {024001} (\bibinfo {year} {2015})},\ \Eprint {http://arxiv.org/abs/1408.3978} {arXiv:1408.3978 [gr-qc]} \BibitemShut {NoStop}%
\bibitem [{\citenamefont {Somiya}(2012)}]{Somiya:2011np}%
  \BibitemOpen
  \bibfield  {author} {\bibinfo {author} {\bibfnamefont {K.}~\bibnamefont {Somiya}} (\bibinfo {collaboration} {KAGRA}),\ }\href {\doibase 10.1088/0264-9381/29/12/124007} {\bibfield  {journal} {\bibinfo  {journal} {Class. Quant. Grav.}\ }\textbf {\bibinfo {volume} {29}},\ \bibinfo {pages} {124007} (\bibinfo {year} {2012})},\ \Eprint {http://arxiv.org/abs/1111.7185} {arXiv:1111.7185 [gr-qc]} \BibitemShut {NoStop}%
\bibitem [{\citenamefont {Abbott}\ \emph {et~al.}(2016{\natexlab{a}})\citenamefont {Abbott} \emph {et~al.}}]{TheLIGOScientific:2016pea}%
  \BibitemOpen
  \bibfield  {author} {\bibinfo {author} {\bibfnamefont {B.~P.}\ \bibnamefont {Abbott}} \emph {et~al.} (\bibinfo {collaboration} {LIGO Scientific, Virgo}),\ }\href {\doibase 10.1103/PhysRevX.6.041015, 10.1103/PhysRevX.8.039903} {\bibfield  {journal} {\bibinfo  {journal} {Phys. Rev.}\ }\textbf {\bibinfo {volume} {X6}},\ \bibinfo {pages} {041015} (\bibinfo {year} {2016}{\natexlab{a}})},\ \bibinfo {note} {[erratum: Phys. Rev.X8,no.3,039903(2018)]},\ \Eprint {http://arxiv.org/abs/1606.04856} {arXiv:1606.04856 [gr-qc]} \BibitemShut {NoStop}%
\bibitem [{\citenamefont {Abbott}\ \emph {et~al.}(2019{\natexlab{a}})\citenamefont {Abbott} \emph {et~al.}}]{LIGOScientific:2018mvr}%
  \BibitemOpen
  \bibfield  {author} {\bibinfo {author} {\bibfnamefont {B.~P.}\ \bibnamefont {Abbott}} \emph {et~al.} (\bibinfo {collaboration} {LIGO Scientific, Virgo}),\ }\href {\doibase 10.1103/PhysRevX.9.031040} {\bibfield  {journal} {\bibinfo  {journal} {Phys. Rev.}\ }\textbf {\bibinfo {volume} {X9}},\ \bibinfo {pages} {031040} (\bibinfo {year} {2019}{\natexlab{a}})},\ \Eprint {http://arxiv.org/abs/1811.12907} {arXiv:1811.12907 [astro-ph.HE]} \BibitemShut {NoStop}%
\bibitem [{\citenamefont {Abbott}\ \emph {et~al.}(2021{\natexlab{a}})\citenamefont {Abbott} \emph {et~al.}}]{LIGOScientific:2020ibl}%
  \BibitemOpen
  \bibfield  {author} {\bibinfo {author} {\bibfnamefont {R.}~\bibnamefont {Abbott}} \emph {et~al.} (\bibinfo {collaboration} {LIGO Scientific, Virgo}),\ }\href {\doibase 10.1103/PhysRevX.11.021053} {\bibfield  {journal} {\bibinfo  {journal} {Phys. Rev. X}\ }\textbf {\bibinfo {volume} {11}},\ \bibinfo {pages} {021053} (\bibinfo {year} {2021}{\natexlab{a}})},\ \Eprint {http://arxiv.org/abs/2010.14527} {arXiv:2010.14527 [gr-qc]} \BibitemShut {NoStop}%
\bibitem [{\citenamefont {Abbott}\ \emph {et~al.}(2021{\natexlab{b}})\citenamefont {Abbott} \emph {et~al.}}]{LIGOScientific:2021djp}%
  \BibitemOpen
  \bibfield  {author} {\bibinfo {author} {\bibfnamefont {R.}~\bibnamefont {Abbott}} \emph {et~al.} (\bibinfo {collaboration} {LIGO Scientific, VIRGO, KAGRA}),\ }\href@noop {} {\  (\bibinfo {year} {2021}{\natexlab{b}})},\ \Eprint {http://arxiv.org/abs/2111.03606} {arXiv:2111.03606 [gr-qc]} \BibitemShut {NoStop}%
\bibitem [{\citenamefont {Abbott}\ \emph {et~al.}(2016{\natexlab{b}})\citenamefont {Abbott} \emph {et~al.}}]{LIGOScientific:2016lio}%
  \BibitemOpen
  \bibfield  {author} {\bibinfo {author} {\bibfnamefont {B.~P.}\ \bibnamefont {Abbott}} \emph {et~al.} (\bibinfo {collaboration} {LIGO Scientific, Virgo}),\ }\href {\doibase 10.1103/PhysRevLett.116.221101} {\bibfield  {journal} {\bibinfo  {journal} {Phys. Rev. Lett.}\ }\textbf {\bibinfo {volume} {116}},\ \bibinfo {pages} {221101} (\bibinfo {year} {2016}{\natexlab{b}})},\ \bibinfo {note} {[Erratum: Phys.Rev.Lett. 121, 129902 (2018)]},\ \Eprint {http://arxiv.org/abs/1602.03841} {arXiv:1602.03841 [gr-qc]} \BibitemShut {NoStop}%
\bibitem [{\citenamefont {Will}(2014)}]{Will:2014kxa}%
  \BibitemOpen
  \bibfield  {author} {\bibinfo {author} {\bibfnamefont {C.~M.}\ \bibnamefont {Will}},\ }\href {\doibase 10.12942/lrr-2014-4} {\bibfield  {journal} {\bibinfo  {journal} {Living Rev. Rel.}\ }\textbf {\bibinfo {volume} {17}},\ \bibinfo {pages} {4} (\bibinfo {year} {2014})},\ \Eprint {http://arxiv.org/abs/1403.7377} {arXiv:1403.7377 [gr-qc]} \BibitemShut {NoStop}%
\bibitem [{\citenamefont {Yunes}\ and\ \citenamefont {Siemens}(2013)}]{Yunes:2013dva}%
  \BibitemOpen
  \bibfield  {author} {\bibinfo {author} {\bibfnamefont {N.}~\bibnamefont {Yunes}}\ and\ \bibinfo {author} {\bibfnamefont {X.}~\bibnamefont {Siemens}},\ }\href {\doibase 10.12942/lrr-2013-9} {\bibfield  {journal} {\bibinfo  {journal} {Living Rev. Rel.}\ }\textbf {\bibinfo {volume} {16}},\ \bibinfo {pages} {9} (\bibinfo {year} {2013})},\ \Eprint {http://arxiv.org/abs/1304.3473} {arXiv:1304.3473 [gr-qc]} \BibitemShut {NoStop}%
\bibitem [{\citenamefont {Yunes}\ \emph {et~al.}(2016)\citenamefont {Yunes}, \citenamefont {Yagi},\ and\ \citenamefont {Pretorius}}]{Yunes:2016jcc}%
  \BibitemOpen
  \bibfield  {author} {\bibinfo {author} {\bibfnamefont {N.}~\bibnamefont {Yunes}}, \bibinfo {author} {\bibfnamefont {K.}~\bibnamefont {Yagi}}, \ and\ \bibinfo {author} {\bibfnamefont {F.}~\bibnamefont {Pretorius}},\ }\href {\doibase 10.1103/PhysRevD.94.084002} {\bibfield  {journal} {\bibinfo  {journal} {Phys. Rev. D}\ }\textbf {\bibinfo {volume} {94}},\ \bibinfo {pages} {084002} (\bibinfo {year} {2016})},\ \Eprint {http://arxiv.org/abs/1603.08955} {arXiv:1603.08955 [gr-qc]} \BibitemShut {NoStop}%
\bibitem [{\citenamefont {Berti}\ \emph {et~al.}(2015)\citenamefont {Berti} \emph {et~al.}}]{Berti:2015itd}%
  \BibitemOpen
  \bibfield  {author} {\bibinfo {author} {\bibfnamefont {E.}~\bibnamefont {Berti}} \emph {et~al.},\ }\href {\doibase 10.1088/0264-9381/32/24/243001} {\bibfield  {journal} {\bibinfo  {journal} {Class. Quant. Grav.}\ }\textbf {\bibinfo {volume} {32}},\ \bibinfo {pages} {243001} (\bibinfo {year} {2015})},\ \Eprint {http://arxiv.org/abs/1501.07274} {arXiv:1501.07274 [gr-qc]} \BibitemShut {NoStop}%
\bibitem [{\citenamefont {Abbott}\ \emph {et~al.}(2019{\natexlab{b}})\citenamefont {Abbott} \emph {et~al.}}]{LIGOScientific:2018dkp}%
  \BibitemOpen
  \bibfield  {author} {\bibinfo {author} {\bibfnamefont {B.~P.}\ \bibnamefont {Abbott}} \emph {et~al.} (\bibinfo {collaboration} {LIGO Scientific, Virgo}),\ }\href {\doibase 10.1103/PhysRevLett.123.011102} {\bibfield  {journal} {\bibinfo  {journal} {Phys. Rev. Lett.}\ }\textbf {\bibinfo {volume} {123}},\ \bibinfo {pages} {011102} (\bibinfo {year} {2019}{\natexlab{b}})},\ \Eprint {http://arxiv.org/abs/1811.00364} {arXiv:1811.00364 [gr-qc]} \BibitemShut {NoStop}%
\bibitem [{\citenamefont {Krolak}\ \emph {et~al.}(1995)\citenamefont {Krolak}, \citenamefont {Kokkotas},\ and\ \citenamefont {Schaefer}}]{Krolak:1995md}%
  \BibitemOpen
  \bibfield  {author} {\bibinfo {author} {\bibfnamefont {A.}~\bibnamefont {Krolak}}, \bibinfo {author} {\bibfnamefont {K.~D.}\ \bibnamefont {Kokkotas}}, \ and\ \bibinfo {author} {\bibfnamefont {G.}~\bibnamefont {Schaefer}},\ }\href {\doibase 10.1103/PhysRevD.52.2089} {\bibfield  {journal} {\bibinfo  {journal} {Phys. Rev. D}\ }\textbf {\bibinfo {volume} {52}},\ \bibinfo {pages} {2089} (\bibinfo {year} {1995})},\ \Eprint {http://arxiv.org/abs/gr-qc/9503013} {arXiv:gr-qc/9503013} \BibitemShut {NoStop}%
\bibitem [{\citenamefont {Yagi}\ and\ \citenamefont {Tanaka}(2010{\natexlab{a}})}]{Yagi:2009zz}%
  \BibitemOpen
  \bibfield  {author} {\bibinfo {author} {\bibfnamefont {K.}~\bibnamefont {Yagi}}\ and\ \bibinfo {author} {\bibfnamefont {T.}~\bibnamefont {Tanaka}},\ }\href {\doibase 10.1143/PTP.123.1069} {\bibfield  {journal} {\bibinfo  {journal} {Prog. Theor. Phys.}\ }\textbf {\bibinfo {volume} {123}},\ \bibinfo {pages} {1069} (\bibinfo {year} {2010}{\natexlab{a}})},\ \Eprint {http://arxiv.org/abs/0908.3283} {arXiv:0908.3283 [gr-qc]} \BibitemShut {NoStop}%
\bibitem [{\citenamefont {Ma}\ and\ \citenamefont {Yunes}(2019)}]{Ma:2019rei}%
  \BibitemOpen
  \bibfield  {author} {\bibinfo {author} {\bibfnamefont {S.}~\bibnamefont {Ma}}\ and\ \bibinfo {author} {\bibfnamefont {N.}~\bibnamefont {Yunes}},\ }\href {\doibase 10.1103/PhysRevD.100.124032} {\bibfield  {journal} {\bibinfo  {journal} {Phys. Rev. D}\ }\textbf {\bibinfo {volume} {100}},\ \bibinfo {pages} {124032} (\bibinfo {year} {2019})},\ \Eprint {http://arxiv.org/abs/1908.07089} {arXiv:1908.07089 [gr-qc]} \BibitemShut {NoStop}%
\bibitem [{\citenamefont {Carson}\ \emph {et~al.}(2020)\citenamefont {Carson}, \citenamefont {Seymour},\ and\ \citenamefont {Yagi}}]{Carson:2019fxr}%
  \BibitemOpen
  \bibfield  {author} {\bibinfo {author} {\bibfnamefont {Z.}~\bibnamefont {Carson}}, \bibinfo {author} {\bibfnamefont {B.~C.}\ \bibnamefont {Seymour}}, \ and\ \bibinfo {author} {\bibfnamefont {K.}~\bibnamefont {Yagi}},\ }\href {\doibase 10.1088/1361-6382/ab6a1f} {\bibfield  {journal} {\bibinfo  {journal} {Class. Quant. Grav.}\ }\textbf {\bibinfo {volume} {37}},\ \bibinfo {pages} {065008} (\bibinfo {year} {2020})},\ \Eprint {http://arxiv.org/abs/1907.03897} {arXiv:1907.03897 [gr-qc]} \BibitemShut {NoStop}%
\bibitem [{\citenamefont {Sampson}\ \emph {et~al.}(2014)\citenamefont {Sampson}, \citenamefont {Yunes}, \citenamefont {Cornish}, \citenamefont {Ponce}, \citenamefont {Barausse}, \citenamefont {Klein}, \citenamefont {Palenzuela},\ and\ \citenamefont {Lehner}}]{Sampson:2014qqa}%
  \BibitemOpen
  \bibfield  {author} {\bibinfo {author} {\bibfnamefont {L.}~\bibnamefont {Sampson}}, \bibinfo {author} {\bibfnamefont {N.}~\bibnamefont {Yunes}}, \bibinfo {author} {\bibfnamefont {N.}~\bibnamefont {Cornish}}, \bibinfo {author} {\bibfnamefont {M.}~\bibnamefont {Ponce}}, \bibinfo {author} {\bibfnamefont {E.}~\bibnamefont {Barausse}}, \bibinfo {author} {\bibfnamefont {A.}~\bibnamefont {Klein}}, \bibinfo {author} {\bibfnamefont {C.}~\bibnamefont {Palenzuela}}, \ and\ \bibinfo {author} {\bibfnamefont {L.}~\bibnamefont {Lehner}},\ }\href {\doibase 10.1103/PhysRevD.90.124091} {\bibfield  {journal} {\bibinfo  {journal} {Phys. Rev. D}\ }\textbf {\bibinfo {volume} {90}},\ \bibinfo {pages} {124091} (\bibinfo {year} {2014})},\ \Eprint {http://arxiv.org/abs/1407.7038} {arXiv:1407.7038 [gr-qc]} \BibitemShut {NoStop}%
\bibitem [{\citenamefont {Scharre}\ and\ \citenamefont {Will}(2002)}]{Scharre:2001hn}%
  \BibitemOpen
  \bibfield  {author} {\bibinfo {author} {\bibfnamefont {P.~D.}\ \bibnamefont {Scharre}}\ and\ \bibinfo {author} {\bibfnamefont {C.~M.}\ \bibnamefont {Will}},\ }\href {\doibase 10.1103/PhysRevD.65.042002} {\bibfield  {journal} {\bibinfo  {journal} {Phys. Rev. D}\ }\textbf {\bibinfo {volume} {65}},\ \bibinfo {pages} {042002} (\bibinfo {year} {2002})},\ \Eprint {http://arxiv.org/abs/gr-qc/0109044} {arXiv:gr-qc/0109044} \BibitemShut {NoStop}%
\bibitem [{\citenamefont {Will}\ and\ \citenamefont {Yunes}(2004)}]{Will:2004xi}%
  \BibitemOpen
  \bibfield  {author} {\bibinfo {author} {\bibfnamefont {C.~M.}\ \bibnamefont {Will}}\ and\ \bibinfo {author} {\bibfnamefont {N.}~\bibnamefont {Yunes}},\ }\href {\doibase 10.1088/0264-9381/21/18/006} {\bibfield  {journal} {\bibinfo  {journal} {Class. Quant. Grav.}\ }\textbf {\bibinfo {volume} {21}},\ \bibinfo {pages} {4367} (\bibinfo {year} {2004})},\ \Eprint {http://arxiv.org/abs/gr-qc/0403100} {arXiv:gr-qc/0403100} \BibitemShut {NoStop}%
\bibitem [{\citenamefont {Berti}\ \emph {et~al.}(2005{\natexlab{a}})\citenamefont {Berti}, \citenamefont {Buonanno},\ and\ \citenamefont {Will}}]{Berti:2005qd}%
  \BibitemOpen
  \bibfield  {author} {\bibinfo {author} {\bibfnamefont {E.}~\bibnamefont {Berti}}, \bibinfo {author} {\bibfnamefont {A.}~\bibnamefont {Buonanno}}, \ and\ \bibinfo {author} {\bibfnamefont {C.~M.}\ \bibnamefont {Will}},\ }\href {\doibase 10.1088/0264-9381/22/18/S08} {\bibfield  {journal} {\bibinfo  {journal} {Class. Quant. Grav.}\ }\textbf {\bibinfo {volume} {22}},\ \bibinfo {pages} {S943} (\bibinfo {year} {2005}{\natexlab{a}})},\ \Eprint {http://arxiv.org/abs/gr-qc/0504017} {arXiv:gr-qc/0504017} \BibitemShut {NoStop}%
\bibitem [{\citenamefont {Berti}\ \emph {et~al.}(2005{\natexlab{b}})\citenamefont {Berti}, \citenamefont {Buonanno},\ and\ \citenamefont {Will}}]{Berti:2004bd}%
  \BibitemOpen
  \bibfield  {author} {\bibinfo {author} {\bibfnamefont {E.}~\bibnamefont {Berti}}, \bibinfo {author} {\bibfnamefont {A.}~\bibnamefont {Buonanno}}, \ and\ \bibinfo {author} {\bibfnamefont {C.~M.}\ \bibnamefont {Will}},\ }\href {\doibase 10.1103/PhysRevD.71.084025} {\bibfield  {journal} {\bibinfo  {journal} {Phys. Rev. D}\ }\textbf {\bibinfo {volume} {71}},\ \bibinfo {pages} {084025} (\bibinfo {year} {2005}{\natexlab{b}})},\ \Eprint {http://arxiv.org/abs/gr-qc/0411129} {arXiv:gr-qc/0411129} \BibitemShut {NoStop}%
\bibitem [{\citenamefont {Yagi}\ and\ \citenamefont {Tanaka}(2010{\natexlab{b}})}]{Yagi:2009zm}%
  \BibitemOpen
  \bibfield  {author} {\bibinfo {author} {\bibfnamefont {K.}~\bibnamefont {Yagi}}\ and\ \bibinfo {author} {\bibfnamefont {T.}~\bibnamefont {Tanaka}},\ }\href {\doibase 10.1103/PhysRevD.81.109902} {\bibfield  {journal} {\bibinfo  {journal} {Phys. Rev. D}\ }\textbf {\bibinfo {volume} {81}},\ \bibinfo {pages} {064008} (\bibinfo {year} {2010}{\natexlab{b}})},\ \bibinfo {note} {[Erratum: Phys.Rev.D 81, 109902 (2010)]},\ \Eprint {http://arxiv.org/abs/0906.4269} {arXiv:0906.4269 [gr-qc]} \BibitemShut {NoStop}%
\bibitem [{\citenamefont {Arun}(2012)}]{Arun:2012hf}%
  \BibitemOpen
  \bibfield  {author} {\bibinfo {author} {\bibfnamefont {K.~G.}\ \bibnamefont {Arun}},\ }\href {\doibase 10.1088/0264-9381/29/7/075011} {\bibfield  {journal} {\bibinfo  {journal} {Class. Quant. Grav.}\ }\textbf {\bibinfo {volume} {29}},\ \bibinfo {pages} {075011} (\bibinfo {year} {2012})},\ \Eprint {http://arxiv.org/abs/1202.5911} {arXiv:1202.5911 [gr-qc]} \BibitemShut {NoStop}%
\bibitem [{\citenamefont {Cardoso}\ \emph {et~al.}(2011)\citenamefont {Cardoso}, \citenamefont {Chakrabarti}, \citenamefont {Pani}, \citenamefont {Berti},\ and\ \citenamefont {Gualtieri}}]{Cardoso:2011xi}%
  \BibitemOpen
  \bibfield  {author} {\bibinfo {author} {\bibfnamefont {V.}~\bibnamefont {Cardoso}}, \bibinfo {author} {\bibfnamefont {S.}~\bibnamefont {Chakrabarti}}, \bibinfo {author} {\bibfnamefont {P.}~\bibnamefont {Pani}}, \bibinfo {author} {\bibfnamefont {E.}~\bibnamefont {Berti}}, \ and\ \bibinfo {author} {\bibfnamefont {L.}~\bibnamefont {Gualtieri}},\ }\href {\doibase 10.1103/PhysRevLett.107.241101} {\bibfield  {journal} {\bibinfo  {journal} {Phys. Rev. Lett.}\ }\textbf {\bibinfo {volume} {107}},\ \bibinfo {pages} {241101} (\bibinfo {year} {2011})},\ \Eprint {http://arxiv.org/abs/1109.6021} {arXiv:1109.6021 [gr-qc]} \BibitemShut {NoStop}%
\bibitem [{\citenamefont {Yunes}\ \emph {et~al.}(2012)\citenamefont {Yunes}, \citenamefont {Pani},\ and\ \citenamefont {Cardoso}}]{Yunes:2011aa}%
  \BibitemOpen
  \bibfield  {author} {\bibinfo {author} {\bibfnamefont {N.}~\bibnamefont {Yunes}}, \bibinfo {author} {\bibfnamefont {P.}~\bibnamefont {Pani}}, \ and\ \bibinfo {author} {\bibfnamefont {V.}~\bibnamefont {Cardoso}},\ }\href {\doibase 10.1103/PhysRevD.85.102003} {\bibfield  {journal} {\bibinfo  {journal} {Phys. Rev. D}\ }\textbf {\bibinfo {volume} {85}},\ \bibinfo {pages} {102003} (\bibinfo {year} {2012})},\ \Eprint {http://arxiv.org/abs/1112.3351} {arXiv:1112.3351 [gr-qc]} \BibitemShut {NoStop}%
\bibitem [{\citenamefont {Berti}\ \emph {et~al.}(2012)\citenamefont {Berti}, \citenamefont {Gualtieri}, \citenamefont {Horbatsch},\ and\ \citenamefont {Alsing}}]{Berti:2012bp}%
  \BibitemOpen
  \bibfield  {author} {\bibinfo {author} {\bibfnamefont {E.}~\bibnamefont {Berti}}, \bibinfo {author} {\bibfnamefont {L.}~\bibnamefont {Gualtieri}}, \bibinfo {author} {\bibfnamefont {M.}~\bibnamefont {Horbatsch}}, \ and\ \bibinfo {author} {\bibfnamefont {J.}~\bibnamefont {Alsing}},\ }\href {\doibase 10.1103/PhysRevD.85.122005} {\bibfield  {journal} {\bibinfo  {journal} {Phys. Rev. D}\ }\textbf {\bibinfo {volume} {85}},\ \bibinfo {pages} {122005} (\bibinfo {year} {2012})},\ \Eprint {http://arxiv.org/abs/1204.4340} {arXiv:1204.4340 [gr-qc]} \BibitemShut {NoStop}%
\bibitem [{\citenamefont {Tuna}\ \emph {et~al.}(2022)\citenamefont {Tuna}, \citenamefont {\"Unl\"ut\"urk},\ and\ \citenamefont {Ramazano\u{g}lu}}]{Tuna:2022qqr}%
  \BibitemOpen
  \bibfield  {author} {\bibinfo {author} {\bibfnamefont {S.}~\bibnamefont {Tuna}}, \bibinfo {author} {\bibfnamefont {K.~I.}\ \bibnamefont {\"Unl\"ut\"urk}}, \ and\ \bibinfo {author} {\bibfnamefont {F.~M.}\ \bibnamefont {Ramazano\u{g}lu}},\ }\href {\doibase 10.1103/PhysRevD.105.124070} {\bibfield  {journal} {\bibinfo  {journal} {Phys. Rev. D}\ }\textbf {\bibinfo {volume} {105}},\ \bibinfo {pages} {124070} (\bibinfo {year} {2022})},\ \Eprint {http://arxiv.org/abs/2204.02138} {arXiv:2204.02138 [gr-qc]} \BibitemShut {NoStop}%
\bibitem [{\citenamefont {Pretorius}(2005)}]{Pretorius:2005gq}%
  \BibitemOpen
  \bibfield  {author} {\bibinfo {author} {\bibfnamefont {F.}~\bibnamefont {Pretorius}},\ }\href {\doibase 10.1103/PhysRevLett.95.121101} {\bibfield  {journal} {\bibinfo  {journal} {Phys. Rev. Lett.}\ }\textbf {\bibinfo {volume} {95}},\ \bibinfo {pages} {121101} (\bibinfo {year} {2005})},\ \Eprint {http://arxiv.org/abs/gr-qc/0507014} {arXiv:gr-qc/0507014 [gr-qc]} \BibitemShut {NoStop}%
\bibitem [{\citenamefont {Ripley}(2022)}]{Ripley:2022cdh}%
  \BibitemOpen
  \bibfield  {author} {\bibinfo {author} {\bibfnamefont {J.~L.}\ \bibnamefont {Ripley}},\ }\href {\doibase 10.1142/S0218271822300178} {\bibfield  {journal} {\bibinfo  {journal} {Int. J. Mod. Phys. D}\ }\textbf {\bibinfo {volume} {31}},\ \bibinfo {pages} {2230017} (\bibinfo {year} {2022})},\ \Eprint {http://arxiv.org/abs/2207.13074} {arXiv:2207.13074 [gr-qc]} \BibitemShut {NoStop}%
\bibitem [{\citenamefont {East}\ and\ \citenamefont {Ripley}(2021{\natexlab{a}})}]{East:2020hgw}%
  \BibitemOpen
  \bibfield  {author} {\bibinfo {author} {\bibfnamefont {W.~E.}\ \bibnamefont {East}}\ and\ \bibinfo {author} {\bibfnamefont {J.~L.}\ \bibnamefont {Ripley}},\ }\href {\doibase 10.1103/PhysRevD.103.044040} {\bibfield  {journal} {\bibinfo  {journal} {Phys. Rev. D}\ }\textbf {\bibinfo {volume} {103}},\ \bibinfo {pages} {044040} (\bibinfo {year} {2021}{\natexlab{a}})},\ \Eprint {http://arxiv.org/abs/2011.03547} {arXiv:2011.03547 [gr-qc]} \BibitemShut {NoStop}%
\bibitem [{\citenamefont {East}\ and\ \citenamefont {Ripley}(2021{\natexlab{b}})}]{East:2021bqk}%
  \BibitemOpen
  \bibfield  {author} {\bibinfo {author} {\bibfnamefont {W.~E.}\ \bibnamefont {East}}\ and\ \bibinfo {author} {\bibfnamefont {J.~L.}\ \bibnamefont {Ripley}},\ }\href {\doibase 10.1103/PhysRevLett.127.101102} {\bibfield  {journal} {\bibinfo  {journal} {Phys. Rev. Lett.}\ }\textbf {\bibinfo {volume} {127}},\ \bibinfo {pages} {101102} (\bibinfo {year} {2021}{\natexlab{b}})},\ \Eprint {http://arxiv.org/abs/2105.08571} {arXiv:2105.08571 [gr-qc]} \BibitemShut {NoStop}%
\bibitem [{\citenamefont {East}\ and\ \citenamefont {Pretorius}(2022)}]{East:2022rqi}%
  \BibitemOpen
  \bibfield  {author} {\bibinfo {author} {\bibfnamefont {W.~E.}\ \bibnamefont {East}}\ and\ \bibinfo {author} {\bibfnamefont {F.}~\bibnamefont {Pretorius}},\ }\href {\doibase 10.1103/PhysRevD.106.104055} {\bibfield  {journal} {\bibinfo  {journal} {Phys. Rev. D}\ }\textbf {\bibinfo {volume} {106}},\ \bibinfo {pages} {104055} (\bibinfo {year} {2022})},\ \Eprint {http://arxiv.org/abs/2208.09488} {arXiv:2208.09488 [gr-qc]} \BibitemShut {NoStop}%
\bibitem [{\citenamefont {Corman}\ \emph {et~al.}(2023)\citenamefont {Corman}, \citenamefont {Ripley},\ and\ \citenamefont {East}}]{Corman:2022xqg}%
  \BibitemOpen
  \bibfield  {author} {\bibinfo {author} {\bibfnamefont {M.}~\bibnamefont {Corman}}, \bibinfo {author} {\bibfnamefont {J.~L.}\ \bibnamefont {Ripley}}, \ and\ \bibinfo {author} {\bibfnamefont {W.~E.}\ \bibnamefont {East}},\ }\href {\doibase 10.1103/PhysRevD.107.024014} {\bibfield  {journal} {\bibinfo  {journal} {Phys. Rev. D}\ }\textbf {\bibinfo {volume} {107}},\ \bibinfo {pages} {024014} (\bibinfo {year} {2023})},\ \Eprint {http://arxiv.org/abs/2210.09235} {arXiv:2210.09235 [gr-qc]} \BibitemShut {NoStop}%
\bibitem [{\citenamefont {Corman}\ and\ \citenamefont {East}(2024)}]{Corman:2024vlk}%
  \BibitemOpen
  \bibfield  {author} {\bibinfo {author} {\bibfnamefont {M.}~\bibnamefont {Corman}}\ and\ \bibinfo {author} {\bibfnamefont {W.~E.}\ \bibnamefont {East}},\ }\href@noop {} {\  (\bibinfo {year} {2024})},\ \Eprint {http://arxiv.org/abs/2405.18496} {arXiv:2405.18496 [gr-qc]} \BibitemShut {NoStop}%
\bibitem [{\citenamefont {Kov\'acs}\ and\ \citenamefont {Reall}(2020{\natexlab{a}})}]{Kovacs:2020pns}%
  \BibitemOpen
  \bibfield  {author} {\bibinfo {author} {\bibfnamefont {A.~D.}\ \bibnamefont {Kov\'acs}}\ and\ \bibinfo {author} {\bibfnamefont {H.~S.}\ \bibnamefont {Reall}},\ }\href {\doibase 10.1103/PhysRevLett.124.221101} {\bibfield  {journal} {\bibinfo  {journal} {Phys. Rev. Lett.}\ }\textbf {\bibinfo {volume} {124}},\ \bibinfo {pages} {221101} (\bibinfo {year} {2020}{\natexlab{a}})},\ \Eprint {http://arxiv.org/abs/2003.04327} {arXiv:2003.04327 [gr-qc]} \BibitemShut {NoStop}%
\bibitem [{\citenamefont {Kov\'acs}\ and\ \citenamefont {Reall}(2020{\natexlab{b}})}]{Kovacs:2020ywu}%
  \BibitemOpen
  \bibfield  {author} {\bibinfo {author} {\bibfnamefont {A.~D.}\ \bibnamefont {Kov\'acs}}\ and\ \bibinfo {author} {\bibfnamefont {H.~S.}\ \bibnamefont {Reall}},\ }\href {\doibase 10.1103/PhysRevD.101.124003} {\bibfield  {journal} {\bibinfo  {journal} {Phys. Rev. D}\ }\textbf {\bibinfo {volume} {101}},\ \bibinfo {pages} {124003} (\bibinfo {year} {2020}{\natexlab{b}})},\ \Eprint {http://arxiv.org/abs/2003.08398} {arXiv:2003.08398 [gr-qc]} \BibitemShut {NoStop}%
\bibitem [{\citenamefont {Healy}\ \emph {et~al.}(2012)\citenamefont {Healy}, \citenamefont {Bode}, \citenamefont {Haas}, \citenamefont {Pazos}, \citenamefont {Laguna}, \citenamefont {Shoemaker},\ and\ \citenamefont {Yunes}}]{Healy:2011ef}%
  \BibitemOpen
  \bibfield  {author} {\bibinfo {author} {\bibfnamefont {J.}~\bibnamefont {Healy}}, \bibinfo {author} {\bibfnamefont {T.}~\bibnamefont {Bode}}, \bibinfo {author} {\bibfnamefont {R.}~\bibnamefont {Haas}}, \bibinfo {author} {\bibfnamefont {E.}~\bibnamefont {Pazos}}, \bibinfo {author} {\bibfnamefont {P.}~\bibnamefont {Laguna}}, \bibinfo {author} {\bibfnamefont {D.}~\bibnamefont {Shoemaker}}, \ and\ \bibinfo {author} {\bibfnamefont {N.}~\bibnamefont {Yunes}},\ }\href {\doibase 10.1088/0264-9381/29/23/232002} {\bibfield  {journal} {\bibinfo  {journal} {Class. Quant. Grav.}\ }\textbf {\bibinfo {volume} {29}},\ \bibinfo {pages} {232002} (\bibinfo {year} {2012})},\ \Eprint {http://arxiv.org/abs/1112.3928} {arXiv:1112.3928 [gr-qc]} \BibitemShut {NoStop}%
\bibitem [{\citenamefont {Barausse}\ \emph {et~al.}(2013)\citenamefont {Barausse}, \citenamefont {Palenzuela}, \citenamefont {Ponce},\ and\ \citenamefont {Lehner}}]{Barausse:2012da}%
  \BibitemOpen
  \bibfield  {author} {\bibinfo {author} {\bibfnamefont {E.}~\bibnamefont {Barausse}}, \bibinfo {author} {\bibfnamefont {C.}~\bibnamefont {Palenzuela}}, \bibinfo {author} {\bibfnamefont {M.}~\bibnamefont {Ponce}}, \ and\ \bibinfo {author} {\bibfnamefont {L.}~\bibnamefont {Lehner}},\ }\href {\doibase 10.1103/PhysRevD.87.081506} {\bibfield  {journal} {\bibinfo  {journal} {Phys. Rev. D}\ }\textbf {\bibinfo {volume} {87}},\ \bibinfo {pages} {081506} (\bibinfo {year} {2013})},\ \Eprint {http://arxiv.org/abs/1212.5053} {arXiv:1212.5053 [gr-qc]} \BibitemShut {NoStop}%
\bibitem [{\citenamefont {Shibata}\ \emph {et~al.}(2014)\citenamefont {Shibata}, \citenamefont {Taniguchi}, \citenamefont {Okawa},\ and\ \citenamefont {Buonanno}}]{Shibata:2013pra}%
  \BibitemOpen
  \bibfield  {author} {\bibinfo {author} {\bibfnamefont {M.}~\bibnamefont {Shibata}}, \bibinfo {author} {\bibfnamefont {K.}~\bibnamefont {Taniguchi}}, \bibinfo {author} {\bibfnamefont {H.}~\bibnamefont {Okawa}}, \ and\ \bibinfo {author} {\bibfnamefont {A.}~\bibnamefont {Buonanno}},\ }\href {\doibase 10.1103/PhysRevD.89.084005} {\bibfield  {journal} {\bibinfo  {journal} {Phys. Rev. D}\ }\textbf {\bibinfo {volume} {89}},\ \bibinfo {pages} {084005} (\bibinfo {year} {2014})},\ \Eprint {http://arxiv.org/abs/1310.0627} {arXiv:1310.0627 [gr-qc]} \BibitemShut {NoStop}%
\bibitem [{\citenamefont {Bezares}\ \emph {et~al.}(2022)\citenamefont {Bezares}, \citenamefont {Aguilera-Miret}, \citenamefont {ter Haar}, \citenamefont {Crisostomi}, \citenamefont {Palenzuela},\ and\ \citenamefont {Barausse}}]{Bezares:2021dma}%
  \BibitemOpen
  \bibfield  {author} {\bibinfo {author} {\bibfnamefont {M.}~\bibnamefont {Bezares}}, \bibinfo {author} {\bibfnamefont {R.}~\bibnamefont {Aguilera-Miret}}, \bibinfo {author} {\bibfnamefont {L.}~\bibnamefont {ter Haar}}, \bibinfo {author} {\bibfnamefont {M.}~\bibnamefont {Crisostomi}}, \bibinfo {author} {\bibfnamefont {C.}~\bibnamefont {Palenzuela}}, \ and\ \bibinfo {author} {\bibfnamefont {E.}~\bibnamefont {Barausse}},\ }\href {\doibase 10.1103/PhysRevLett.128.091103} {\bibfield  {journal} {\bibinfo  {journal} {Phys. Rev. Lett.}\ }\textbf {\bibinfo {volume} {128}},\ \bibinfo {pages} {091103} (\bibinfo {year} {2022})},\ \Eprint {http://arxiv.org/abs/2107.05648} {arXiv:2107.05648 [gr-qc]} \BibitemShut {NoStop}%
\bibitem [{\citenamefont {Ma}\ \emph {et~al.}(2023)\citenamefont {Ma}, \citenamefont {Varma}, \citenamefont {Stein}, \citenamefont {Foucart}, \citenamefont {Duez}, \citenamefont {Kidder}, \citenamefont {Pfeiffer},\ and\ \citenamefont {Scheel}}]{Ma:2023sok}%
  \BibitemOpen
  \bibfield  {author} {\bibinfo {author} {\bibfnamefont {S.}~\bibnamefont {Ma}}, \bibinfo {author} {\bibfnamefont {V.}~\bibnamefont {Varma}}, \bibinfo {author} {\bibfnamefont {L.~C.}\ \bibnamefont {Stein}}, \bibinfo {author} {\bibfnamefont {F.}~\bibnamefont {Foucart}}, \bibinfo {author} {\bibfnamefont {M.~D.}\ \bibnamefont {Duez}}, \bibinfo {author} {\bibfnamefont {L.~E.}\ \bibnamefont {Kidder}}, \bibinfo {author} {\bibfnamefont {H.~P.}\ \bibnamefont {Pfeiffer}}, \ and\ \bibinfo {author} {\bibfnamefont {M.~A.}\ \bibnamefont {Scheel}},\ }\href {\doibase 10.1103/PhysRevD.107.124051} {\bibfield  {journal} {\bibinfo  {journal} {Phys. Rev. D}\ }\textbf {\bibinfo {volume} {107}},\ \bibinfo {pages} {124051} (\bibinfo {year} {2023})},\ \Eprint {http://arxiv.org/abs/2304.11836} {arXiv:2304.11836 [gr-qc]} \BibitemShut {NoStop}%
\bibitem [{\citenamefont {Okounkova}\ \emph {et~al.}(2023)\citenamefont {Okounkova}, \citenamefont {Isi}, \citenamefont {Chatziioannou},\ and\ \citenamefont {Farr}}]{Okounkova:2022grv}%
  \BibitemOpen
  \bibfield  {author} {\bibinfo {author} {\bibfnamefont {M.}~\bibnamefont {Okounkova}}, \bibinfo {author} {\bibfnamefont {M.}~\bibnamefont {Isi}}, \bibinfo {author} {\bibfnamefont {K.}~\bibnamefont {Chatziioannou}}, \ and\ \bibinfo {author} {\bibfnamefont {W.~M.}\ \bibnamefont {Farr}},\ }\href {\doibase 10.1103/PhysRevD.107.024046} {\bibfield  {journal} {\bibinfo  {journal} {Phys. Rev. D}\ }\textbf {\bibinfo {volume} {107}},\ \bibinfo {pages} {024046} (\bibinfo {year} {2023})},\ \Eprint {http://arxiv.org/abs/2208.02805} {arXiv:2208.02805 [gr-qc]} \BibitemShut {NoStop}%
\bibitem [{\citenamefont {Okounkova}\ \emph {et~al.}(2017)\citenamefont {Okounkova}, \citenamefont {Stein}, \citenamefont {Scheel},\ and\ \citenamefont {Hemberger}}]{Okounkova:2017yby}%
  \BibitemOpen
  \bibfield  {author} {\bibinfo {author} {\bibfnamefont {M.}~\bibnamefont {Okounkova}}, \bibinfo {author} {\bibfnamefont {L.~C.}\ \bibnamefont {Stein}}, \bibinfo {author} {\bibfnamefont {M.~A.}\ \bibnamefont {Scheel}}, \ and\ \bibinfo {author} {\bibfnamefont {D.~A.}\ \bibnamefont {Hemberger}},\ }\href {\doibase 10.1103/PhysRevD.96.044020} {\bibfield  {journal} {\bibinfo  {journal} {Phys. Rev. D}\ }\textbf {\bibinfo {volume} {96}},\ \bibinfo {pages} {044020} (\bibinfo {year} {2017})},\ \Eprint {http://arxiv.org/abs/1705.07924} {arXiv:1705.07924 [gr-qc]} \BibitemShut {NoStop}%
\bibitem [{\citenamefont {Okounkova}\ \emph {et~al.}(2020)\citenamefont {Okounkova}, \citenamefont {Stein}, \citenamefont {Moxon}, \citenamefont {Scheel},\ and\ \citenamefont {Teukolsky}}]{Okounkova:2019zjf}%
  \BibitemOpen
  \bibfield  {author} {\bibinfo {author} {\bibfnamefont {M.}~\bibnamefont {Okounkova}}, \bibinfo {author} {\bibfnamefont {L.~C.}\ \bibnamefont {Stein}}, \bibinfo {author} {\bibfnamefont {J.}~\bibnamefont {Moxon}}, \bibinfo {author} {\bibfnamefont {M.~A.}\ \bibnamefont {Scheel}}, \ and\ \bibinfo {author} {\bibfnamefont {S.~A.}\ \bibnamefont {Teukolsky}},\ }\href {\doibase 10.1103/PhysRevD.101.104016} {\bibfield  {journal} {\bibinfo  {journal} {Phys. Rev. D}\ }\textbf {\bibinfo {volume} {101}},\ \bibinfo {pages} {104016} (\bibinfo {year} {2020})},\ \Eprint {http://arxiv.org/abs/1911.02588} {arXiv:1911.02588 [gr-qc]} \BibitemShut {NoStop}%
\bibitem [{\citenamefont {Witek}\ \emph {et~al.}(2019)\citenamefont {Witek}, \citenamefont {Gualtieri}, \citenamefont {Pani},\ and\ \citenamefont {Sotiriou}}]{Witek:2018dmd}%
  \BibitemOpen
  \bibfield  {author} {\bibinfo {author} {\bibfnamefont {H.}~\bibnamefont {Witek}}, \bibinfo {author} {\bibfnamefont {L.}~\bibnamefont {Gualtieri}}, \bibinfo {author} {\bibfnamefont {P.}~\bibnamefont {Pani}}, \ and\ \bibinfo {author} {\bibfnamefont {T.~P.}\ \bibnamefont {Sotiriou}},\ }\href {\doibase 10.1103/PhysRevD.99.064035} {\bibfield  {journal} {\bibinfo  {journal} {Phys. Rev. D}\ }\textbf {\bibinfo {volume} {99}},\ \bibinfo {pages} {064035} (\bibinfo {year} {2019})},\ \Eprint {http://arxiv.org/abs/1810.05177} {arXiv:1810.05177 [gr-qc]} \BibitemShut {NoStop}%
\bibitem [{\citenamefont {Okounkova}(2020)}]{Okounkova:2020rqw}%
  \BibitemOpen
  \bibfield  {author} {\bibinfo {author} {\bibfnamefont {M.}~\bibnamefont {Okounkova}},\ }\href {\doibase 10.1103/PhysRevD.102.084046} {\bibfield  {journal} {\bibinfo  {journal} {Phys. Rev. D}\ }\textbf {\bibinfo {volume} {102}},\ \bibinfo {pages} {084046} (\bibinfo {year} {2020})},\ \Eprint {http://arxiv.org/abs/2001.03571} {arXiv:2001.03571 [gr-qc]} \BibitemShut {NoStop}%
\bibitem [{\citenamefont {Cayuso}\ \emph {et~al.}(2017)\citenamefont {Cayuso}, \citenamefont {Ortiz},\ and\ \citenamefont {Lehner}}]{Cayuso:2017iqc}%
  \BibitemOpen
  \bibfield  {author} {\bibinfo {author} {\bibfnamefont {J.}~\bibnamefont {Cayuso}}, \bibinfo {author} {\bibfnamefont {N.}~\bibnamefont {Ortiz}}, \ and\ \bibinfo {author} {\bibfnamefont {L.}~\bibnamefont {Lehner}},\ }\href {\doibase 10.1103/PhysRevD.96.084043} {\bibfield  {journal} {\bibinfo  {journal} {Phys. Rev. D}\ }\textbf {\bibinfo {volume} {96}},\ \bibinfo {pages} {084043} (\bibinfo {year} {2017})},\ \Eprint {http://arxiv.org/abs/1706.07421} {arXiv:1706.07421 [gr-qc]} \BibitemShut {NoStop}%
\bibitem [{\citenamefont {Franchini}\ \emph {et~al.}(2022)\citenamefont {Franchini}, \citenamefont {Bezares}, \citenamefont {Barausse},\ and\ \citenamefont {Lehner}}]{Franchini:2022ukz}%
  \BibitemOpen
  \bibfield  {author} {\bibinfo {author} {\bibfnamefont {N.}~\bibnamefont {Franchini}}, \bibinfo {author} {\bibfnamefont {M.}~\bibnamefont {Bezares}}, \bibinfo {author} {\bibfnamefont {E.}~\bibnamefont {Barausse}}, \ and\ \bibinfo {author} {\bibfnamefont {L.}~\bibnamefont {Lehner}},\ }\href {\doibase 10.1103/PhysRevD.106.064061} {\bibfield  {journal} {\bibinfo  {journal} {Phys. Rev. D}\ }\textbf {\bibinfo {volume} {106}},\ \bibinfo {pages} {064061} (\bibinfo {year} {2022})},\ \Eprint {http://arxiv.org/abs/2206.00014} {arXiv:2206.00014 [gr-qc]} \BibitemShut {NoStop}%
\bibitem [{\citenamefont {Lara}\ \emph {et~al.}(2024)\citenamefont {Lara}, \citenamefont {Pfeiffer}, \citenamefont {Wittek}, \citenamefont {Vu}, \citenamefont {Nelli}, \citenamefont {Carpenter}, \citenamefont {Lovelace}, \citenamefont {Scheel},\ and\ \citenamefont {Throwe}}]{Lara:2024rwa}%
  \BibitemOpen
  \bibfield  {author} {\bibinfo {author} {\bibfnamefont {G.}~\bibnamefont {Lara}}, \bibinfo {author} {\bibfnamefont {H.~P.}\ \bibnamefont {Pfeiffer}}, \bibinfo {author} {\bibfnamefont {N.~A.}\ \bibnamefont {Wittek}}, \bibinfo {author} {\bibfnamefont {N.~L.}\ \bibnamefont {Vu}}, \bibinfo {author} {\bibfnamefont {K.~C.}\ \bibnamefont {Nelli}}, \bibinfo {author} {\bibfnamefont {A.}~\bibnamefont {Carpenter}}, \bibinfo {author} {\bibfnamefont {G.}~\bibnamefont {Lovelace}}, \bibinfo {author} {\bibfnamefont {M.~A.}\ \bibnamefont {Scheel}}, \ and\ \bibinfo {author} {\bibfnamefont {W.}~\bibnamefont {Throwe}},\ }\href {\doibase 10.1103/PhysRevD.110.024033} {\bibfield  {journal} {\bibinfo  {journal} {Phys. Rev. D}\ }\textbf {\bibinfo {volume} {110}},\ \bibinfo {pages} {024033} (\bibinfo {year} {2024})},\ \Eprint {http://arxiv.org/abs/2403.08705} {arXiv:2403.08705 [gr-qc]} \BibitemShut {NoStop}%
\bibitem [{\citenamefont {Corman}\ \emph {et~al.}(2024)\citenamefont {Corman}, \citenamefont {Lehner}, \citenamefont {East},\ and\ \citenamefont {Dideron}}]{Corman:2024cdr}%
  \BibitemOpen
  \bibfield  {author} {\bibinfo {author} {\bibfnamefont {M.}~\bibnamefont {Corman}}, \bibinfo {author} {\bibfnamefont {L.}~\bibnamefont {Lehner}}, \bibinfo {author} {\bibfnamefont {W.~E.}\ \bibnamefont {East}}, \ and\ \bibinfo {author} {\bibfnamefont {G.}~\bibnamefont {Dideron}},\ }\href@noop {} {\  (\bibinfo {year} {2024})},\ \Eprint {http://arxiv.org/abs/2405.15581} {arXiv:2405.15581 [gr-qc]} \BibitemShut {NoStop}%
\bibitem [{\citenamefont {Iozzo}\ \emph {et~al.}(2021)\citenamefont {Iozzo}, \citenamefont {Boyle}, \citenamefont {Deppe}, \citenamefont {Moxon}, \citenamefont {Scheel}, \citenamefont {Kidder}, \citenamefont {Pfeiffer},\ and\ \citenamefont {Teukolsky}}]{Iozzo:2020jcu}%
  \BibitemOpen
  \bibfield  {author} {\bibinfo {author} {\bibfnamefont {D.~A.~B.}\ \bibnamefont {Iozzo}}, \bibinfo {author} {\bibfnamefont {M.}~\bibnamefont {Boyle}}, \bibinfo {author} {\bibfnamefont {N.}~\bibnamefont {Deppe}}, \bibinfo {author} {\bibfnamefont {J.}~\bibnamefont {Moxon}}, \bibinfo {author} {\bibfnamefont {M.~A.}\ \bibnamefont {Scheel}}, \bibinfo {author} {\bibfnamefont {L.~E.}\ \bibnamefont {Kidder}}, \bibinfo {author} {\bibfnamefont {H.~P.}\ \bibnamefont {Pfeiffer}}, \ and\ \bibinfo {author} {\bibfnamefont {S.~A.}\ \bibnamefont {Teukolsky}},\ }\href {\doibase 10.1103/PhysRevD.103.024039} {\bibfield  {journal} {\bibinfo  {journal} {Phys. Rev. D}\ }\textbf {\bibinfo {volume} {103}},\ \bibinfo {pages} {024039} (\bibinfo {year} {2021})},\ \Eprint {http://arxiv.org/abs/2010.15200} {arXiv:2010.15200 [gr-qc]} \BibitemShut {NoStop}%
\bibitem [{\citenamefont {Mitman}\ \emph {et~al.}(2024)\citenamefont {Mitman} \emph {et~al.}}]{Mitman:2024uss}%
  \BibitemOpen
  \bibfield  {author} {\bibinfo {author} {\bibfnamefont {K.}~\bibnamefont {Mitman}} \emph {et~al.},\ }\href@noop {} {\  (\bibinfo {year} {2024})},\ \Eprint {http://arxiv.org/abs/2405.08868} {arXiv:2405.08868 [gr-qc]} \BibitemShut {NoStop}%
\bibitem [{\citenamefont {Peterson}\ \emph {et~al.}(2024)\citenamefont {Peterson}, \citenamefont {Gautam}, \citenamefont {Va\~n\'o Vi\~nuales},\ and\ \citenamefont {Hilditch}}]{Peterson:2024bxk}%
  \BibitemOpen
  \bibfield  {author} {\bibinfo {author} {\bibfnamefont {C.}~\bibnamefont {Peterson}}, \bibinfo {author} {\bibfnamefont {S.}~\bibnamefont {Gautam}}, \bibinfo {author} {\bibfnamefont {A.}~\bibnamefont {Va\~n\'o Vi\~nuales}}, \ and\ \bibinfo {author} {\bibfnamefont {D.}~\bibnamefont {Hilditch}},\ }\href@noop {} {\  (\bibinfo {year} {2024})},\ \Eprint {http://arxiv.org/abs/2409.02994} {arXiv:2409.02994 [gr-qc]} \BibitemShut {NoStop}%
\bibitem [{\citenamefont {Panosso~Macedo}\ and\ \citenamefont {Zenginoglu}(2024)}]{PanossoMacedo:2024nkw}%
  \BibitemOpen
  \bibfield  {author} {\bibinfo {author} {\bibfnamefont {R.}~\bibnamefont {Panosso~Macedo}}\ and\ \bibinfo {author} {\bibfnamefont {A.}~\bibnamefont {Zenginoglu}},\ }\href@noop {} {\  (\bibinfo {year} {2024})},\ \Eprint {http://arxiv.org/abs/2409.11478} {arXiv:2409.11478 [gr-qc]} \BibitemShut {NoStop}%
\bibitem [{\citenamefont {Winicour}(2009)}]{Winicour:2008vpn}%
  \BibitemOpen
  \bibfield  {author} {\bibinfo {author} {\bibfnamefont {J.}~\bibnamefont {Winicour}},\ }\href {\doibase 10.12942/lrr-2009-3} {\bibfield  {journal} {\bibinfo  {journal} {Living Rev. Rel.}\ }\textbf {\bibinfo {volume} {12}},\ \bibinfo {pages} {3} (\bibinfo {year} {2009})},\ \Eprint {http://arxiv.org/abs/0810.1903} {arXiv:0810.1903 [gr-qc]} \BibitemShut {NoStop}%
\bibitem [{\citenamefont {Moxon}\ \emph {et~al.}(2020)\citenamefont {Moxon}, \citenamefont {Scheel},\ and\ \citenamefont {Teukolsky}}]{Moxon:2020gha}%
  \BibitemOpen
  \bibfield  {author} {\bibinfo {author} {\bibfnamefont {J.}~\bibnamefont {Moxon}}, \bibinfo {author} {\bibfnamefont {M.~A.}\ \bibnamefont {Scheel}}, \ and\ \bibinfo {author} {\bibfnamefont {S.~A.}\ \bibnamefont {Teukolsky}},\ }\href {\doibase 10.1103/PhysRevD.102.044052} {\bibfield  {journal} {\bibinfo  {journal} {Phys. Rev. D}\ }\textbf {\bibinfo {volume} {102}},\ \bibinfo {pages} {044052} (\bibinfo {year} {2020})},\ \Eprint {http://arxiv.org/abs/2007.01339} {arXiv:2007.01339 [gr-qc]} \BibitemShut {NoStop}%
\bibitem [{\citenamefont {Moxon}\ \emph {et~al.}(2023)\citenamefont {Moxon}, \citenamefont {Scheel}, \citenamefont {Teukolsky}, \citenamefont {Deppe}, \citenamefont {Fischer}, \citenamefont {H\'ebert}, \citenamefont {Kidder},\ and\ \citenamefont {Throwe}}]{Moxon:2021gbv}%
  \BibitemOpen
  \bibfield  {author} {\bibinfo {author} {\bibfnamefont {J.}~\bibnamefont {Moxon}}, \bibinfo {author} {\bibfnamefont {M.~A.}\ \bibnamefont {Scheel}}, \bibinfo {author} {\bibfnamefont {S.~A.}\ \bibnamefont {Teukolsky}}, \bibinfo {author} {\bibfnamefont {N.}~\bibnamefont {Deppe}}, \bibinfo {author} {\bibfnamefont {N.}~\bibnamefont {Fischer}}, \bibinfo {author} {\bibfnamefont {F.}~\bibnamefont {H\'ebert}}, \bibinfo {author} {\bibfnamefont {L.~E.}\ \bibnamefont {Kidder}}, \ and\ \bibinfo {author} {\bibfnamefont {W.}~\bibnamefont {Throwe}},\ }\href {\doibase 10.1103/PhysRevD.107.064013} {\bibfield  {journal} {\bibinfo  {journal} {Phys. Rev. D}\ }\textbf {\bibinfo {volume} {107}},\ \bibinfo {pages} {064013} (\bibinfo {year} {2023})},\ \Eprint {http://arxiv.org/abs/2110.08635} {arXiv:2110.08635 [gr-qc]} \BibitemShut {NoStop}%
\bibitem [{\citenamefont {Ma}\ \emph {et~al.}(2024)\citenamefont {Ma} \emph {et~al.}}]{Ma:2023qjn}%
  \BibitemOpen
  \bibfield  {author} {\bibinfo {author} {\bibfnamefont {S.}~\bibnamefont {Ma}} \emph {et~al.},\ }\href {\doibase 10.1103/PhysRevD.109.124027} {\bibfield  {journal} {\bibinfo  {journal} {Phys. Rev. D}\ }\textbf {\bibinfo {volume} {109}},\ \bibinfo {pages} {124027} (\bibinfo {year} {2024})},\ \Eprint {http://arxiv.org/abs/2308.10361} {arXiv:2308.10361 [gr-qc]} \BibitemShut {NoStop}%
\bibitem [{\citenamefont {Deppe}\ \emph {et~al.}(2024)\citenamefont {Deppe}, \citenamefont {Throwe}, \citenamefont {Kidder}, \citenamefont {Vu}, \citenamefont {Nelli} \emph {et~al.}}]{deppe_2024_10967177}%
  \BibitemOpen
  \bibfield  {author} {\bibinfo {author} {\bibfnamefont {N.}~\bibnamefont {Deppe}}, \bibinfo {author} {\bibfnamefont {W.}~\bibnamefont {Throwe}}, \bibinfo {author} {\bibfnamefont {L.~E.}\ \bibnamefont {Kidder}}, \bibinfo {author} {\bibfnamefont {N.~L.}\ \bibnamefont {Vu}}, \bibinfo {author} {\bibfnamefont {K.~C.}\ \bibnamefont {Nelli}},  \emph {et~al.},\ }\href {\doibase 10.5281/zenodo.10967177} {\enquote {\bibinfo {title} {Spectre},}\ }\bibinfo {howpublished} {\href{https://doi.org/10.5281/zenodo.10967177}{10.5281/zenodo.10967177}} (\bibinfo {year} {2024})\BibitemShut {NoStop}%
\bibitem [{\citenamefont {Frittelli}(2005)}]{Frittelli:2004pk}%
  \BibitemOpen
  \bibfield  {author} {\bibinfo {author} {\bibfnamefont {S.}~\bibnamefont {Frittelli}},\ }\href {\doibase 10.1103/PhysRevD.71.024021} {\bibfield  {journal} {\bibinfo  {journal} {Phys. Rev. D}\ }\textbf {\bibinfo {volume} {71}},\ \bibinfo {pages} {024021} (\bibinfo {year} {2005})},\ \Eprint {http://arxiv.org/abs/gr-qc/0408035} {arXiv:gr-qc/0408035} \BibitemShut {NoStop}%
\bibitem [{\citenamefont {Giannakopoulos}\ \emph {et~al.}(2020)\citenamefont {Giannakopoulos}, \citenamefont {Hilditch},\ and\ \citenamefont {Zilhao}}]{Giannakopoulos:2020dih}%
  \BibitemOpen
  \bibfield  {author} {\bibinfo {author} {\bibfnamefont {T.}~\bibnamefont {Giannakopoulos}}, \bibinfo {author} {\bibfnamefont {D.}~\bibnamefont {Hilditch}}, \ and\ \bibinfo {author} {\bibfnamefont {M.}~\bibnamefont {Zilhao}},\ }\href {\doibase 10.1103/PhysRevD.102.064035} {\bibfield  {journal} {\bibinfo  {journal} {Phys. Rev. D}\ }\textbf {\bibinfo {volume} {102}},\ \bibinfo {pages} {064035} (\bibinfo {year} {2020})},\ \Eprint {http://arxiv.org/abs/2007.06419} {arXiv:2007.06419 [gr-qc]} \BibitemShut {NoStop}%
\bibitem [{\citenamefont {Giannakopoulos}\ \emph {et~al.}(2023)\citenamefont {Giannakopoulos}, \citenamefont {Bishop}, \citenamefont {Hilditch}, \citenamefont {Pollney},\ and\ \citenamefont {Zilh\~ao}}]{Giannakopoulos:2023zzm}%
  \BibitemOpen
  \bibfield  {author} {\bibinfo {author} {\bibfnamefont {T.}~\bibnamefont {Giannakopoulos}}, \bibinfo {author} {\bibfnamefont {N.~T.}\ \bibnamefont {Bishop}}, \bibinfo {author} {\bibfnamefont {D.}~\bibnamefont {Hilditch}}, \bibinfo {author} {\bibfnamefont {D.}~\bibnamefont {Pollney}}, \ and\ \bibinfo {author} {\bibfnamefont {M.}~\bibnamefont {Zilh\~ao}},\ }\href {\doibase 10.1103/PhysRevD.108.104033} {\bibfield  {journal} {\bibinfo  {journal} {Phys. Rev. D}\ }\textbf {\bibinfo {volume} {108}},\ \bibinfo {pages} {104033} (\bibinfo {year} {2023})},\ \Eprint {http://arxiv.org/abs/2306.13010} {arXiv:2306.13010 [gr-qc]} \BibitemShut {NoStop}%
\bibitem [{\citenamefont {Gundlach}(2024)}]{Gundlach:2024xmo}%
  \BibitemOpen
  \bibfield  {author} {\bibinfo {author} {\bibfnamefont {C.}~\bibnamefont {Gundlach}},\ }\href {\doibase 10.1103/PhysRevD.110.024020} {\bibfield  {journal} {\bibinfo  {journal} {Phys. Rev. D}\ }\textbf {\bibinfo {volume} {110}},\ \bibinfo {pages} {024020} (\bibinfo {year} {2024})},\ \Eprint {http://arxiv.org/abs/2404.16720} {arXiv:2404.16720 [gr-qc]} \BibitemShut {NoStop}%
\bibitem [{\citenamefont {Horndeski}(1974)}]{Horndeski:1974wa}%
  \BibitemOpen
  \bibfield  {author} {\bibinfo {author} {\bibfnamefont {G.~W.}\ \bibnamefont {Horndeski}},\ }\href {\doibase 10.1007/BF01807638} {\bibfield  {journal} {\bibinfo  {journal} {Int. J. Theor. Phys.}\ }\textbf {\bibinfo {volume} {10}},\ \bibinfo {pages} {363} (\bibinfo {year} {1974})}\BibitemShut {NoStop}%
\bibitem [{\citenamefont {Jackiw}\ and\ \citenamefont {Pi}(2003)}]{Jackiw:2003pm}%
  \BibitemOpen
  \bibfield  {author} {\bibinfo {author} {\bibfnamefont {R.}~\bibnamefont {Jackiw}}\ and\ \bibinfo {author} {\bibfnamefont {S.~Y.}\ \bibnamefont {Pi}},\ }\href {\doibase 10.1103/PhysRevD.68.104012} {\bibfield  {journal} {\bibinfo  {journal} {Phys. Rev. D}\ }\textbf {\bibinfo {volume} {68}},\ \bibinfo {pages} {104012} (\bibinfo {year} {2003})},\ \Eprint {http://arxiv.org/abs/gr-qc/0308071} {arXiv:gr-qc/0308071} \BibitemShut {NoStop}%
\bibitem [{\citenamefont {Alexander}\ and\ \citenamefont {Yunes}(2009)}]{Alexander:2009tp}%
  \BibitemOpen
  \bibfield  {author} {\bibinfo {author} {\bibfnamefont {S.}~\bibnamefont {Alexander}}\ and\ \bibinfo {author} {\bibfnamefont {N.}~\bibnamefont {Yunes}},\ }\href {\doibase 10.1016/j.physrep.2009.07.002} {\bibfield  {journal} {\bibinfo  {journal} {Phys. Rept.}\ }\textbf {\bibinfo {volume} {480}},\ \bibinfo {pages} {1} (\bibinfo {year} {2009})},\ \Eprint {http://arxiv.org/abs/0907.2562} {arXiv:0907.2562 [hep-th]} \BibitemShut {NoStop}%
\bibitem [{\citenamefont {Price}(1972{\natexlab{a}})}]{PhysRevD.5.2419}%
  \BibitemOpen
  \bibfield  {author} {\bibinfo {author} {\bibfnamefont {R.~H.}\ \bibnamefont {Price}},\ }\href {\doibase 10.1103/PhysRevD.5.2419} {\bibfield  {journal} {\bibinfo  {journal} {Phys. Rev. D}\ }\textbf {\bibinfo {volume} {5}},\ \bibinfo {pages} {2419} (\bibinfo {year} {1972}{\natexlab{a}})}\BibitemShut {NoStop}%
\bibitem [{\citenamefont {Price}(1972{\natexlab{b}})}]{PhysRevD.5.2439}%
  \BibitemOpen
  \bibfield  {author} {\bibinfo {author} {\bibfnamefont {R.~H.}\ \bibnamefont {Price}},\ }\href {\doibase 10.1103/PhysRevD.5.2439} {\bibfield  {journal} {\bibinfo  {journal} {Phys. Rev. D}\ }\textbf {\bibinfo {volume} {5}},\ \bibinfo {pages} {2439} (\bibinfo {year} {1972}{\natexlab{b}})}\BibitemShut {NoStop}%
\bibitem [{\citenamefont {Damour}\ and\ \citenamefont {Esposito-Farese}(1992)}]{Damour:1992we}%
  \BibitemOpen
  \bibfield  {author} {\bibinfo {author} {\bibfnamefont {T.}~\bibnamefont {Damour}}\ and\ \bibinfo {author} {\bibfnamefont {G.}~\bibnamefont {Esposito-Farese}},\ }\href {\doibase 10.1088/0264-9381/9/9/015} {\bibfield  {journal} {\bibinfo  {journal} {Class. Quant. Grav.}\ }\textbf {\bibinfo {volume} {9}},\ \bibinfo {pages} {2093} (\bibinfo {year} {1992})}\BibitemShut {NoStop}%
\bibitem [{\citenamefont {Bergmann}(1968)}]{bergmann1968comments}%
  \BibitemOpen
  \bibfield  {author} {\bibinfo {author} {\bibfnamefont {P.~G.}\ \bibnamefont {Bergmann}},\ }\href@noop {} {\bibfield  {journal} {\bibinfo  {journal} {International Journal of Theoretical Physics}\ }\textbf {\bibinfo {volume} {1}},\ \bibinfo {pages} {25} (\bibinfo {year} {1968})}\BibitemShut {NoStop}%
\bibitem [{\citenamefont {Wagoner}(1970)}]{Wagoner:1970vr}%
  \BibitemOpen
  \bibfield  {author} {\bibinfo {author} {\bibfnamefont {R.~V.}\ \bibnamefont {Wagoner}},\ }\href {\doibase 10.1103/PhysRevD.1.3209} {\bibfield  {journal} {\bibinfo  {journal} {Phys. Rev. D}\ }\textbf {\bibinfo {volume} {1}},\ \bibinfo {pages} {3209} (\bibinfo {year} {1970})}\BibitemShut {NoStop}%
\bibitem [{\citenamefont {{Bondi}}\ \emph {et~al.}(1962)\citenamefont {{Bondi}}, \citenamefont {{van der Burg}},\ and\ \citenamefont {{Metzner}}}]{1962RSPSA.269...21B}%
  \BibitemOpen
  \bibfield  {author} {\bibinfo {author} {\bibfnamefont {H.}~\bibnamefont {{Bondi}}}, \bibinfo {author} {\bibfnamefont {M.~G.~J.}\ \bibnamefont {{van der Burg}}}, \ and\ \bibinfo {author} {\bibfnamefont {A.~W.~K.}\ \bibnamefont {{Metzner}}},\ }\href {\doibase 10.1098/rspa.1962.0161} {\bibfield  {journal} {\bibinfo  {journal} {Proceedings of the Royal Society of London Series A}\ }\textbf {\bibinfo {volume} {269}},\ \bibinfo {pages} {21} (\bibinfo {year} {1962})}\BibitemShut {NoStop}%
\bibitem [{\citenamefont {Barnich}\ and\ \citenamefont {Troessaert}(2010)}]{Barnich:2009se}%
  \BibitemOpen
  \bibfield  {author} {\bibinfo {author} {\bibfnamefont {G.}~\bibnamefont {Barnich}}\ and\ \bibinfo {author} {\bibfnamefont {C.}~\bibnamefont {Troessaert}},\ }\href {\doibase 10.1103/PhysRevLett.105.111103} {\bibfield  {journal} {\bibinfo  {journal} {Phys. Rev. Lett.}\ }\textbf {\bibinfo {volume} {105}},\ \bibinfo {pages} {111103} (\bibinfo {year} {2010})},\ \Eprint {http://arxiv.org/abs/0909.2617} {arXiv:0909.2617 [gr-qc]} \BibitemShut {NoStop}%
\bibitem [{\citenamefont {Flanagan}\ and\ \citenamefont {Nichols}(2017)}]{Flanagan:2015pxa}%
  \BibitemOpen
  \bibfield  {author} {\bibinfo {author} {\bibfnamefont {E.~E.}\ \bibnamefont {Flanagan}}\ and\ \bibinfo {author} {\bibfnamefont {D.~A.}\ \bibnamefont {Nichols}},\ }\href {\doibase 10.1103/PhysRevD.95.044002} {\bibfield  {journal} {\bibinfo  {journal} {Phys. Rev. D}\ }\textbf {\bibinfo {volume} {95}},\ \bibinfo {pages} {044002} (\bibinfo {year} {2017})},\ \bibinfo {note} {[Erratum: Phys.Rev.D 108, 069902 (2023)]},\ \Eprint {http://arxiv.org/abs/1510.03386} {arXiv:1510.03386 [hep-th]} \BibitemShut {NoStop}%
\bibitem [{\citenamefont {Barreto}\ \emph {et~al.}(2005)\citenamefont {Barreto}, \citenamefont {Da~Silva}, \citenamefont {Gomez}, \citenamefont {Lehner}, \citenamefont {Rosales},\ and\ \citenamefont {Winicour}}]{Barreto:2004fn}%
  \BibitemOpen
  \bibfield  {author} {\bibinfo {author} {\bibfnamefont {W.}~\bibnamefont {Barreto}}, \bibinfo {author} {\bibfnamefont {A.}~\bibnamefont {Da~Silva}}, \bibinfo {author} {\bibfnamefont {R.}~\bibnamefont {Gomez}}, \bibinfo {author} {\bibfnamefont {L.}~\bibnamefont {Lehner}}, \bibinfo {author} {\bibfnamefont {L.}~\bibnamefont {Rosales}}, \ and\ \bibinfo {author} {\bibfnamefont {J.}~\bibnamefont {Winicour}},\ }\href {\doibase 10.1103/PhysRevD.71.064028} {\bibfield  {journal} {\bibinfo  {journal} {Phys. Rev. D}\ }\textbf {\bibinfo {volume} {71}},\ \bibinfo {pages} {064028} (\bibinfo {year} {2005})},\ \Eprint {http://arxiv.org/abs/gr-qc/0412066} {arXiv:gr-qc/0412066} \BibitemShut {NoStop}%
\bibitem [{\citenamefont {Tahura}\ \emph {et~al.}(2021{\natexlab{a}})\citenamefont {Tahura}, \citenamefont {Nichols}, \citenamefont {Saffer}, \citenamefont {Stein},\ and\ \citenamefont {Yagi}}]{Tahura:2020vsa}%
  \BibitemOpen
  \bibfield  {author} {\bibinfo {author} {\bibfnamefont {S.}~\bibnamefont {Tahura}}, \bibinfo {author} {\bibfnamefont {D.~A.}\ \bibnamefont {Nichols}}, \bibinfo {author} {\bibfnamefont {A.}~\bibnamefont {Saffer}}, \bibinfo {author} {\bibfnamefont {L.~C.}\ \bibnamefont {Stein}}, \ and\ \bibinfo {author} {\bibfnamefont {K.}~\bibnamefont {Yagi}},\ }\href {\doibase 10.1103/PhysRevD.103.104026} {\bibfield  {journal} {\bibinfo  {journal} {Phys. Rev. D}\ }\textbf {\bibinfo {volume} {103}},\ \bibinfo {pages} {104026} (\bibinfo {year} {2021}{\natexlab{a}})},\ \Eprint {http://arxiv.org/abs/2007.13799} {arXiv:2007.13799 [gr-qc]} \BibitemShut {NoStop}%
\bibitem [{\citenamefont {Reinecke}\ and\ \citenamefont {Seljebotn}(2013)}]{Reinecke:2013}%
  \BibitemOpen
  \bibfield  {author} {\bibinfo {author} {\bibfnamefont {M.}~\bibnamefont {Reinecke}}\ and\ \bibinfo {author} {\bibfnamefont {D.~S.}\ \bibnamefont {Seljebotn}},\ }\href {\doibase 10.1051/0004-6361/201321494} {\bibfield  {journal} {\bibinfo  {journal} {Astronomy \& Astrophysics}\ }\textbf {\bibinfo {volume} {554}},\ \bibinfo {pages} {A112} (\bibinfo {year} {2013})}\BibitemShut {NoStop}%
\bibitem [{Lib()}]{Libsharp}%
  \BibitemOpen
  \href@noop {} {\enquote {\bibinfo {title} {libsharp},}\ }\bibinfo {howpublished} {\url{https://github.com/Libsharp/libsharp}},\ \bibinfo {note} {accessed: 2021-08-25}\BibitemShut {NoStop}%
\bibitem [{\citenamefont {Scheel}\ \emph {et~al.}(2004)\citenamefont {Scheel}, \citenamefont {Erickcek}, \citenamefont {Burko}, \citenamefont {Kidder}, \citenamefont {Pfeiffer},\ and\ \citenamefont {Teukolsky}}]{Scheel:2003vs}%
  \BibitemOpen
  \bibfield  {author} {\bibinfo {author} {\bibfnamefont {M.~A.}\ \bibnamefont {Scheel}}, \bibinfo {author} {\bibfnamefont {A.~L.}\ \bibnamefont {Erickcek}}, \bibinfo {author} {\bibfnamefont {L.~M.}\ \bibnamefont {Burko}}, \bibinfo {author} {\bibfnamefont {L.~E.}\ \bibnamefont {Kidder}}, \bibinfo {author} {\bibfnamefont {H.~P.}\ \bibnamefont {Pfeiffer}}, \ and\ \bibinfo {author} {\bibfnamefont {S.~A.}\ \bibnamefont {Teukolsky}},\ }\href {\doibase 10.1103/PhysRevD.69.104006} {\bibfield  {journal} {\bibinfo  {journal} {Phys. Rev. D}\ }\textbf {\bibinfo {volume} {69}},\ \bibinfo {pages} {104006} (\bibinfo {year} {2004})},\ \Eprint {http://arxiv.org/abs/gr-qc/0305027} {arXiv:gr-qc/0305027} \BibitemShut {NoStop}%
\bibitem [{\citenamefont {Chandrasekhar}(1998)}]{Chandrasekhar:BHs}%
  \BibitemOpen
  \bibfield  {author} {\bibinfo {author} {\bibfnamefont {S.}~\bibnamefont {Chandrasekhar}},\ }\href@noop {} {\emph {\bibinfo {title} {The Mathematical Theory of Black Holes}}}\ (\bibinfo  {publisher} {Clarendon Press},\ \bibinfo {year} {1998})\BibitemShut {NoStop}%
\bibitem [{\citenamefont {Gasper\'\i{}n}\ \emph {et~al.}(2025)\citenamefont {Gasper\'\i{}n}, \citenamefont {Mohamed},\ and\ \citenamefont {Mena}}]{Gasperin:2024bfc}%
  \BibitemOpen
  \bibfield  {author} {\bibinfo {author} {\bibfnamefont {E.}~\bibnamefont {Gasper\'\i{}n}}, \bibinfo {author} {\bibfnamefont {M.~M.~A.}\ \bibnamefont {Mohamed}}, \ and\ \bibinfo {author} {\bibfnamefont {F.~C.}\ \bibnamefont {Mena}},\ }\href {\doibase 10.1016/j.geomphys.2024.105389} {\bibfield  {journal} {\bibinfo  {journal} {J. Geom. Phys.}\ }\textbf {\bibinfo {volume} {209}},\ \bibinfo {pages} {105389} (\bibinfo {year} {2025})},\ \Eprint {http://arxiv.org/abs/2408.03389} {arXiv:2408.03389 [gr-qc]} \BibitemShut {NoStop}%
\bibitem [{\citenamefont {Stephani}\ \emph {et~al.}(2009)\citenamefont {Stephani}, \citenamefont {Kramer}, \citenamefont {MacCallum}, \citenamefont {Hoenselaers},\ and\ \citenamefont {Herlt}}]{stephani2009exact}%
  \BibitemOpen
  \bibfield  {author} {\bibinfo {author} {\bibfnamefont {H.}~\bibnamefont {Stephani}}, \bibinfo {author} {\bibfnamefont {D.}~\bibnamefont {Kramer}}, \bibinfo {author} {\bibfnamefont {M.}~\bibnamefont {MacCallum}}, \bibinfo {author} {\bibfnamefont {C.}~\bibnamefont {Hoenselaers}}, \ and\ \bibinfo {author} {\bibfnamefont {E.}~\bibnamefont {Herlt}},\ }\href@noop {} {\emph {\bibinfo {title} {Exact solutions of Einstein's field equations}}}\ (\bibinfo  {publisher} {Cambridge university press},\ \bibinfo {year} {2009})\BibitemShut {NoStop}%
\bibitem [{\citenamefont {Adamo}\ \emph {et~al.}(2009)\citenamefont {Adamo}, \citenamefont {Kozameh},\ and\ \citenamefont {Newman}}]{Adamo:2009vu}%
  \BibitemOpen
  \bibfield  {author} {\bibinfo {author} {\bibfnamefont {T.~M.}\ \bibnamefont {Adamo}}, \bibinfo {author} {\bibfnamefont {C.~N.}\ \bibnamefont {Kozameh}}, \ and\ \bibinfo {author} {\bibfnamefont {E.~T.}\ \bibnamefont {Newman}},\ }\href {\doibase 10.12942/lrr-2009-6} {\bibfield  {journal} {\bibinfo  {journal} {Living Rev. Rel.}\ }\textbf {\bibinfo {volume} {12}},\ \bibinfo {pages} {6} (\bibinfo {year} {2009})},\ \Eprint {http://arxiv.org/abs/0906.2155} {arXiv:0906.2155 [gr-qc]} \BibitemShut {NoStop}%
\bibitem [{\citenamefont {Scholtz}\ and\ \citenamefont {Holka}(2014)}]{Scholtz:2013toa}%
  \BibitemOpen
  \bibfield  {author} {\bibinfo {author} {\bibfnamefont {M.}~\bibnamefont {Scholtz}}\ and\ \bibinfo {author} {\bibfnamefont {L.}~\bibnamefont {Holka}},\ }\href {\doibase 10.1007/s10714-014-1665-7} {\bibfield  {journal} {\bibinfo  {journal} {Gen. Rel. Grav.}\ }\textbf {\bibinfo {volume} {46}},\ \bibinfo {pages} {1665} (\bibinfo {year} {2014})},\ \Eprint {http://arxiv.org/abs/1312.7084} {arXiv:1312.7084 [gr-qc]} \BibitemShut {NoStop}%
\bibitem [{\citenamefont {Boyle}()}]{scri}%
  \BibitemOpen
  \bibfield  {author} {\bibinfo {author} {\bibfnamefont {M.}~\bibnamefont {Boyle}},\ }\href@noop {} {\enquote {\bibinfo {title} {Scri},}\ }\bibinfo {note} {\url{https://github.com/moble/scri}}\BibitemShut {NoStop}%
\bibitem [{\citenamefont {Boyle}\ \emph {et~al.}(2020)\citenamefont {Boyle}, \citenamefont {Iozzo},\ and\ \citenamefont {Stein}}]{mike_boyle_2020_4041972}%
  \BibitemOpen
  \bibfield  {author} {\bibinfo {author} {\bibfnamefont {M.}~\bibnamefont {Boyle}}, \bibinfo {author} {\bibfnamefont {D.}~\bibnamefont {Iozzo}}, \ and\ \bibinfo {author} {\bibfnamefont {L.~C.}\ \bibnamefont {Stein}},\ }\href {\doibase 10.5281/zenodo.4041972} {\enquote {\bibinfo {title} {moble/scri: v1.2},}\ } (\bibinfo {year} {2020})\BibitemShut {NoStop}%
\bibitem [{\citenamefont {Boyle}(2013)}]{Boyle:2013nka}%
  \BibitemOpen
  \bibfield  {author} {\bibinfo {author} {\bibfnamefont {M.}~\bibnamefont {Boyle}},\ }\href {\doibase 10.1103/PhysRevD.87.104006} {\bibfield  {journal} {\bibinfo  {journal} {Phys. Rev.}\ }\textbf {\bibinfo {volume} {D87}},\ \bibinfo {pages} {104006} (\bibinfo {year} {2013})},\ \Eprint {http://arxiv.org/abs/1302.2919} {arXiv:1302.2919 [gr-qc]} \BibitemShut {NoStop}%
\bibitem [{\citenamefont {Boyle}\ \emph {et~al.}(2014)\citenamefont {Boyle}, \citenamefont {Kidder}, \citenamefont {Ossokine},\ and\ \citenamefont {Pfeiffer}}]{Boyle:2014ioa}%
  \BibitemOpen
  \bibfield  {author} {\bibinfo {author} {\bibfnamefont {M.}~\bibnamefont {Boyle}}, \bibinfo {author} {\bibfnamefont {L.~E.}\ \bibnamefont {Kidder}}, \bibinfo {author} {\bibfnamefont {S.}~\bibnamefont {Ossokine}}, \ and\ \bibinfo {author} {\bibfnamefont {H.~P.}\ \bibnamefont {Pfeiffer}},\ }\href@noop {} {\  (\bibinfo {year} {2014})},\ \Eprint {http://arxiv.org/abs/1409.4431} {arXiv:1409.4431 [gr-qc]} \BibitemShut {NoStop}%
\bibitem [{\citenamefont {Boyle}(2016)}]{Boyle:2015nqa}%
  \BibitemOpen
  \bibfield  {author} {\bibinfo {author} {\bibfnamefont {M.}~\bibnamefont {Boyle}},\ }\href {\doibase 10.1103/PhysRevD.93.084031} {\bibfield  {journal} {\bibinfo  {journal} {Phys. Rev.}\ }\textbf {\bibinfo {volume} {D93}},\ \bibinfo {pages} {084031} (\bibinfo {year} {2016})},\ \Eprint {http://arxiv.org/abs/1509.00862} {arXiv:1509.00862 [gr-qc]} \BibitemShut {NoStop}%
\bibitem [{\citenamefont {Lehner}\ and\ \citenamefont {Moreschi}(2007)}]{Lehner:2007ip}%
  \BibitemOpen
  \bibfield  {author} {\bibinfo {author} {\bibfnamefont {L.}~\bibnamefont {Lehner}}\ and\ \bibinfo {author} {\bibfnamefont {O.~M.}\ \bibnamefont {Moreschi}},\ }\href {\doibase 10.1103/PhysRevD.76.124040} {\bibfield  {journal} {\bibinfo  {journal} {Phys. Rev. D}\ }\textbf {\bibinfo {volume} {76}},\ \bibinfo {pages} {124040} (\bibinfo {year} {2007})},\ \Eprint {http://arxiv.org/abs/0706.1319} {arXiv:0706.1319 [gr-qc]} \BibitemShut {NoStop}%
\bibitem [{\citenamefont {Helfer}(1993)}]{helfer1993null}%
  \BibitemOpen
  \bibfield  {author} {\bibinfo {author} {\bibfnamefont {A.~D.}\ \bibnamefont {Helfer}},\ }\href@noop {} {\bibfield  {journal} {\bibinfo  {journal} {Journal of mathematical physics}\ }\textbf {\bibinfo {volume} {34}},\ \bibinfo {pages} {3478} (\bibinfo {year} {1993})}\BibitemShut {NoStop}%
\bibitem [{\citenamefont {Mitman}\ \emph {et~al.}(2020)\citenamefont {Mitman}, \citenamefont {Moxon}, \citenamefont {Scheel}, \citenamefont {Teukolsky}, \citenamefont {Boyle}, \citenamefont {Deppe}, \citenamefont {Kidder},\ and\ \citenamefont {Throwe}}]{Mitman:2020pbt}%
  \BibitemOpen
  \bibfield  {author} {\bibinfo {author} {\bibfnamefont {K.}~\bibnamefont {Mitman}}, \bibinfo {author} {\bibfnamefont {J.}~\bibnamefont {Moxon}}, \bibinfo {author} {\bibfnamefont {M.~A.}\ \bibnamefont {Scheel}}, \bibinfo {author} {\bibfnamefont {S.~A.}\ \bibnamefont {Teukolsky}}, \bibinfo {author} {\bibfnamefont {M.}~\bibnamefont {Boyle}}, \bibinfo {author} {\bibfnamefont {N.}~\bibnamefont {Deppe}}, \bibinfo {author} {\bibfnamefont {L.~E.}\ \bibnamefont {Kidder}}, \ and\ \bibinfo {author} {\bibfnamefont {W.}~\bibnamefont {Throwe}},\ }\href {\doibase 10.1103/PhysRevD.102.104007} {\bibfield  {journal} {\bibinfo  {journal} {Phys. Rev. D}\ }\textbf {\bibinfo {volume} {102}},\ \bibinfo {pages} {104007} (\bibinfo {year} {2020})},\ \Eprint {http://arxiv.org/abs/2007.11562} {arXiv:2007.11562 [gr-qc]} \BibitemShut {NoStop}%
\bibitem [{\citenamefont {Schnittman}\ \emph {et~al.}(2008)\citenamefont {Schnittman}, \citenamefont {Buonanno}, \citenamefont {van Meter}, \citenamefont {Baker}, \citenamefont {Boggs}, \citenamefont {Centrella}, \citenamefont {Kelly},\ and\ \citenamefont {McWilliams}}]{Schnittman:2007ij}%
  \BibitemOpen
  \bibfield  {author} {\bibinfo {author} {\bibfnamefont {J.~D.}\ \bibnamefont {Schnittman}}, \bibinfo {author} {\bibfnamefont {A.}~\bibnamefont {Buonanno}}, \bibinfo {author} {\bibfnamefont {J.~R.}\ \bibnamefont {van Meter}}, \bibinfo {author} {\bibfnamefont {J.~G.}\ \bibnamefont {Baker}}, \bibinfo {author} {\bibfnamefont {W.~D.}\ \bibnamefont {Boggs}}, \bibinfo {author} {\bibfnamefont {J.}~\bibnamefont {Centrella}}, \bibinfo {author} {\bibfnamefont {B.~J.}\ \bibnamefont {Kelly}}, \ and\ \bibinfo {author} {\bibfnamefont {S.~T.}\ \bibnamefont {McWilliams}},\ }\href {\doibase 10.1103/PhysRevD.77.044031} {\bibfield  {journal} {\bibinfo  {journal} {Phys. Rev.}\ }\textbf {\bibinfo {volume} {D77}},\ \bibinfo {pages} {044031} (\bibinfo {year} {2008})},\ \Eprint {http://arxiv.org/abs/0707.0301} {arXiv:0707.0301 [gr-qc]} \BibitemShut {NoStop}%
\bibitem [{\citenamefont {Ma}\ \emph {et~al.}(2021)\citenamefont {Ma}, \citenamefont {Giesler}, \citenamefont {Varma}, \citenamefont {Scheel},\ and\ \citenamefont {Chen}}]{Ma:2021znq}%
  \BibitemOpen
  \bibfield  {author} {\bibinfo {author} {\bibfnamefont {S.}~\bibnamefont {Ma}}, \bibinfo {author} {\bibfnamefont {M.}~\bibnamefont {Giesler}}, \bibinfo {author} {\bibfnamefont {V.}~\bibnamefont {Varma}}, \bibinfo {author} {\bibfnamefont {M.~A.}\ \bibnamefont {Scheel}}, \ and\ \bibinfo {author} {\bibfnamefont {Y.}~\bibnamefont {Chen}},\ }\href {\doibase 10.1103/PhysRevD.104.084003} {\bibfield  {journal} {\bibinfo  {journal} {Phys. Rev. D}\ }\textbf {\bibinfo {volume} {104}},\ \bibinfo {pages} {084003} (\bibinfo {year} {2021})},\ \Eprint {http://arxiv.org/abs/2107.04890} {arXiv:2107.04890 [gr-qc]} \BibitemShut {NoStop}%
\bibitem [{\citenamefont {Ma}\ and\ \citenamefont {Yang}(2024)}]{Ma:2024qcv}%
  \BibitemOpen
  \bibfield  {author} {\bibinfo {author} {\bibfnamefont {S.}~\bibnamefont {Ma}}\ and\ \bibinfo {author} {\bibfnamefont {H.}~\bibnamefont {Yang}},\ }\href {\doibase 10.1103/PhysRevD.109.104070} {\bibfield  {journal} {\bibinfo  {journal} {Phys. Rev. D}\ }\textbf {\bibinfo {volume} {109}},\ \bibinfo {pages} {104070} (\bibinfo {year} {2024})},\ \Eprint {http://arxiv.org/abs/2401.15516} {arXiv:2401.15516 [gr-qc]} \BibitemShut {NoStop}%
\bibitem [{\citenamefont {Bourg}\ \emph {et~al.}(2024)\citenamefont {Bourg}, \citenamefont {Panosso~Macedo}, \citenamefont {Spiers}, \citenamefont {Leather}, \citenamefont {Bonga},\ and\ \citenamefont {Pound}}]{Bourg:2024jme}%
  \BibitemOpen
  \bibfield  {author} {\bibinfo {author} {\bibfnamefont {P.}~\bibnamefont {Bourg}}, \bibinfo {author} {\bibfnamefont {R.}~\bibnamefont {Panosso~Macedo}}, \bibinfo {author} {\bibfnamefont {A.}~\bibnamefont {Spiers}}, \bibinfo {author} {\bibfnamefont {B.}~\bibnamefont {Leather}}, \bibinfo {author} {\bibfnamefont {B.}~\bibnamefont {Bonga}}, \ and\ \bibinfo {author} {\bibfnamefont {A.}~\bibnamefont {Pound}},\ }\href@noop {} {\  (\bibinfo {year} {2024})},\ \Eprint {http://arxiv.org/abs/2405.10270} {arXiv:2405.10270 [gr-qc]} \BibitemShut {NoStop}%
\bibitem [{\citenamefont {Baumgarte}\ and\ \citenamefont {Shapiro}(2010)}]{Baumgarte_Shaprio:NumRel}%
  \BibitemOpen
  \bibfield  {author} {\bibinfo {author} {\bibfnamefont {T.~W.}\ \bibnamefont {Baumgarte}}\ and\ \bibinfo {author} {\bibfnamefont {S.~L.}\ \bibnamefont {Shapiro}},\ }\href {\doibase 10.1017/CBO9781139193344} {\emph {\bibinfo {title} {{Numerical Relativity: Solving Einstein's Equations on the Computer}}}}\ (\bibinfo  {publisher} {Cambridge University Press},\ \bibinfo {year} {2010})\BibitemShut {NoStop}%
\bibitem [{\citenamefont {Barkett}\ \emph {et~al.}(2020)\citenamefont {Barkett}, \citenamefont {Moxon}, \citenamefont {Scheel},\ and\ \citenamefont {Szil\'agyi}}]{Barkett:2019uae}%
  \BibitemOpen
  \bibfield  {author} {\bibinfo {author} {\bibfnamefont {K.}~\bibnamefont {Barkett}}, \bibinfo {author} {\bibfnamefont {J.}~\bibnamefont {Moxon}}, \bibinfo {author} {\bibfnamefont {M.~A.}\ \bibnamefont {Scheel}}, \ and\ \bibinfo {author} {\bibfnamefont {B.}~\bibnamefont {Szil\'agyi}},\ }\href {\doibase 10.1103/PhysRevD.102.024004} {\bibfield  {journal} {\bibinfo  {journal} {Phys. Rev. D}\ }\textbf {\bibinfo {volume} {102}},\ \bibinfo {pages} {024004} (\bibinfo {year} {2020})},\ \Eprint {http://arxiv.org/abs/1910.09677} {arXiv:1910.09677 [gr-qc]} \BibitemShut {NoStop}%
\bibitem [{\citenamefont {Bishop}(2005)}]{Bishop:2004ug}%
  \BibitemOpen
  \bibfield  {author} {\bibinfo {author} {\bibfnamefont {N.~T.}\ \bibnamefont {Bishop}},\ }\href {\doibase 10.1088/0264-9381/22/12/006} {\bibfield  {journal} {\bibinfo  {journal} {Class. Quant. Grav.}\ }\textbf {\bibinfo {volume} {22}},\ \bibinfo {pages} {2393} (\bibinfo {year} {2005})},\ \Eprint {http://arxiv.org/abs/gr-qc/0412006} {arXiv:gr-qc/0412006} \BibitemShut {NoStop}%
\bibitem [{\citenamefont {Pfeiffer}\ \emph {et~al.}(2003)\citenamefont {Pfeiffer}, \citenamefont {Kidder}, \citenamefont {Scheel},\ and\ \citenamefont {Teukolsky}}]{Pfeiffer:2002wt}%
  \BibitemOpen
  \bibfield  {author} {\bibinfo {author} {\bibfnamefont {H.~P.}\ \bibnamefont {Pfeiffer}}, \bibinfo {author} {\bibfnamefont {L.~E.}\ \bibnamefont {Kidder}}, \bibinfo {author} {\bibfnamefont {M.~A.}\ \bibnamefont {Scheel}}, \ and\ \bibinfo {author} {\bibfnamefont {S.~A.}\ \bibnamefont {Teukolsky}},\ }\href {\doibase 10.1016/S0010-4655(02)00847-0} {\bibfield  {journal} {\bibinfo  {journal} {Comput. Phys. Commun.}\ }\textbf {\bibinfo {volume} {152}},\ \bibinfo {pages} {253} (\bibinfo {year} {2003})},\ \Eprint {http://arxiv.org/abs/gr-qc/0202096} {arXiv:gr-qc/0202096 [gr-qc]} \BibitemShut {NoStop}%
\bibitem [{SpE()}]{SpECwebsite}%
  \BibitemOpen
  \href@noop {} {\enquote {\bibinfo {title} {The {S}pectral {E}instein {C}ode},}\ }\bibinfo {note} {\url{http://www.black-holes.org/SpEC.html}}\BibitemShut {NoStop}%
\bibitem [{\citenamefont {Lindblom}\ \emph {et~al.}(2006)\citenamefont {Lindblom}, \citenamefont {Scheel}, \citenamefont {Kidder}, \citenamefont {Owen},\ and\ \citenamefont {Rinne}}]{Lindblom:2005qh}%
  \BibitemOpen
  \bibfield  {author} {\bibinfo {author} {\bibfnamefont {L.}~\bibnamefont {Lindblom}}, \bibinfo {author} {\bibfnamefont {M.~A.}\ \bibnamefont {Scheel}}, \bibinfo {author} {\bibfnamefont {L.~E.}\ \bibnamefont {Kidder}}, \bibinfo {author} {\bibfnamefont {R.}~\bibnamefont {Owen}}, \ and\ \bibinfo {author} {\bibfnamefont {O.}~\bibnamefont {Rinne}},\ }\href {\doibase 10.1088/0264-9381/23/16/S09} {\bibfield  {journal} {\bibinfo  {journal} {Class. Quant. Grav.}\ }\textbf {\bibinfo {volume} {23}},\ \bibinfo {pages} {S447} (\bibinfo {year} {2006})},\ \Eprint {http://arxiv.org/abs/gr-qc/0512093} {arXiv:gr-qc/0512093 [gr-qc]} \BibitemShut {NoStop}%
\bibitem [{\citenamefont {Leaver}(1986)}]{PhysRevD.34.384}%
  \BibitemOpen
  \bibfield  {author} {\bibinfo {author} {\bibfnamefont {E.~W.}\ \bibnamefont {Leaver}},\ }\href {\doibase 10.1103/PhysRevD.34.384} {\bibfield  {journal} {\bibinfo  {journal} {Phys. Rev. D}\ }\textbf {\bibinfo {volume} {34}},\ \bibinfo {pages} {384} (\bibinfo {year} {1986})}\BibitemShut {NoStop}%
\bibitem [{\citenamefont {Gundlach}\ \emph {et~al.}(1994{\natexlab{a}})\citenamefont {Gundlach}, \citenamefont {Price},\ and\ \citenamefont {Pullin}}]{PhysRevD.49.883}%
  \BibitemOpen
  \bibfield  {author} {\bibinfo {author} {\bibfnamefont {C.}~\bibnamefont {Gundlach}}, \bibinfo {author} {\bibfnamefont {R.~H.}\ \bibnamefont {Price}}, \ and\ \bibinfo {author} {\bibfnamefont {J.}~\bibnamefont {Pullin}},\ }\href {\doibase 10.1103/PhysRevD.49.883} {\bibfield  {journal} {\bibinfo  {journal} {Phys. Rev. D}\ }\textbf {\bibinfo {volume} {49}},\ \bibinfo {pages} {883} (\bibinfo {year} {1994}{\natexlab{a}})}\BibitemShut {NoStop}%
\bibitem [{\citenamefont {Gundlach}\ \emph {et~al.}(1994{\natexlab{b}})\citenamefont {Gundlach}, \citenamefont {Price},\ and\ \citenamefont {Pullin}}]{PhysRevD.49.890}%
  \BibitemOpen
  \bibfield  {author} {\bibinfo {author} {\bibfnamefont {C.}~\bibnamefont {Gundlach}}, \bibinfo {author} {\bibfnamefont {R.~H.}\ \bibnamefont {Price}}, \ and\ \bibinfo {author} {\bibfnamefont {J.}~\bibnamefont {Pullin}},\ }\href {\doibase 10.1103/PhysRevD.49.890} {\bibfield  {journal} {\bibinfo  {journal} {Phys. Rev. D}\ }\textbf {\bibinfo {volume} {49}},\ \bibinfo {pages} {890} (\bibinfo {year} {1994}{\natexlab{b}})}\BibitemShut {NoStop}%
\bibitem [{\citenamefont {Virtanen}\ \emph {et~al.}(2020)\citenamefont {Virtanen}, \citenamefont {Gommers}, \citenamefont {Oliphant}, \citenamefont {Haberland}, \citenamefont {Reddy}, \citenamefont {Cournapeau}, \citenamefont {Burovski}, \citenamefont {Peterson}, \citenamefont {Weckesser}, \citenamefont {Bright}, \citenamefont {{van der Walt}}, \citenamefont {Brett}, \citenamefont {Wilson}, \citenamefont {Millman}, \citenamefont {Mayorov}, \citenamefont {Nelson}, \citenamefont {Jones}, \citenamefont {Kern}, \citenamefont {Larson}, \citenamefont {Carey}, \citenamefont {Polat}, \citenamefont {Feng}, \citenamefont {Moore}, \citenamefont {{VanderPlas}}, \citenamefont {Laxalde}, \citenamefont {Perktold}, \citenamefont {Cimrman}, \citenamefont {Henriksen}, \citenamefont {Quintero}, \citenamefont {Harris}, \citenamefont {Archibald}, \citenamefont {Ribeiro}, \citenamefont {Pedregosa}, \citenamefont {{van Mulbregt}},\ and\ \citenamefont {{SciPy 1.0 Contributors}}}]{2020SciPy-NMeth}%
  \BibitemOpen
  \bibfield  {author} {\bibinfo {author} {\bibfnamefont {P.}~\bibnamefont {Virtanen}}, \bibinfo {author} {\bibfnamefont {R.}~\bibnamefont {Gommers}}, \bibinfo {author} {\bibfnamefont {T.~E.}\ \bibnamefont {Oliphant}}, \bibinfo {author} {\bibfnamefont {M.}~\bibnamefont {Haberland}}, \bibinfo {author} {\bibfnamefont {T.}~\bibnamefont {Reddy}}, \bibinfo {author} {\bibfnamefont {D.}~\bibnamefont {Cournapeau}}, \bibinfo {author} {\bibfnamefont {E.}~\bibnamefont {Burovski}}, \bibinfo {author} {\bibfnamefont {P.}~\bibnamefont {Peterson}}, \bibinfo {author} {\bibfnamefont {W.}~\bibnamefont {Weckesser}}, \bibinfo {author} {\bibfnamefont {J.}~\bibnamefont {Bright}}, \bibinfo {author} {\bibfnamefont {S.~J.}\ \bibnamefont {{van der Walt}}}, \bibinfo {author} {\bibfnamefont {M.}~\bibnamefont {Brett}}, \bibinfo {author} {\bibfnamefont {J.}~\bibnamefont {Wilson}}, \bibinfo {author} {\bibfnamefont {K.~J.}\ \bibnamefont {Millman}}, \bibinfo {author} {\bibfnamefont {N.}~\bibnamefont {Mayorov}}, \bibinfo {author} {\bibfnamefont {A.~R.~J.}\ \bibnamefont {Nelson}}, \bibinfo {author} {\bibfnamefont {E.}~\bibnamefont {Jones}}, \bibinfo {author} {\bibfnamefont {R.}~\bibnamefont {Kern}}, \bibinfo {author} {\bibfnamefont {E.}~\bibnamefont {Larson}}, \bibinfo {author} {\bibfnamefont {C.~J.}\ \bibnamefont {Carey}}, \bibinfo {author} {\bibfnamefont {{\.I}.}~\bibnamefont {Polat}}, \bibinfo {author} {\bibfnamefont {Y.}~\bibnamefont {Feng}}, \bibinfo {author} {\bibfnamefont {E.~W.}\ \bibnamefont {Moore}}, \bibinfo {author} {\bibfnamefont {J.}~\bibnamefont {{VanderPlas}}}, \bibinfo {author} {\bibfnamefont {D.}~\bibnamefont {Laxalde}}, \bibinfo {author} {\bibfnamefont {J.}~\bibnamefont {Perktold}}, \bibinfo {author} {\bibfnamefont {R.}~\bibnamefont {Cimrman}}, \bibinfo {author} {\bibfnamefont {I.}~\bibnamefont {Henriksen}}, \bibinfo {author} {\bibfnamefont {E.~A.}\ \bibnamefont {Quintero}}, \bibinfo {author} {\bibfnamefont {C.~R.}\ \bibnamefont {Harris}}, \bibinfo {author} {\bibfnamefont {A.~M.}\ \bibnamefont {Archibald}}, \bibinfo {author} {\bibfnamefont {A.~H.}\ \bibnamefont {Ribeiro}}, \bibinfo {author} {\bibfnamefont {F.}~\bibnamefont {Pedregosa}}, \bibinfo {author} {\bibfnamefont {P.}~\bibnamefont {{van Mulbregt}}}, \ and\ \bibinfo {author} {\bibnamefont {{SciPy 1.0 Contributors}}},\ }\href {\doibase 10.1038/s41592-019-0686-2} {\bibfield  {journal} {\bibinfo  {journal} {Nature Methods}\ }\textbf {\bibinfo {volume} {17}},\ \bibinfo {pages} {261} (\bibinfo {year} {2020})}\BibitemShut {NoStop}%
\bibitem [{\citenamefont {{Thorne}}\ and\ \citenamefont {{Dykla}}(1971)}]{1971ApJ...166L..35T}%
  \BibitemOpen
  \bibfield  {author} {\bibinfo {author} {\bibfnamefont {K.~S.}\ \bibnamefont {{Thorne}}}\ and\ \bibinfo {author} {\bibfnamefont {J.~J.}\ \bibnamefont {{Dykla}}},\ }\href {\doibase 10.1086/180734} {\bibfield  {journal} {\bibinfo  {journal} {\apjl}\ }\textbf {\bibinfo {volume} {166}},\ \bibinfo {pages} {L35} (\bibinfo {year} {1971})}\BibitemShut {NoStop}%
\bibitem [{\citenamefont {Sotiriou}\ and\ \citenamefont {Faraoni}(2012)}]{Sotiriou:2011dz}%
  \BibitemOpen
  \bibfield  {author} {\bibinfo {author} {\bibfnamefont {T.~P.}\ \bibnamefont {Sotiriou}}\ and\ \bibinfo {author} {\bibfnamefont {V.}~\bibnamefont {Faraoni}},\ }\href {\doibase 10.1103/PhysRevLett.108.081103} {\bibfield  {journal} {\bibinfo  {journal} {Phys. Rev. Lett.}\ }\textbf {\bibinfo {volume} {108}},\ \bibinfo {pages} {081103} (\bibinfo {year} {2012})},\ \Eprint {http://arxiv.org/abs/1109.6324} {arXiv:1109.6324 [gr-qc]} \BibitemShut {NoStop}%
\bibitem [{\citenamefont {Hawking}(1972)}]{Hawking:1972qk}%
  \BibitemOpen
  \bibfield  {author} {\bibinfo {author} {\bibfnamefont {S.~W.}\ \bibnamefont {Hawking}},\ }\href {\doibase 10.1007/BF01877518} {\bibfield  {journal} {\bibinfo  {journal} {Commun. Math. Phys.}\ }\textbf {\bibinfo {volume} {25}},\ \bibinfo {pages} {167} (\bibinfo {year} {1972})}\BibitemShut {NoStop}%
\bibitem [{\citenamefont {Pasterski}\ \emph {et~al.}(2016)\citenamefont {Pasterski}, \citenamefont {Strominger},\ and\ \citenamefont {Zhiboedov}}]{Pasterski:2015tva}%
  \BibitemOpen
  \bibfield  {author} {\bibinfo {author} {\bibfnamefont {S.}~\bibnamefont {Pasterski}}, \bibinfo {author} {\bibfnamefont {A.}~\bibnamefont {Strominger}}, \ and\ \bibinfo {author} {\bibfnamefont {A.}~\bibnamefont {Zhiboedov}},\ }\href {\doibase 10.1007/JHEP12(2016)053} {\bibfield  {journal} {\bibinfo  {journal} {JHEP}\ }\textbf {\bibinfo {volume} {12}},\ \bibinfo {pages} {053} (\bibinfo {year} {2016})},\ \Eprint {http://arxiv.org/abs/1502.06120} {arXiv:1502.06120 [hep-th]} \BibitemShut {NoStop}%
\bibitem [{\citenamefont {Du}\ and\ \citenamefont {Nishizawa}(2016)}]{Du:2016hww}%
  \BibitemOpen
  \bibfield  {author} {\bibinfo {author} {\bibfnamefont {S.~M.}\ \bibnamefont {Du}}\ and\ \bibinfo {author} {\bibfnamefont {A.}~\bibnamefont {Nishizawa}},\ }\href {\doibase 10.1103/PhysRevD.94.104063} {\bibfield  {journal} {\bibinfo  {journal} {Phys. Rev. D}\ }\textbf {\bibinfo {volume} {94}},\ \bibinfo {pages} {104063} (\bibinfo {year} {2016})},\ \Eprint {http://arxiv.org/abs/1609.09825} {arXiv:1609.09825 [gr-qc]} \BibitemShut {NoStop}%
\bibitem [{\citenamefont {Koyama}(2020)}]{Koyama:2020vfc}%
  \BibitemOpen
  \bibfield  {author} {\bibinfo {author} {\bibfnamefont {K.}~\bibnamefont {Koyama}},\ }\href {\doibase 10.1103/PhysRevD.102.021502} {\bibfield  {journal} {\bibinfo  {journal} {Phys. Rev. D}\ }\textbf {\bibinfo {volume} {102}},\ \bibinfo {pages} {021502} (\bibinfo {year} {2020})},\ \Eprint {http://arxiv.org/abs/2006.15914} {arXiv:2006.15914 [gr-qc]} \BibitemShut {NoStop}%
\bibitem [{\citenamefont {Heisenberg}\ \emph {et~al.}(2023)\citenamefont {Heisenberg}, \citenamefont {Yunes},\ and\ \citenamefont {Zosso}}]{Heisenberg:2023prj}%
  \BibitemOpen
  \bibfield  {author} {\bibinfo {author} {\bibfnamefont {L.}~\bibnamefont {Heisenberg}}, \bibinfo {author} {\bibfnamefont {N.}~\bibnamefont {Yunes}}, \ and\ \bibinfo {author} {\bibfnamefont {J.}~\bibnamefont {Zosso}},\ }\href {\doibase 10.1103/PhysRevD.108.024010} {\bibfield  {journal} {\bibinfo  {journal} {Phys. Rev. D}\ }\textbf {\bibinfo {volume} {108}},\ \bibinfo {pages} {024010} (\bibinfo {year} {2023})},\ \Eprint {http://arxiv.org/abs/2303.02021} {arXiv:2303.02021 [gr-qc]} \BibitemShut {NoStop}%
\bibitem [{\citenamefont {Lang}(2014)}]{Lang:2013fna}%
  \BibitemOpen
  \bibfield  {author} {\bibinfo {author} {\bibfnamefont {R.~N.}\ \bibnamefont {Lang}},\ }\href {\doibase 10.1103/PhysRevD.89.084014} {\bibfield  {journal} {\bibinfo  {journal} {Phys. Rev. D}\ }\textbf {\bibinfo {volume} {89}},\ \bibinfo {pages} {084014} (\bibinfo {year} {2014})},\ \Eprint {http://arxiv.org/abs/1310.3320} {arXiv:1310.3320 [gr-qc]} \BibitemShut {NoStop}%
\bibitem [{\citenamefont {Lang}(2015)}]{Lang:2014osa}%
  \BibitemOpen
  \bibfield  {author} {\bibinfo {author} {\bibfnamefont {R.~N.}\ \bibnamefont {Lang}},\ }\href {\doibase 10.1103/PhysRevD.91.084027} {\bibfield  {journal} {\bibinfo  {journal} {Phys. Rev. D}\ }\textbf {\bibinfo {volume} {91}},\ \bibinfo {pages} {084027} (\bibinfo {year} {2015})},\ \Eprint {http://arxiv.org/abs/1411.3073} {arXiv:1411.3073 [gr-qc]} \BibitemShut {NoStop}%
\bibitem [{\citenamefont {Tahura}\ \emph {et~al.}(2021{\natexlab{b}})\citenamefont {Tahura}, \citenamefont {Nichols},\ and\ \citenamefont {Yagi}}]{Tahura:2021hbk}%
  \BibitemOpen
  \bibfield  {author} {\bibinfo {author} {\bibfnamefont {S.}~\bibnamefont {Tahura}}, \bibinfo {author} {\bibfnamefont {D.~A.}\ \bibnamefont {Nichols}}, \ and\ \bibinfo {author} {\bibfnamefont {K.}~\bibnamefont {Yagi}},\ }\href {\doibase 10.1103/PhysRevD.104.104010} {\bibfield  {journal} {\bibinfo  {journal} {Phys. Rev. D}\ }\textbf {\bibinfo {volume} {104}},\ \bibinfo {pages} {104010} (\bibinfo {year} {2021}{\natexlab{b}})},\ \Eprint {http://arxiv.org/abs/2107.02208} {arXiv:2107.02208 [gr-qc]} \BibitemShut {NoStop}%
\bibitem [{\citenamefont {Hou}\ and\ \citenamefont {Zhu}(2021{\natexlab{a}})}]{Hou:2020tnd}%
  \BibitemOpen
  \bibfield  {author} {\bibinfo {author} {\bibfnamefont {S.}~\bibnamefont {Hou}}\ and\ \bibinfo {author} {\bibfnamefont {Z.-H.}\ \bibnamefont {Zhu}},\ }\href {\doibase 10.1007/JHEP01(2021)083} {\bibfield  {journal} {\bibinfo  {journal} {JHEP}\ }\textbf {\bibinfo {volume} {01}},\ \bibinfo {pages} {083} (\bibinfo {year} {2021}{\natexlab{a}})},\ \Eprint {http://arxiv.org/abs/2005.01310} {arXiv:2005.01310 [gr-qc]} \BibitemShut {NoStop}%
\bibitem [{\citenamefont {Hou}\ and\ \citenamefont {Zhu}(2021{\natexlab{b}})}]{Hou:2020wbo}%
  \BibitemOpen
  \bibfield  {author} {\bibinfo {author} {\bibfnamefont {S.}~\bibnamefont {Hou}}\ and\ \bibinfo {author} {\bibfnamefont {Z.-H.}\ \bibnamefont {Zhu}},\ }\href {\doibase 10.1088/1674-1137/abd087} {\bibfield  {journal} {\bibinfo  {journal} {Chin. Phys. C}\ }\textbf {\bibinfo {volume} {45}},\ \bibinfo {pages} {023122} (\bibinfo {year} {2021}{\natexlab{b}})},\ \Eprint {http://arxiv.org/abs/2008.05154} {arXiv:2008.05154 [gr-qc]} \BibitemShut {NoStop}%
\bibitem [{\citenamefont {Hou}(2021)}]{Hou:2020xme}%
  \BibitemOpen
  \bibfield  {author} {\bibinfo {author} {\bibfnamefont {S.}~\bibnamefont {Hou}},\ }\href {\doibase 10.1002/asna.202113887} {\bibfield  {journal} {\bibinfo  {journal} {Astron. Nachr.}\ }\textbf {\bibinfo {volume} {342}},\ \bibinfo {pages} {96} (\bibinfo {year} {2021})},\ \Eprint {http://arxiv.org/abs/2011.02087} {arXiv:2011.02087 [gr-qc]} \BibitemShut {NoStop}%
\bibitem [{\citenamefont {De~Amicis}\ \emph {et~al.}(2024)\citenamefont {De~Amicis}, \citenamefont {Albanesi},\ and\ \citenamefont {Carullo}}]{DeAmicis:2024not}%
  \BibitemOpen
  \bibfield  {author} {\bibinfo {author} {\bibfnamefont {M.}~\bibnamefont {De~Amicis}}, \bibinfo {author} {\bibfnamefont {S.}~\bibnamefont {Albanesi}}, \ and\ \bibinfo {author} {\bibfnamefont {G.}~\bibnamefont {Carullo}},\ }\href@noop {} {\  (\bibinfo {year} {2024})},\ \Eprint {http://arxiv.org/abs/2406.17018} {arXiv:2406.17018 [gr-qc]} \BibitemShut {NoStop}%
\bibitem [{\citenamefont {Islam}\ \emph {et~al.}(2024)\citenamefont {Islam}, \citenamefont {Faggioli}, \citenamefont {Khanna}, \citenamefont {Field}, \citenamefont {van~de Meent},\ and\ \citenamefont {Buonanno}}]{Islam:2024vro}%
  \BibitemOpen
  \bibfield  {author} {\bibinfo {author} {\bibfnamefont {T.}~\bibnamefont {Islam}}, \bibinfo {author} {\bibfnamefont {G.}~\bibnamefont {Faggioli}}, \bibinfo {author} {\bibfnamefont {G.}~\bibnamefont {Khanna}}, \bibinfo {author} {\bibfnamefont {S.~E.}\ \bibnamefont {Field}}, \bibinfo {author} {\bibfnamefont {M.}~\bibnamefont {van~de Meent}}, \ and\ \bibinfo {author} {\bibfnamefont {A.}~\bibnamefont {Buonanno}},\ }\href@noop {} {\  (\bibinfo {year} {2024})},\ \Eprint {http://arxiv.org/abs/2407.04682} {arXiv:2407.04682 [gr-qc]} \BibitemShut {NoStop}%
\bibitem [{\citenamefont {Carullo}\ and\ \citenamefont {De~Amicis}(2023)}]{Carullo:2023tff}%
  \BibitemOpen
  \bibfield  {author} {\bibinfo {author} {\bibfnamefont {G.}~\bibnamefont {Carullo}}\ and\ \bibinfo {author} {\bibfnamefont {M.}~\bibnamefont {De~Amicis}},\ }\href@noop {} {\  (\bibinfo {year} {2023})},\ \Eprint {http://arxiv.org/abs/2310.12968} {arXiv:2310.12968 [gr-qc]} \BibitemShut {NoStop}%
\bibitem [{\citenamefont {Allen}\ \emph {et~al.}(2004)\citenamefont {Allen}, \citenamefont {Buckmiller}, \citenamefont {Burko},\ and\ \citenamefont {Price}}]{Allen:2004js}%
  \BibitemOpen
  \bibfield  {author} {\bibinfo {author} {\bibfnamefont {E.~W.}\ \bibnamefont {Allen}}, \bibinfo {author} {\bibfnamefont {E.}~\bibnamefont {Buckmiller}}, \bibinfo {author} {\bibfnamefont {L.~M.}\ \bibnamefont {Burko}}, \ and\ \bibinfo {author} {\bibfnamefont {R.~H.}\ \bibnamefont {Price}},\ }\href {\doibase 10.1103/PhysRevD.70.044038} {\bibfield  {journal} {\bibinfo  {journal} {Phys. Rev. D}\ }\textbf {\bibinfo {volume} {70}},\ \bibinfo {pages} {044038} (\bibinfo {year} {2004})},\ \Eprint {http://arxiv.org/abs/gr-qc/0401092} {arXiv:gr-qc/0401092} \BibitemShut {NoStop}%
\bibitem [{\citenamefont {Wittek}\ \emph {et~al.}(2023)\citenamefont {Wittek} \emph {et~al.}}]{Wittek:2023nyi}%
  \BibitemOpen
  \bibfield  {author} {\bibinfo {author} {\bibfnamefont {N.~A.}\ \bibnamefont {Wittek}} \emph {et~al.},\ }\href {\doibase 10.1103/PhysRevD.108.024041} {\bibfield  {journal} {\bibinfo  {journal} {Phys. Rev. D}\ }\textbf {\bibinfo {volume} {108}},\ \bibinfo {pages} {024041} (\bibinfo {year} {2023})},\ \Eprint {http://arxiv.org/abs/2304.05329} {arXiv:2304.05329 [gr-qc]} \BibitemShut {NoStop}%
\bibitem [{\citenamefont {Wittek}\ \emph {et~al.}(2024)\citenamefont {Wittek}, \citenamefont {Pound}, \citenamefont {Pfeiffer},\ and\ \citenamefont {Barack}}]{Wittek:2024gxn}%
  \BibitemOpen
  \bibfield  {author} {\bibinfo {author} {\bibfnamefont {N.~A.}\ \bibnamefont {Wittek}}, \bibinfo {author} {\bibfnamefont {A.}~\bibnamefont {Pound}}, \bibinfo {author} {\bibfnamefont {H.~P.}\ \bibnamefont {Pfeiffer}}, \ and\ \bibinfo {author} {\bibfnamefont {L.}~\bibnamefont {Barack}},\ }\href@noop {} {\  (\bibinfo {year} {2024})},\ \Eprint {http://arxiv.org/abs/2403.08864} {arXiv:2403.08864 [gr-qc]} \BibitemShut {NoStop}%
\end{thebibliography}%

\end{document}